\documentclass[12pt,fleqn]{ucithesis}

\usepackage{amsmath}
\usepackage{amsthm}
\usepackage{amssymb}
\usepackage{array}
\usepackage{graphicx}
\usepackage[numbers]{natbib}
\usepackage{relsize}
\usepackage[titletoc]{appendix}
\usepackage[table]{xcolor} 

\usepackage{caption}
\usepackage{subcaption}  

\usepackage{algorithmicx}
\usepackage{algorithm,algpseudocode}
\algnewcommand{\Initialize}[1]{%
  \State \textbf{Initialization in design time:}
  \Statex \hspace*{\algorithmicindent}\parbox[t]{.7\linewidth}{\raggedright #1}
}

\algnewcommand{\Initializes}[1]{%
  \State \textbf{Initialization:}
  \Statex \hspace*{\algorithmicindent}\parbox[t]{.5\linewidth}{\raggedright #1}
}
\algnewcommand{\Observe}[1]{%
  \State \textbf{From the \textit{Observe}:}
  \Statex \hspace*{\algorithmicindent}\hspace*{\algorithmicindent}\hspace*{\algorithmicindent}\parbox[t]{.99\linewidth}{\raggedright #1}
}

\algnewcommand{\Act}[1]{%
  \State \textbf{To the \textit{Act}:}
  \Statex \hspace*{\algorithmicindent}\hspace*{\algorithmicindent}\parbox[t]{.99\linewidth}{\raggedright #1}
}

\algnewcommand{\RMO}[1]{%
  \State \textbf{From \textit{Resource Monitoring}:}
  \Statex \hspace*{\algorithmicindent}\hspace*{\algorithmicindent}\parbox[t]{0.5\linewidth}{\raggedright #1}
}

\algnewcommand{\UQ}[1]{%
  \State \textbf{To \textit{Updating Qtable}:}
  \Statex \hspace*{\algorithmicindent}\hspace*{\algorithmicindent}\parbox[t]{0.9\linewidth}{\raggedright #1}
}

\algnewcommand{\UQN}[1]{%
  \State \textbf{To \textit{Updating Q-Network}:}
  \Statex \hspace*{\algorithmicindent}\hspace*{\algorithmicindent}\parbox[t]{0.9\linewidth}{\raggedright #1}
}

\usepackage{multirow}
\usepackage{tabularx}
\usepackage{xcolor}
\usepackage{enumitem}
\usepackage{pifont}

\usepackage{rotating}
\usepackage{booktabs}
\usepackage{pgfplots}
\usetikzlibrary{patterns}
\pgfplotsset{width=6cm, compat=1.6}
\pgfplotscreateplotcyclelist{my black white}{%
solid, every mark/.append style={solid, fill=gray}, mark=*\\%
dotted, every mark/.append style={solid, fill=gray}, mark=square*\\%
densely dotted, every mark/.append style={solid, fill=gray}, mark=otimes*\\%
loosely dotted, every mark/.append style={solid, fill=gray}, mark=triangle*\\%
dashed, every mark/.append style={solid, fill=gray},mark=diamond*\\%
loosely dashed, every mark/.append style={solid, fill=gray},mark=*\\%
densely dashed, every mark/.append style={solid, fill=gray},mark=square*\\%
dashdotted, every mark/.append style={solid, fill=gray},mark=otimes*\\%
dashdotdotted, every mark/.append style={solid},mark=star\\%
densely dashdotted,every mark/.append style={solid, fill=gray},mark=diamond*\\%
}
\pgfplotscreateplotcyclelist{my color}{%
blue,every mark/.append style={fill=blue!80!black},mark=*\\%
red,every mark/.append style={fill=red!80!black},mark=square*\\%
brown!60!black,every mark/.append style={fill=brown!80!black},mark=otimes*\\%
black,mark=star\\%
blue,every mark/.append style={fill=blue!80!black},mark=diamond*\\%
red,densely dashed,every mark/.append style={solid,fill=red!80!black},mark=*\\%
brown!60!black,densely dashed,every mark/.append style={
solid,fill=brown!80!black},mark=square*\\%
black,densely dashed,every mark/.append style={solid,fill=gray},mark=otimes*\\%
blue,densely dashed,mark=star,every mark/.append style=solid\\%
red,densely dashed,every mark/.append style={solid,fill=red!80!black},mark=diamond*\\%
}

\definecolor{rd1}{rgb}{.65,0,.15}
\definecolor{rd2}{rgb}{.84,.19,.15}
\definecolor{orng1}{rgb}{.96,.43,.26}
\definecolor{orng2}{rgb}{.99,.68,.004}
\definecolor{yl1}{rgb}{.99,.88,.55}
\definecolor{yl2}{rgb}{1,1,.75}
\definecolor{gr1}{rgb}{.4,.74,.39}
\definecolor{gr2}{rgb}{.65,.85,.42}
\definecolor{grr1}{rgb}{0,.41,.22}
\definecolor{grr2}{rgb}{.1,.6,.31}
\usetikzlibrary{pgfplots.colorbrewer,}
\pgfplotsset{
    cycle list/.define={my marks}{
        every mark/.append style={solid,fill=\pgfkeysvalueof{/pgfplots/mark list fill}},mark=*\\
        every mark/.append style={solid,fill=\pgfkeysvalueof{/pgfplots/mark list fill}},mark=square*\\
        every mark/.append style={solid,fill=\pgfkeysvalueof{/pgfplots/mark list fill}},mark=triangle*\\
        every mark/.append style={solid,fill=\pgfkeysvalueof{/pgfplots/mark list fill}},mark=diamond*\\
    },
}

\usepackage[plainpages=false,pdfborder={0 0 0}]{hyperref}
\thesistitle{An Integrated Framework for Contextual Personalized LLM-Based Food Recommendation}

\documenttitle{Dissertation}

\degreename{Doctor of Philosophy}

\degreefield{Computer Science}

\authorname{Ali Rostami}

\committeechair{Professor Amir M. Rahmani}
\othercommitteemembers
{
  Professor Ramesh Jain\\
  Professor Nikil Dutt
  
}

\degreeyear{2024}

\copyrightdeclaration
{
  {\copyright} {\Degreeyear} \Authorname
}

\prepublishedcopyrightdeclaration
{
	Partial content {\copyright} 2020 ACM \\
	Partial content {\copyright} 2021 Springer \\
	Partial content {\copyright} 2021 ACM \\
	Partial content {\copyright} 2022 ACM \\
        Partial content {\copyright} 2024 IEEE \\
	All other materials {\copyright} {\Degreeyear} \Authorname
}

\dedications
{
  I dedicate this thesis to my lovely parents and my brother Mohammad for always being there to support me, motivate me, and give me sincere advice. Last but not least I dedicate this thesis to my true role model Imam Al-Zaman Hujjat ibn Hasan and his pure mother lady Fatima al-Seddiqah al-Zahra peace be upon them both. I'm looking forward to meeting them sooner. Undeniably, I truly owe all my academic achievements to them, and this dedication is the smallest gratitude that I could express to them.
}

\acknowledgments
{
    I was co-advised by two of the best advisors I could ask for. So first and foremost, I would like to double thank both of them: Prof. Ramesh Jain and Prof. Amir M. Rahmani for their consistent support from the beginning of my PhD. journey up to its very last moments. 
    
    Prof. Ramesh Jain's unwavering assistance, indispensable advice, and technical instructions showed me the proper direction to excel in this project. My sincere gratitude goes to him because he devotedly cooperated throughout my work.
    
  Also special thanks to Prof. Amir M. Rahmani not only for his priceless time, patience and kindness, but also the brilliant opinions through our meetings which have been undoubtedly effective.

  My appreciation also extends to committee member, Prof. Nikil Dutt for his participation in the thesis committee and providing helpful feedback, which directly or indirectly contributed to this project.

  I would like to shout out to my colleagues at HST Labs, Iman Azimi, Mahyar Abbasian, Emad Kasaeyan Naeini and Sina Shahhosseini for their friendship and support.
}

\newcommand{\mypubentry}[3]{
  \begin{tabular*}{1\textwidth}{@{\extracolsep{\fill}}p{4.5in}r}
    \textbf{#1} & \textbf{#2} \\ 
    \multicolumn{2}{@{\extracolsep{\fill}}p{.95\textwidth}}{#3}\vspace{6pt} \\
  \end{tabular*}
}
\newcommand{\mysoftentry}[3]{
  \begin{tabular*}{1\textwidth}{@{\extracolsep{\fill}}lr}
    \textbf{#1} & \url{#2} \\
    \multicolumn{2}{@{\extracolsep{\fill}}p{.95\textwidth}}
    {\emph{#3}}\vspace{-6pt} \\
  \end{tabular*}
}

\curriculumvitae
{

\textbf{EDUCATION}
  
  \begin{tabular*}{1\textwidth}{@{\extracolsep{\fill}}lr}
    \textbf{Doctor of Philosophy in Computer Science} & \textbf{2024} \\
    \vspace{6pt}
    University of California, Irvine & \emph{Irvine, CA} \\
    \textbf{Master of Science in Computer Science} & \textbf{2021} \\
    \vspace{6pt}
    University of California, Irvine & \emph{Irvine, CA} \\
    \textbf{Bachelor of Science in Computer Engineering} & \textbf{2017} \\
    \vspace{6pt}
    Sharif University of Technology & \emph{Tehran, Iran} \\
  \end{tabular*}

\vspace{12pt}
\textbf{RESEARCH EXPERIENCE}

  \begin{tabular*}{1\textwidth}{@{\extracolsep{\fill}}lr}
    \textbf{Graduate Student Researcher} & \textbf{2017--2024} \\
    \vspace{6pt}
    University of California, Irvine & \emph{Irvine, California} \\
  \end{tabular*}

\vspace{12pt}
\textbf{TEACHING EXPERIENCE}

  \begin{tabular*}{1\textwidth}{@{\extracolsep{\fill}}lr}
    \textbf{Teaching Assistant} & \textbf{2018--2024} \\
    \vspace{6pt}
    University of California, Irvine & \emph{Irvine, California} \\
  \end{tabular*}

\pagebreak

\vspace{12pt}
\textbf{REFEREED CONFERENCE PUBLICATIONS}

  \mypubentry{Personal food model}{Oct 2020}{ACM International Conference on Multimedia}
  \mypubentry{Multimedia food logger}{Oct 2020}{ACM International Conference on Multimedia}
    \mypubentry{Event mining driven context-aware personal food preference modelling}{Feb 2021}{ICPR International Workshops and Challenges}
    \mypubentry{World Food Atlas Project}{Aug 2021}{International Workshop on Multimedia for Cooking and Eating Activities}
    \mypubentry{World Food Atlas for Food Navigation}{Oct 2022}{International Workshop on Multimedia Assisted Dietary Management}

\mypubentry{Food Recommendation as Language Processing (F-RLP): A Personalized and Contextual Paradigm}{Under Review}{International Conference of the IEEE Engineering in Medicine and Biology Society}

\vspace{12pt}
\textbf{SOFTWARE}

  \mysoftentry{T5}{https://huggingface.co/google-t5/t5-base}
  {Text-To-Text Transfer Transformer}
  
    \mysoftentry{P5}{https://github.com/jeykigung/P5}
  {Large language model (LLM) for recommendation}

}

\thesisabstract
{
Personalized food recommendation systems (Food-RecSys) are recognized for their potential to enhance dietary practices and address pressing health issues. Despite their promise, current food recommendation methodologies, predominantly rule-based and classification-driven, encounter substantial barriers in achieving practical applicability. This challenge is largely attributed to two main factors: the poor understanding of food recommendation specific components as a whole and inadequate research on a holistic integration between all the key components which enables the final methodology to achieve effective results, and also, the limitations of conventional machine learning models, which falter in the face of a virtually limitless array of food categories and the inherent imbalance within datasets. In this context, the advent of Large Language Models (LLMs) presents an intriguing alternative, suggesting a path forward for recommendation systems. Nevertheless, existing approaches that employ LLMs generally adopt a generic Recommendation as Language Processing (RLP) strategy, which falls short of incorporating the nuanced components essential for successful Food-RecSys.

To bridge this research gap, this thesis provides a comprehensive overview on an intensely complex big picture, identifies the key components and provides innovative exploratory models for each the multiplex key components required for personalized contextual food recommendation, additionally, articulates a novel feasibility study of a comprehensive integrated personal contextual Food-RecSys framework, underpinned by several pivotal innovations aimed at fortifying the effectiveness of personalized food recommendations. Key among these is the development of a multimedia food logging platform and the World Food Atlas, the latter of which facilitates geolocation-based food queries—a capability notably absent in current offerings. Moreover, this work pioneers the Food Recommendation as Language Processing (F-RLP) framework. F-RLP represents a bespoke solution that adeptly harnesses the strengths of LLMs, furnishing a food-specific recommendation infrastructure that transcends the limitations of generic models.
}

\hypersetup{
	pdftitle={\Thesistitle},
	pdfauthor={\Authorname},
	pdfsubject={\Degreefield},
}

\begin{document}
\preliminarypages

\chapter{Introduction}

\section{Overview}


In the contemporary landscape, characterized by the prevalence of digital assistants and customized user experiences, the domain of artificial intelligence in personalized Food Recommendation Systems (Food-RecSys) has arisen as a significant influence on our dietary explorations \cite{min2019_food_recommendation}. These systems extend beyond the realm of mere convenience, harboring the capacity to enhance nutritional decisions, mitigate nutritional deficits, and contribute to the prevention and management of chronic health conditions \cite{min2019survey}, \cite{min2019_food_recommendation}.

The umbrella of food encompasses an immensely broad spectrum of categories, each with its own distinct facets, essentially rendering the space of potential classes virtually boundless. This diversity poses a significant challenge to classification-based methodologies in crafting effective and practical Food-RecSys that account for the extensive variety of food classes, their characteristics, and the crucial element of personalization \cite{min2019survey}, \cite{rostami2020_personal_food_model}.

Current Food-RecSys encompasses a diverse array of classification based methods, from content-based filtering relying on user-rated recipes to collaborative filtering leveraging shared preferences \cite{ornab2017empirical}. While intriguing in their conceptualization, these methodologies encounter substantial limitations when confronted with the expansive variety of categories and the multidimensional nature of Food-RecSys, rendering them inefficient for practical, daily applications. The large number of classes and complicated dimensions in FR systems make it clear that we need more advanced methods that can handle these difficulties. This makes us wonder if these methods can really be used effectively in real life \cite{jain2022personalized}.

However, progress in Large Language Models (LLMs) has shown that they are good at recommendation tasks that involve a huge number of categories \cite{geng2022recommendation}. This is because they are good at dealing with a lot of options. Despite the potential of LLM-based recommendations in navigating the intricate web of food categories, a number of barriers prevent their application in providing useful, real-world food recommendations \cite{t5}. Notably, these include limitations in logical reasoning, real-time geographic awareness, and the incorporation of unique, domain-specific components vital for the accuracy and performance of Food-RecSys \cite{wu2023computing}. Among the novel components introduced in this work are Multimedia Food Logger (MFL) \cite{Rostami2020FL} and the World Food Atlas (WFA) \cite{Rostami21WFA}. Moreover, the absence of a comprehensive framework tailored to LLM-based food recommendations remains a notable gap in the literature. Some projects, like the Food Recommendation Language Processing (F-RLP) approach, are taking steps toward using LLMs in this way, but they usually train their models using methods that are not LLM-specific, so they do not fully use LLMs for food recommendation tasks \cite{geng2022recommendation}.

In response, in this thesis we propose a holistic framework for LLM-based food recommendations that meticulously integrates domain-specific elements essential for food recommendations while leveraging the strengths of LLM machine learning models to manage an extensive range of food categories effectively as a comprehensive framework. We additionally show the depth of this huge problem and also present a novel feasibility study at the end as a prove of concept of an inherently almost infinitely complex \cite{Rostami20PFM} problem which interacts in every dimension with reality that we know of. Our framework encompasses a detailed exploration of all requisite components, offering a clear delineation of each element's role and significance, alongside a pre-training mechanism that harmonizes the needs of food recommendation with the advantages of LLM-based approaches.

\section{Research aim and objective}
The research aim of this scholarly inquiry is two-fold.

One is to elucidate the essential prerequisites for efficacious Food-RecSys, elements that have been hitherto underemphasized across various dimensions of study. This endeavor seeks to elucidate a coherent and actionable conceptualization of these integral components, such as the food logger and the world food atlas, delineating their pivotal functions and contributions within the architecture of Food-RecSys \cite{rostami2020_personal_food_model}. Additionally, this research aims to furnish a comprehensive exposition on the synergistic interplay among the diverse constituents of personal food recommendation platforms. It proposes an innovative framework that harmonizes these disparate elements, thereby elucidating the substantial advantages derivable from the meticulous design and application of these components within a unified system \cite{min2019survey}.

An additional scholarly pursuit of this thesis is to tackle a prevalent challenge within the realm of personalized Food-RecSys: the inadequacy of conventional classification techniques to effectively manage the categorization within an extensively large and multidimensional space of food-related attributes \cite{ali2013classification}. This investigation aspires to harness the capabilities of Large Language Models (LLMs), renowned for their proficiency in navigating vast expanses of possibilities inherent to tasks like language generation, to surmount this hurdle \cite{zhou2020laplacian}. A critical objective encompasses synthesizing the broad applicability of LLMs in general personal recommendation scenarios with the distinctive features and exigencies of personalized food recommendation domains. The intricacies of food recommendation present a singularly unique challenge, prompting the development of an LLM-based recommendation framework meticulously crafted to not only mitigate inherent drawbacks such as hallucination phenomena and logical inaccuracies but also to transform these issues into strategic advantages \cite{rostami2024food}. This entails the construction of an innovative architecture that synergistically integrates the strengths of LLMs with the nuanced requirements of personalized Food-RecSys, thereby fostering a novel and efficacious methodology as a preliminary prove of concepts which God willing paves the way for much more complex systems in the future.

\section{Thesis Contributions}

This thesis delineates several pivotal contributions to the field of personalized Food-RecSys, employing a comprehensive academic lexicon and integrating novel references and methodologies:

\begin{itemize}
    \item \textbf{Elaboration on Critical System Components:} This research identifies and elaborates on the key components indispensable for the efficacy and precision of Food-RecSys. Unlike general-purpose LLM-based recommendations, which may not account for the nuanced requirements of the food recommendation domain, this study highlights and details the most critical elements that underpin high-performing food recommendation services \cite{min2019survey}.
    \item \textbf{Development of a Specialized Food Logging Platform:} The thesis introduces an advanced specification for food logging, a component vital for personalized food recommendations. The proposed platform offers a refined approach to capturing dietary preferences and consumption patterns, tailored specifically for enhancing personalized recommendation accuracy.

    \item \textbf{Inception of the World Food Atlas:} Acknowledging the necessity of a geospatial database for practical food recommendation, this research pioneers the World Food Atlas. This innovative server aggregates diverse food data sources into a geospatial atlas, enabling user queries based on geographical location, thereby addressing a significant gap in existing recommendation frameworks.

    \item \textbf{Formulation of a Holistic Framework:} Prior works have not presented a comprehensive framework for LLM-based food recommendations. This thesis critiques existing methodologies for their partial utilization of food recommendation components and proposes a prove of concept version of a holistic framework that paves the way for fully leveraging LLM potential while detailing the integral structure of the recommendation system as a comprehensive integrated system., including data organization and processing methodologies to preclude hallucination phenomena.

    \item \textbf{Integration of Food-Specific Models with LLM:} Recognizing the unique challenges of food recommendations, this work integrates specialized models with LLMs to navigate the complex personalization dimensions inherent in food preferences, involving diverse data streams and processed information.

    \item \textbf{Ensuring LLM Recommendation Reliability:} Addressing the limitations of LLMs in understanding logical coherence and real-time geographical context, this thesis introduces mechanisms such as food decoding and option list generation interfaces. These tools constrain LLM recommendations to a reliable set of options, enhancing result relevancy and mitigating the risk of erroneous outputs.

    \item \textbf{Incorporation of Essential External Food Servers:} Highlighting the challenges LLMs face in accessing and processing data from external food servers, such as injecting knowledge from the World Food Atlas and the Food Knowledge Graph, this research explores innovative solutions for integrating external server data into the LLM recommendation process, marking a significant advancement in the field.

    \item \textbf{Pioneering Research in LLM-based Food Recommendations:} By venturing into the relatively unexplored territory of LLM-based food recommendation, this thesis sets the groundwork for a novel research domain. It showcases the suitability of LLMs for handling the vast complexity and variety inherent in personalized food recommendation tasks, thereby laying a foundation for future scholarly endeavors in this promising intersection of technology and gastronomy.
\end{itemize}


\section{Organization}
\begin{figure*}[ht]
 \centering
\includegraphics[width=1\linewidth]{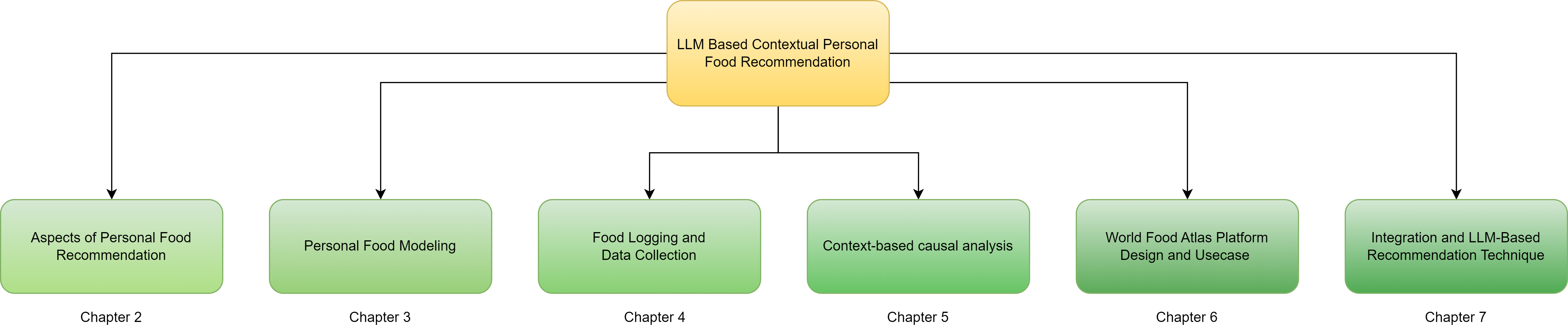}
\caption{Structure of the thesis}
\label{fig:main_fig}
\end{figure*}

The subsequent sections of this dissertation, as illustrated in Figure \ref{fig:main_fig}, are methodically organized to facilitate a comprehensive understanding of Personal Food Recommendation and LLM integration. The structure is outlined as follows:

\begin{itemize}
    \item \textbf{Chapter 2:} This section introduces the concept of Personal Food Recommendation, situating it within the expansive domain of food computing. It surveys existing scholarly efforts related to personal Food-RecSys and LLM-based recommendation frameworks, concluding with a succinct overview of the key components integral to personal food recommendations.
    \item \textbf{Chapter 3:} A pioneering personal food model is delineated, encapsulating digitized representations of user-specific food attributes. This model bifurcates into the biological and preferential personal food models, elucidating their contributions within the overarching framework and offering a case study to exemplify its application.
    \item \textbf{Chapter 4:}  An elaborate schema for data collection tailored to personal food recommendation is introduced, with an emphasis on a food logging platform as the cornerstone of personal data acquisition. This chapter delves into the spectrum of food-related data requisite for tailoring effective personal food recommendations.
    \item \textbf{Chapter 5:} A discourse on context-based causal analysis in food decision-making processes is presented, alongside a novel schema for comprehensive food event context representation. This analysis examines the multifaceted impact of various factors on food choices, supplemented by an experimental study to substantiate these insights.
    \item \textbf{Chapter 6:} The World Food Atlas is proposed as a crucial component for any genuinely functional personal Food-RecSys. This chapter highlights the necessity of real-time, location-based food queries and outlines a new schema and platform architecture for the World Food Atlas.
    \item \textbf{Chapter 7:} A groundbreaking framework that amalgamates the disparate elements of personal food recommendations with the capabilities of LLMs is proposed. This section scrutinizes the hurdles associated with such integration, particularly the inherent limitations of LLMs, and proposes a hybrid framework to overcome these obstacles.
    \item \textbf{Chapter 8:} The dissertation concludes by reflecting on the strengths and limitations of the presented work and suggesting avenues for future research within the field of personalized food recommendations.

\end{itemize}




\chapter{An Overview on Food Recommendation}




\section{Background}
This segment initiates with a concise delineation of the quadruple stratifications inherent in Food Computing \cite{rostami2020_personal_food_model}, subsequently narrowing the focus to the user-centric stratum and delineating the placement of Personal Food Recommendation within this echelon. An extensive exposition on the multifaceted dimensions of Personal Food Recommendations and LLM (Large Language Model)-based recommendation frameworks ensues, incorporating an in-depth exploration of foundational principles pertinent to Recommender Systems and LLMs \cite{geng2022recommendation}. This is augmented by a rigorous review of scholarly literature addressing both domains of personal food recommendation and LLM-based recommendation methodologies. Conclusively, this section enumerates the essential constituents of Personal Food Recommendation, furnishing a succinct elucidation of each element’s function and significance within the broader recommendation architecture.

\subsection{Food Computing Overview}

Food computing is a multidisciplinary research field that covers various layers of food systems. The research can be categorized into four different layers: user-centric, dish-centric, food chain-centric, and environmental-centric \cite{rostami2020personal}. These layers provide different perspectives on the food system, each with its own set of challenges and opportunities.

Food computing is an emerging interdisciplinary field that combines food science, computer science, and artificial intelligence to develop computational methods for analyzing and modeling food-related data. It aims to create novel applications that could transform the food industry and benefit consumers by providing more efficient and effective ways of managing food production, distribution, and consumption \cite{min2019survey}.

One of the main challenges in food computing is the complexity and diversity of food-related data. This data can include information on food ingredients, recipes, nutrition, cooking techniques, and cultural factors. To make sense of this data, food computing researchers use various computational techniques such as machine learning, data mining, and natural language processing \cite{min2019survey}.

Food computing has a wide range of potential applications, from personalized nutrition and food recommendation systems to food safety and traceability. For example, personalized nutrition systems can use data on an individual's health status, lifestyle, and food preferences to generate personalized diet plans that could help prevent or manage chronic diseases. Food recommendation systems can use location-based and context-aware information to suggest restaurants and food options that match a user's taste, budget, and dietary requirements.

Some recent research in food computing has focused on developing methods for analyzing the impact of food consumption on the environment and sustainability. For example, the Food and Agriculture Organization of the United Nations (FAO) has proposed the concept of "food systems intelligence" that aims to integrate data from various sources to enable a more holistic and evidence-based approach to food systems analysis and decision-making \cite{von2023food}, \cite{how2020predictive}.
\begin{figure}[!ht]
  \centering
  
  \includegraphics[width=1\linewidth]{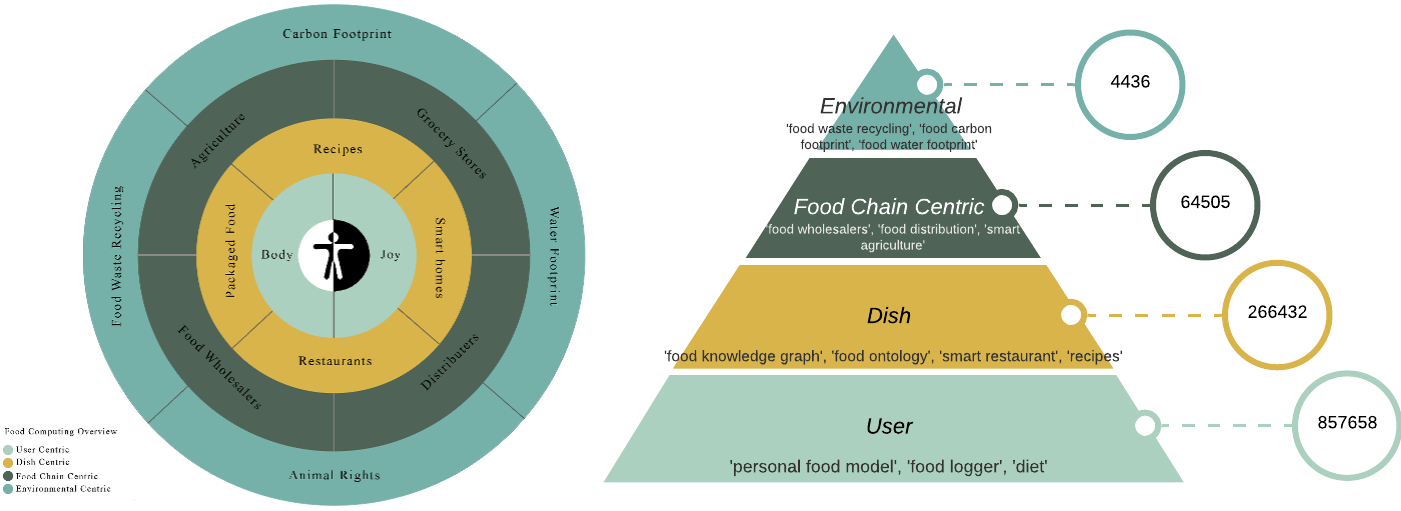}
    \caption{The food computing layers and the pyramid of the body of work. A simple search of publications in each layer by a few trivial keywords from each layer shows the focus of research tends to be more active in areas closer the centric layer demonstrating its more immediate importance in a human centric society.}
    \label{fig:pyramid}
\end{figure}

User-centric research focuses on the user's behavior, preferences, and health. This layer aims to personalize the food recommendations based on the user's goals, such as weight loss, health improvement, or cultural preferences. User-centric research also explores the use of sensors and wearables to monitor the user's food intake, eating patterns, and physiological responses to food. This layer involves several challenges, such as the accuracy of the sensor data, privacy concerns, and the user's willingness to adopt new technologies.

Dish-centric research focuses on the food itself, its ingredients, nutrition, and culinary properties. This layer aims to understand the dish's flavor, texture, and appearance and how they affect the user's enjoyment and satisfaction. Dish-centric research also explores the use of computer vision and machine learning techniques to recognize the dish's ingredients, estimate its nutritional value, and predict its sensory properties. This layer involves several challenges, such as the variability of the ingredients, the complexity of the recipes, and the subjectivity of the sensory evaluation.

Food chain-centric research focuses on the food production, distribution, and consumption chain. This layer aims to improve the efficiency, sustainability, and safety of the food system. Food chain-centric research also explores the use of blockchain, IoT, and AI technologies to track and trace the food products, monitor the environmental impact of food production, and prevent food fraud and contamination. This layer involves several challenges, such as the interoperability of the data sources, the trustworthiness of the information, and the scalability of the solutions.

Environmental-centric research focuses on the environmental impact of food systems, such as greenhouse gas emissions, water consumption, and land use. This layer aims to promote sustainable and regenerative food systems that minimize the negative impact on the environment and maximize the positive impact on the ecosystem. Environmental-centric research also explores the use of circular economy, agroforestry, and biodiversity conservation strategies to create resilient and diversified food systems. This layer involves several challenges, such as the trade-offs between different environmental indicators, the complexity of the food system, and the need for interdisciplinary collaboration.

Overall, the different layers of food computing research provide a holistic view of the food system, each with its own set of challenges and opportunities. Integrating these layers can lead to more comprehensive and personalized food solutions that promote health, sustainability, and social well-being.

\subsection{The User Centric Layer}
\begin{figure*}[t]
    \centering
    \includegraphics[width=1\textwidth]{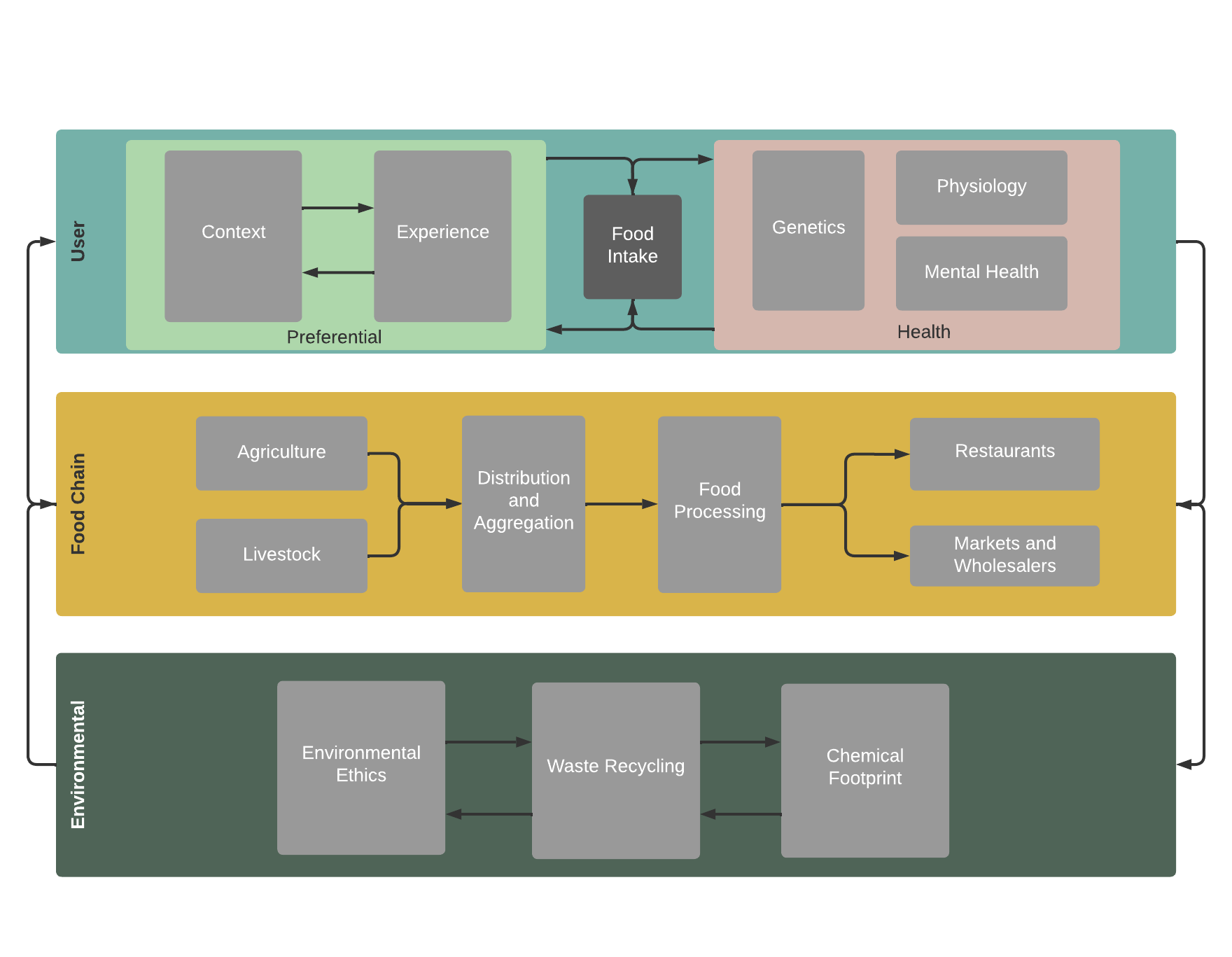}
    \caption{The interaction between the different layers of food computing shows that these layers are not isolated and will inevitably affect each other as they are interconnected. So a positive impact on all layer may start from the user layer's food intake choice based on preference and health.}
    \label{fig:layers}
\end{figure*}
The user-centric layer is the most critical layer of food computing, as it focuses on the individual's needs and preferences. In personalized food recommendation systems, the user-centric layer involves collecting data on an individual's dietary habits, food preferences, and lifestyle. This data is then used to develop a personal model that can generate personalized recommendations based on an individual's unique characteristics.

The user-centric layer is a personal model that forms the core of personalized food recommendation systems. It is based on the idea that each individual has unique food preferences, dietary restrictions, and health goals that must be taken into account when making food recommendations. By collecting and analyzing data about a user's food habits, the user-centric layer can generate personalized recommendations that are tailored to the user's needs and preferences.

Research has shown that personalization is a key factor in the success of food recommendation systems \cite{POTTER2021579}. Many studies have found that personalized food recommendations are more effective than generic recommendations in encouraging users to make healthy food choices \cite{Majjodi2022NudgingTH}, \cite{Starke2021PromotingHF}. Similarly, a study by Mustu et al. (2020) found that users were more likely to use a food recommendation system that provided personalized recommendations based on their individual tastes and preferences \cite{Musto2020TowardsAK}.

The user-centric layer typically includes several components, including a user profile, a food ontology, and a recommendation engine. The user profile contains information about the user's food preferences, dietary restrictions, and health goals, as well as their location and other contextual information. The food ontology is a structured representation of the types of food that are available, including their nutritional content and other relevant information. The recommendation engine uses algorithms and machine learning techniques to generate personalized recommendations based on the user's profile and the food ontology.

Several challenges must be addressed in order to build an effective user-centric layer. One of the main challenges is data collection and analysis. Collecting accurate and relevant data about a user's food habits and preferences can be difficult, and requires the use of various sensors and other data collection techniques. Analyzing this data and generating personalized recommendations also requires the use of advanced machine learning and artificial intelligence techniques.

Despite these challenges, the user-centric layer holds great promise for improving the effectiveness of food recommendation systems. By taking into account the unique needs and preferences of each individual user, these systems can help promote healthy eating habits and prevent chronic diseases such as obesity and diabetes. Further research is needed to explore the potential of the user-centric layer and to develop more effective techniques for collecting and analyzing data about food habits and preferences.

\subsection{Recommender Systems (RecSys)}
Recommender systems constitute a subclass of information filtering technologies designed to address the challenge of information overload in contemporary digital environments \cite{Roy2022}. These systems employ diverse algorithmic approaches to curate and suggest items, services, or content deemed most relevant to individual users \cite{DBLP:journals/corr/ZhangYS17aa}. Central to their function is the analysis of historical user-item interaction data, which may encompass explicit feedback (e.g., ratings, reviews) \cite{recsys2023} or implicit signals (e.g., browsing patterns, purchase history) \cite{signals}.  Broadly speaking, recommender systems leverage techniques such as collaborative filtering \cite{colabfilter}, which  searches for similarities in behavior across users, and content-based methods \cite{contentbased}, which analyze attributes and descriptions of both users and items.  Recent advancements incorporate deep learning models, including graph neural networks (GNNs) \cite{gnn} for superior representation learning, and large language models (LLMs) to enable natural language understanding and generation in the recommendation process \cite{fan2023recommender}.

Deep learning has revolutionized recommender systems due to its exceptional representation learning abilities \cite{10.1145/3285029}, \cite{Selma2021DeepLF}. Models like DSER demonstrate this by converting sequential user–item interactions with deep neural networks (DNNs) to uncover complex user-item relationships \cite{deepnetw}. Graph neural networks (GNNs) have also proven potent for recommender systems, skillfully modeling the inherently graph-structured nature of recommendation data. Textual knowledge further enhances these systems; DeepCoNN, for instance, employs CNNs to process user reviews, boosting the accuracy of rating predictions \cite{kaur2023deepconn}. 

The proliferation of language models within recommender systems underscores their adeptness at processing and generating natural language, heralding a paradigm shift in personalized recommendation capabilities \cite{10.1145/3624918.3629550}. These models demonstrate a nuanced understanding of human communication, thereby unlocking avenues for highly tailored recommendations across various domains, including but not limited to news dissemination and pharmaceuticals \cite{https://doi.org/10.48550/arxiv.2305.19860}. An exemplary manifestation of this trend is observed in BERT4Rec \cite{sun2019bert4rec}, wherein the formidable prowess of Bidirectional Encoder Representations from Transformers (BERT) \cite{devlin2018bert} is harnessed to model the intricate sequential patterns inherent in user behavior. Moreover, Transformer-based frameworks exhibit the remarkable capacity to not only recommend items but also provide cogent explanations concurrently, thereby enriching user trust and comprehension in the recommendation process.

\subsection{Large Language Models (LLMs)}
Large language models (LLMs), pivotal exemplars of advanced artificial intelligence (AI), undergo training on vast textual corpora to discern the intricate structures and relationships embedded within natural language.  Pre-trained models like BERT (Bidirectional Encoder Representations from Transformers) \cite{devlin2018bert}, GPT (Generative Pre-trained Transformer) \cite{DBLP:journals/corr/abs-2005-14165}, and T5 (Text-To-Text Transfer Transformer) \cite{raffel2020exploring} serve as foundational building blocks, offering diverse capabilities and architectures. Encoder-only models, exemplified by BERT, leverage bidirectional attention to glean contextual meaning from both preceding and subsequent tokens. This makes them adept at capturing subtle semantic nuances, critical for tasks like text classification and natural language inference. Decoder-only models, such as GPT, employ unidirectional attention to generate human-quality text, proving invaluable for tasks like text summarization and machine translation. Encoder-decoder models, like T5, offer exceptional versatility, reformulating diverse natural language processing (NLP) tasks as text-to-text problems, enabling solutions across a spectrum of domains \cite{fan2023recommender}.

The exponential growth in LLM scale has propelled a revolution in NLP, with models like GPT-3 \cite{DBLP:journals/corr/abs-2005-14165}, Jurassic-1 Jumbo \cite{lieber2021jurassic}, and Megatron-Turing NLG \cite{smith2022using} demonstrating previously unattainable levels of linguistic fluency and knowledge representation. Trained on massive datasets, these LLMs showcase remarkable in-context learning (ICL) capabilities  [48], understanding and responding dynamically to provided prompts rather than solely relying on pre-trained knowledge. This adaptability sets them apart from earlier NLP models. Techniques like chain-of-thought prompting (CoT) \cite{zhang2022automatic} further augment LLMs' reasoning capacity. By providing step-by-step demonstrations within the prompt itself, CoT guides the model's problem-solving process.  Recent extensions like self-consistency \cite{ahmed2023better}, which use majority voting across multiple generated responses, yield further improvements in LLM accuracy and reliability.

LLMs' far-reaching applications extend to fields as varied as biomedical simulators \cite{schaefer2023large}, legal analysis \cite{cui2023chatlaw}, and software development \cite{belzner2023large} and various other scientific analysis \cite{zheng2023large}. Within recommender systems (RecSys), LLMs like ChatGPT play transformative roles. Their ability to analyze user histories and predict item preferences [57], [58]  enables more accurate and personalized recommendations. LLMs power  sequential recommender systems \cite{harte2023leveraging}, \cite{boz2024improving}, forecasting user interactions based on past behavior. Importantly, LLMs are central to explainable recommendations \cite{gao2023chat}, where they articulate the reasoning behind suggested items, fostering user trust such as the educational chat bot presented in \cite{abu2024supporting} supporting student decisions.

The conversational nature of LLMs drives innovation in user-centric recommendation experiences.  Frameworks like MACRS \cite{fang2024multi} and CPR \cite{zhang2023user} integrate conversational interfaces with recommendation capabilities, employing prompt-based strategies to guide dialogue.  Additionally, researchers actively investigate the potential of LLMs in graph-based recommendation, recognizing their capacity to process complex relational data \cite{guo2024integrating}.

\section{Related Work}
This section contains a thorough assessment of the literature on a range of topics related to LLM-based recommenders and personalized food recommendations. We begin a thorough exploration to clarify the changing terrain of personalized recommendation systems. I begin by examining the historical development of recommendation systems.

 After conducting an in-depth investigation in the field of N of 1 food recommendation systems, we explore the current landscape of non-LLM-based food recommendations. I examine the unique features, methods, and constraints of existing meal recommendation systems. Moreover, we also investigate the complex elements incorporated in modern food recommendation systems, illuminating the various aspects that come together to create comprehensive recommendation systems.
 
 Moving on to the field of Language Model based recommendation systems, this section summarizes how recommendation systems have evolved from traditional models to the present paradigm shift that utilizes Language Model technologies. We specifically discuss the development and advancement of LM-based recommendation systems, highlighting their crucial function in contemporary recommendation frameworks. We study the underdeveloped field of LLM-based meal recommendations, noting a significant lack of focused research in this area.
This comprehensive analysis aims to outline the developmental path of personalized recommendation systems, emphasize the importance of LLM-based recommendations, and highlight the complexities of current non-LLM-based food recommendations and their various elements.

Lastly, we provide the present research gap and problems to wrap up this part.

\subsection{history of RecSys Evolution}
Recommendation systems have become ubiquitous in our daily lives, from recommending movies on Netflix \cite{gomez2015netflix} to suggesting products on Amazon \cite{linden2003amazon}. The history of recommendation systems dates back to the early days of the internet when researchers began to explore the use of collaborative filtering techniques to personalize recommendations \cite{van1985collaborative}, \cite{meimei1985personalized}. One of the earliest examples of personalized recommendation systems is GroupLens, a research project at the University of Minnesota, which was launched in 1992 \cite{grouplens}. The system used collaborative filtering to make movie recommendations to users based on their past viewing behavior.

In the years that followed, several other researchers began to explore the use of collaborative filtering techniques for personalized recommendations. In 1999, Amazon launched their personalized recommendation system, which was based on collaborative filtering and association rule mining paving the way for this paper \cite{linden2003amazon}. The system was able to make personalized product recommendations to users based on their past purchase behavior.

In the early 2000s, content-based recommendation systems also emerged as a popular approach for personalized recommendations. Content-based recommendation systems use the features of the items being recommended to make personalized recommendations to users. For example, a book recommendation system might use the genre, style, and title of a book to make personalized recommendations to users \cite{schafer2001commerce}, \cite{mooney2000content}.

In recent years, deep learning techniques have emerged as a popular approach for personalized recommendations \cite{10.1145/3285029}. Deep learning models can be used to learn complex patterns in user behavior and make personalized recommendations based on those patterns \cite{Selma2021DeepLF}. One of the most popular deep learning models for personalized recommendations is the neural collaborative filtering model, which was proposed in 2017 \cite{he2017neural}.

Despite the success of personalized recommendation systems, there are also several challenges associated with their development and deployment. One of the biggest challenges is the issue of data privacy and user control. As personalized recommendation systems become more sophisticated, they require access to more user data, which can raise concerns about data privacy and user control \cite{jeckmans2013privacy}, \cite{friedman2015privacy}.

Personalized recommendation systems have come a long way since the early days of the internet. From collaborative filtering to deep learning, researchers have developed a variety of techniques to make personalized recommendations to users. As these systems continue to evolve, it is important to consider the ethical and privacy implications of their development and deployment.

\subsection{N-of-1 Food Recommendation Systems (Food-RecSys)}

Personalized food recommendation systems are engineered to furnish individuals with dietary options meticulously tailored to their distinct preferences, nutritional requisites, and health goals, harnessing diverse datasets encompassing user profiles, dietary records, and situational contexts. These platforms endeavor to facilitate the adoption of healthier dietary practices and the achievement of nutritional equilibrium by meticulously accounting for variables such as caloric content, macronutrient composition, and specific dietary restrictions, thus proposing consumables that are both salubrious and congruent with personal taste predilections. Moreover, they serve to acquaint individuals with novel gastronomic experiences and traditions, thereby enriching culinary diversity and stimulating dietary exploration.

The architecture of these platforms is underpinned by an array of sophisticated machine learning and artificial intelligence paradigms, including but not limited to collaborative filtering, which predicates future preferences on historical user interactions, and content-based filtering, which analyzes the attributes of food items in relation to previously preferred selections \cite{eliyas2022recommendation}. Hybrid methodologies that amalgamate these and other approaches offer enhanced efficacy, drawing upon a multifaceted data corpus to surmount the inherent limitations of singular strategies \cite{melese2021food}.

The genesis and refinement of personalized nutritional recommendation systems present formidable challenges, necessitating meticulous attention to the selection and integration of predictive features and data sources, alongside stringent evaluation of system performance. This process is further complicated by the imperative to navigate ethical considerations pertaining to data privacy, individual autonomy, and the potential for social influence \cite{sanchez2020recommendation}, \cite{qiao2022privacy}.

Notwithstanding these obstacles, personalized nutritional recommendation systems harbinger a paradigm shift in dietary planning and nutritional cognizance \cite{chen2017person}. By delivering individualized dietary suggestions aligned with unique health specifications and gustatory inclinations, these systems pave the way for more enlightened and health-conscious dietary choices. The historical trajectory of recommendation systems reveals a progressive integration of cutting-edge technologies, such as Language Models (LMs) and Large Language Models (LLMs), into the domain of dietary recommendations. However, the exploration of LLM-based recommendation systems in the context of nutrition remains in its infancy, indicating a burgeoning field ripe for academic inquiry and technological innovation, with the potential to significantly advance personalized health and nutritional science.

\subsubsection{Content‑Based Filtering}
n the domain of culinary recommendation systems, a variety of machine learning strategies, notably those under the umbrella of Content-Based Filtering (CBF) methodologies, have been instrumental in curating personalized recipe suggestions. These methodologies primarily leverage matrix factorization techniques, renowned for their capacity to unearth latent attributes that define the dynamics of interactions between users and culinary items \cite{wang2021market2dish}. A notable advancement in this area involves the adaptation of matrix factorization methods to analyze feature matrices—comprising diverse content elements such as ingredients, preparation methods, and nutritional profiles—rather than relying solely on user rating matrices \cite{li2018application}. This innovation aims to integrate a multifaceted array of content information into the recommendation mechanism, thereby enhancing the precision of recipe suggestions\cite{siddik2023collaborative}.

Further, innovative recommender systems have been conceptualized, employing matrix factorization techniques enriched with tags and latent factor analysis to deliver superior, customized recipe recommendations \cite{li2018application}. These systems harness datasets encapsulating user preferences manifested through ratings and tags indicative of specific food attributes or ingredients, facilitating the generation of recommendations that closely align with individual tastes and dietary requirements. Beyond these methods, the field has explored various other techniques for recipe content analysis, including but not limited to decision vector analysis \cite{toledo2019food}, natural language processing \cite{tao2020utilization}, feature vector analysis \cite{trang2018overview}, clustering analysis \cite{manoharan2020patient}, and case-based reasoning \cite{mahardikaa2020case}, \cite{duarte2022blending}. These approaches collectively contribute to the sophisticated comparison of recipes and the subsequent recommendation process, underscoring the dynamic interplay between user preferences and recipe characteristics in the development of tailored culinary suggestions.

\subsubsection{Hybrid Filtering}
In the domain of culinary recommendation engines, the utilization of Hybrid Filtering (HF) strategies, predominantly in conjunction with content-based filtering methodologies, has been a focal point of innovation. Research documented in sources \cite{melese2021food}, \cite{vairale2020recommendation}, \cite{geng2023hybrid} illustrates the application of the N-neighbors algorithm, a predictive model that identifies user preferences by establishing a cohort of N users exhibiting analogous tastes, employing similarity metrics such as Pearson correlation and cosine similarity for this purpose. Further exploration by \cite{guntupalli2020high} and \cite{al2022food} into clustering techniques, including SPARK and enhanced clustering methods, underscores the adaptation of data mining practices for recommendation systems. These techniques aim to aggregate users into homogenous groups, facilitating targeted recommendations.

The contrast between the N-neighbors approach and clustering methodologies is pronounced in their foundational principles \cite{rakhmawati2023halal}; while the former posits that users sharing similarities with their immediate neighbors are inherently similar, the latter categorizes users into clusters based on a central mean, which serves as the criterion for grouping and can be dynamically adjusted to reflect evolving user preferences \cite{jasimfood}. Additionally, the repertoire of hybrid techniques in food recommendation systems is broad, encompassing context modeling, feature-based filtering, knowledge engineering, and matrix factorization. These methodologies collectively enhance the precision and relevance of recommendations, thereby elevating the user experience in navigating culinary choices.

\subsubsection{Collaborative Filtering}

In the interval spanning from 2006 to 2017, a variety of methodologies emerged within the scope of Collaborative Filtering (CF) for the purpose of recommending culinary selections to diverse user groups. These methodologies primarily focused on discerning analogous items or user profiles within a specified aggregation of recipes \cite{thongsri2022implementation}, \cite{siddik2023collaborative}. Techniques such as Term Frequency-Inverse Document Frequency (TF-IDF) for term extraction, the tidal trust algorithm, and the K-nearest neighbors algorithm were instrumental in this regard \cite{Gopalakrishnan2020AFR}, \cite{chhipa2022recipe}. The TF-IDF method, serving as an information retrieval technique, facilitates the quantification of a term's relevance within a corpus, thereby enabling the evaluation of an ingredient's prominence within regional culinary compilations, as elucidated in the study referenced in \cite{Gopalakrishnan2020AFR}, \cite{chhipa2022recipe}. Conversely, the K-nearest neighbors algorithm, a quintessential data mining classifier, predicates the categorization of an item based on the predominant classification among its proximal counterparts, thus supporting the tailoring of recipe recommendations to user preferences.

\subsubsection{Deep Learning Methods}
Deep learning methods are gaining popularity in most fields and the field of food-RecSys is no exception as discussed in this comprehensive survey study \cite{9742919}.
Moreover a study provides food-RecSys which shows promising results in becoming an effective way of propagating healthy weight awareness among common people based on Based on BMI, BMR, k-NN Algorithm, and a BPNN \cite{Gopalakrishnan2020AFR}. Ju (2022) takes a different approach, using a transformer-based deep learning model to recommend restaurant dishes based on nutritional needs presenting a restaurant food recommendation system which recommends food dishes to users according to their special nutritional needs via a transformer-based deep learning model \cite{Ju2022MenuAIRF}. The problem with deep learning-based collaborative filtering methods in food recommendation is that they require the involvement of all users, which may affect the accuracy of the prediction generation process \cite{probdeep}. This paper mentions the problem with classifying food items and discusses a method to improve the classification of classes that are difficult to distinguish in fine-grained food recognition tasks \cite{classprob}.

\subsection{Large Language Model Based Recommender Systems}

Developing and deploying Large Language Models (LLMs) in recommendation tasks typically involve three main approaches: pre-training, fine-tuning, and prompting. Each of these methods plays a crucial role in optimizing LLMs for specific recommendation tasks, enhancing their performance and adaptability and we will take a look at work which has been done in each category \cite{10.1145/3624918.3629550}, \cite{fan2023recommender}.

\subsubsection{Recommendation by Prompt engineering}
Prompting represents a transformative paradigm in the field of natural language processing (NLP), offering an alternative to conventional pre-training and fine-tuning approaches for adapting large language models (LLMs) to specialized tasks. At its core, prompting involves the strategic crafting of task-specific textual templates that guide the input of LLMs. Consider, for instance, the relation extraction task; the prompt "The relationship between X and Y is Z" effectively focuses the LLM's attention on identifying semantic connections within the text. This adaptability aligns with the language generation focus instilled during LLM pre-training \cite{gao2020making}, facilitating a unification of diverse downstream objectives \cite{fan2023recommender}.

Within the domain of recommender systems (RecSys), prompting methodologies are gaining traction due to their ability to enhance LLM performance. Techniques like In-Context Learning (ICL) \cite{gao2020making} and Chain-of-Thought (CoT) \cite{wei2022chain} exemplify the meticulous design of prompts tailored to various recommendation scenarios. Prompt tuning serves as a complementary approach where additional prompt tokens are embedded within the LLM architecture and iteratively refined using recommendation-specific datasets \cite{li2023personalized}.  The recent surge in instruction tuning further demonstrates the potential of prompting; it strategically blends pre-training and fine-tuning with instruction-based prompts to fine-tune LLMs across a multitude of recommendation tasks. This hybrid approach has been shown to bolster zero-shot performance on novel, unseen recommendation challenges \cite{wei2021finetuned}.

To elucidate the distinctions between these prompting techniques, it's crucial to examine how they influence input formatting and parameter modification within LLMs. Comparative studies often highlight whether specific parameters remain static or become trainable during the adaptation process.  This technical analysis, combined with in-depth discussions of prompting, prompt tuning, and instruction tuning,  serves to illuminate the pathways towards optimized LLM performance within recommender systems \cite{wu2024personalized}.  Additionally, the scholarly landscape is often systematized through comprehensive tables that categorize existing research.  These tables typically cross-reference  prompting techniques, specific recommendation use-cases, and the underlying LLM backbones employed, providing a valuable reference for the RecSys community.

\begin{itemize}
    \item \textbf{Prompting:} Early-stage conventional prompting strategies sought to align downstream tasks with language generation paradigms, mirroring the focus of LLM pre-training.  These techniques have since been augmented by the advent of In-Context Learning (ICL) and Chain-of-Thought (CoT) prompting, significantly broadening their applicability \cite{fan2023recommender}. Conventional Prompting encompasses two core approaches: prompt engineering and few-shot prompting. We review the different types of various prompting strategies and methodologies briefly. Prompt engineering involves meticulously structuring prompts to resemble the textual data LLMs encounter during pre-training, facilitating the translation of downstream objectives.  Few-shot prompting \cite{ye2022unreliability} provides select input-output examples to guide the LLM towards generating the desired output format.  These conventional methods demonstrate limited efficacy outside of tasks that fundamentally align with language generation \cite{fan2023recommender}.

In-Context Learning (ICL) emerged alongside GPT-3, marking a significant advancement in the adaptability of LLMs.  ICL harnesses the power of both prompts and in-context demonstrations to teach LLMs new tasks.  It operates in few-shot mode, employing curated input-output examples, and zero-shot mode, relying solely on task descriptions. Within Recommender Systems (RecSys), ICL has proven invaluable, enabling LLMs to perform top-K recommendations, rating predictions, and explainable recommendations \cite{liu2023chatgpt}, \cite{zhiyuli2023bookgpt}.  Zero-shot ICL is particularly apt for conversational recommendation scenarios where user-provided demonstrations are less practical \cite{wang-etal-2023-rethinking-evaluation}.

Chain-of-Thought (CoT) Prompting addresses inherent limitations in LLM reasoning, especially in tasks demanding complex logic.  By strategically embedding intermediate reasoning steps into prompts, CoT bolsters LLM decision-making, encouraging a step-wise breakdown of problems. CoT can function in both few-shot and zero-shot settings.  Recent research demonstrates CoT's potential to enhance LLM graph reasoning capabilities \cite{yao2023beyond}.  This suggests exciting possibilities within RecSys, where recommendations can be framed as link prediction problems within graph-based representations.

    \item \textbf{Prompt Tuning:} Prompt tuning is a technique that enhances the adaptation of Large Language Models (LLMs) to specific downstream tasks by adding and optimizing new prompt tokens based on the task-specific dataset. This approach requires less task-specific knowledge and human effort than traditional manual prompt creation and involves minimal parameter updates, primarily focusing on the tunable prompt and the input layer of LLMs \cite{fan2023recommender}.

Prompt tuning for Large Language Models (LLMs) is categorized into two types: hard prompt tuning and soft prompt tuning \cite{shi2024don}. Hard prompt tuning involves generating and updating discrete text prompts in natural language, guiding LLMs to perform specific tasks by refining these prompts based on task-specific datasets, which can be seen as a discrete optimization challenge \cite{wen2024hard}. On the other hand, soft prompt tuning employs continuous vectors (text embeddings) as prompts, optimized through gradient methods based on task-specific datasets \cite{gu2021ppt}. This optimization happens in a continuous space, allowing for a more flexible adjustment of prompts with minimal parameter updates \cite{lester2021power}. While hard prompts leverage the familiarity and interpretability of natural language, soft prompts utilize the efficiency of continuous space optimization, albeit with less direct human interpretability due to their abstract nature. Both methods aim to enhance the performance of LLMs on downstream tasks with minimal human effort and task-specific knowledge \cite{wen2024hard}.
    \item \textbf{Instruction Tuning:} Instruction tuning addresses limitations in prompt-based methods, specifically their suboptimal zero-shot performance on novel tasks.  It blends prompting with classic pre-training and fine-tuning, teaching LLMs to follow broad instructions instead of memorizing task-specific solutions \cite{zhang2023instruction}. This approach boosts zero-shot capabilities by training LLMs for generalized task understanding. It involves two pivotal stages: instruction (prompt) generation and model tuning. In the first stage, instruction-based prompts are crafted in natural language, outlining task-oriented inputs and desired outputs derived from task-specific datasets \cite{zhang2023instruction}. These prompts act as versatile templates for various tasks, including recommendation systems, by integrating elements like user preferences and intentions \cite{peng2023instruction}, \cite{zhang2023gpt4roi}. The second stage entails fine-tuning the LLMs using these generated instructions across multiple tasks, which can be approached through either full-model tuning or more parameter-efficient methods \cite{zhang2023instruction}. This dual-stage process enhances LLMs' capability to accurately interpret and execute instructions across a wide array of tasks, significantly improving their adaptability and zero-shot performance on tasks they haven't encountered during training \cite{fan2023recommender}.
\end{itemize}

\subsubsection{Recommendation by Fine-Tuning}

In the realm of recommender systems, the adaptation of pre-trained Large Language Models (LLMs) through fine-tuning represents a critical juncture towards achieving task-specific performance enhancements \cite{bakker2022fine}. This refinement process necessitates the infusion of domain-specific intelligence into LLMs, accomplished by training on datasets characterized by user-item interactions and ancillary data pertaining to users and products. The methodologies employed for fine-tuning bifurcate into two distinct paradigms based on the magnitude of weight adjustments required: comprehensive model fine-tuning, which entails a holistic adjustment of the model's weights, and parameter-efficient fine-tuning, which selectively modifies a limited subset of weights or incorporates trainable adapters to achieve task alignment \cite{fan2023recommender}.

The comprehensive model fine-tuning paradigm is exemplified in its application across various recommendation tasks, including conversational recommender systems and professional matchmaking platforms. While this approach offers a direct pathway to task-specific model adaptation, it bears the potential for inducing biases that could disproportionately affect marginalized groups. Efforts to counteract such biases involve innovative techniques like dynamic entity masking and algorithmic neutrality adjustments. On the other hand, the parameter-efficient fine-tuning (PEFT) approach has gained traction as a sustainable alternative amidst escalating computational demands \cite{pu2023empirical}. PEFT strategies, notably the integration of adapter modules within LLM architectures, facilitate targeted modifications to the model's weights, thereby preserving the integrity of the pre-trained parameters \cite{liu2022few}. Advanced techniques such as Low-Rank Adaptation (LoRA) employ low-rank matrix decomposition to simulate parameter modifications efficiently \cite{dettmers2024qlora}, thereby offering a computationally viable solution for model adaptation . These parameter-efficient methodologies not only circumvent the drawbacks associated with comprehensive fine-tuning, such as the risk of catastrophic forgetting, but also align with the operational exigencies of deploying sophisticated LLMs within recommender systems on standard computing platforms \cite{hua2023tutorial}, \cite{yin2023heterogeneous}, \cite{lin2023multi}.

\subsubsection{Recommendation by Pre-Training}
The foundational phase of Large Language Models (LLMs) development, known as pre-training, is instrumental in equipping these models with a comprehensive linguistic comprehension. This process involves the exposition of LLMs to extensive datasets comprising a wide array of unlabeled linguistic content. Such exposure is pivotal for the models to acquire a profound understanding of various linguistic facets, including but not limited to syntax, semantics, and pragmatic nuances, along with an innate capacity for common sense reasoning. The pre-training regimen ensures that LLMs can generate responses that are not only coherent but also pertinent to the given context. The methodologies employed in pre-training are largely contingent upon the architectural design of the model. Specifically, Masked Language Modeling (MLM) \cite{bao2020unilmv2} is favored for structures incorporating either an encoder-only or an encoder-decoder framework \cite{ghazvininejad2019mask}. This method involves the obfuscation of specific tokens or sequences, which the model then endeavors to predict, utilizing the unmasked context as a guide. Conversely, Next Token Prediction (NTP) is adopted for models with a decoder-only architecture, focusing on the prediction of forthcoming tokens predicated on the contextual framework provided by preceding tokens \cite{xue2024repeat}. These methodologies are fundamental in fostering an in-depth linguistic acumen and generative capacity within LLMs \cite{fan2023recommender}.

Adapting these pre-training paradigms to the nuanced domain of recommender systems entails the formulation of methodologies that capture the complex and multifaceted nature of user behavior. Innovations such as PTUM's \cite{wu2020ptum} Masked Behavior Prediction (MBP) and Next K Behavior Prediction (NBP) are reflective of adaptations of MLM and NTP, respectively, albeit tailored to the predictive modeling of user interactions \cite{fu2023robust}. MBP aims to ascertain a singular masked user behavior from a sequence, thus honing the model's precision in identifying individual user actions. NBP, in contrast, seeks to prognosticate a series of future user behaviors drawing upon historical interaction data, crucial for the longitudinal modeling of user preferences. Further augmenting the adaptability of LLMs to recommender systems, the M6 framework \cite{DBLP:journals/corr/abs-2110-03888} introduces objectives analogous to text infilling and auto-regressive language generation. These are conceptually aligned with MLM and NTP but are specifically designed to bolster the model's proficiency in evaluating text or event relevance within recommendation contexts. Additionally, the P5 \cite{geng2022recommendation} model which is a follow up on the T5 model \cite{li2023prompt}, leverages multi-mask modeling across heterogeneous recommendation task datasets, aiming for a model that not only generalizes across distinct tasks but also exhibits an innate zero-shot generation capability for tasks it has yet to encounter. Such pre-training strategies are imperative for imbuing LLMs with the sophisticated understanding requisite for delivering nuanced and personalized recommendations within the complex ecosystem of recommender systems. In this work we will use such framework, specially T5 and P5 to adopt a hybrid approach which leverages from Pre-Training.


\subsection{Research gaps}

\subsubsection{Gaps in Personal Food Recommendation}
While substantial progress has been made in the field of personal food recommendation, certain critical gaps persist that current research has yet to fully address. Firstly, existing systems often lack a detailed design for all essential components specific to the food recommendation domain. These components, vital for the high performance and accuracy of recommendations, remain underdeveloped or overlooked in generalized LLM-based recommendation systems. Additionally, the area of food logging within personal food recommendations has seen advancements; however, there remains a need for more comprehensive and clearly defined platforms which are specifically designed for personal food-RecSys. Current models do not adequately cater to the specific requirements necessary for effective personal food logging as either non open-source design or not supporting a comprehensive logging model, thus limiting the potential for personalized dietary guidance.

Moreover, the absence of frameworks which are based on a geospatial database containing accessible food options based on user location significantly hinders the practicality of food recommendations. The need for a system that consolidates food data from various sources into a geospatially aware atlas, facilitating location-based food queries, is a notable deficiency in the current literature.

\subsubsection{Gaps in LLM-Based Food Recommendation}
Advancements in the utilization of Large Language Models (LLMs) for recommendations have introduced novel capabilities; however, the field lacks specific applications to food recommendation. General-purpose LLM recommenders often fall short in addressing the nuanced requirements of food recommendation systems. There is a distinct gap in research focusing on the development of LLM-based frameworks specifically tailored for the food recommendation domain. Existing works fail to present a holistic approach that comprehensively integrates the necessary components for food recommendations, including the organization and processing of data, thus leaving the potential of LLMs underutilized.

Furthermore, the integration of food-specific models into LLM frameworks remains inadequately explored. The unique challenges of food recommendations, which involve complex personalization dimensions and require the processing of diverse data streams, demand specialized models that current general-purpose LLMs cannot provide. Additionally, the reliability of LLM logic in food recommendations faces obstacles such as inadequate understanding of user logic and real-time geographic awareness, areas where existing solutions are lacking.

Lastly, the interaction with essential external food servers poses a significant challenge for LLM-based systems. The necessity for access to varied external servers and resources, crucial for enriching the recommendation process with relevant and diverse food options, has not been effectively addressed. Current methodologies do not offer viable mechanisms for LLMs to integrate data from food-specific external servers, highlighting a critical gap in the field.

\subsubsection{Summary}
In summary, while there have been strides in both personal food recommendation and LLM-based recommendation technologies, significant research gaps remain, particularly in areas that require domain-specific adjustments and integrations to fully leverage the capabilities of LLMs for personalized food recommendations.

\section{Components of Food Recommendation Systems}

\begin{figure*}[t]
    \centering
    \includegraphics[width=1\textwidth]{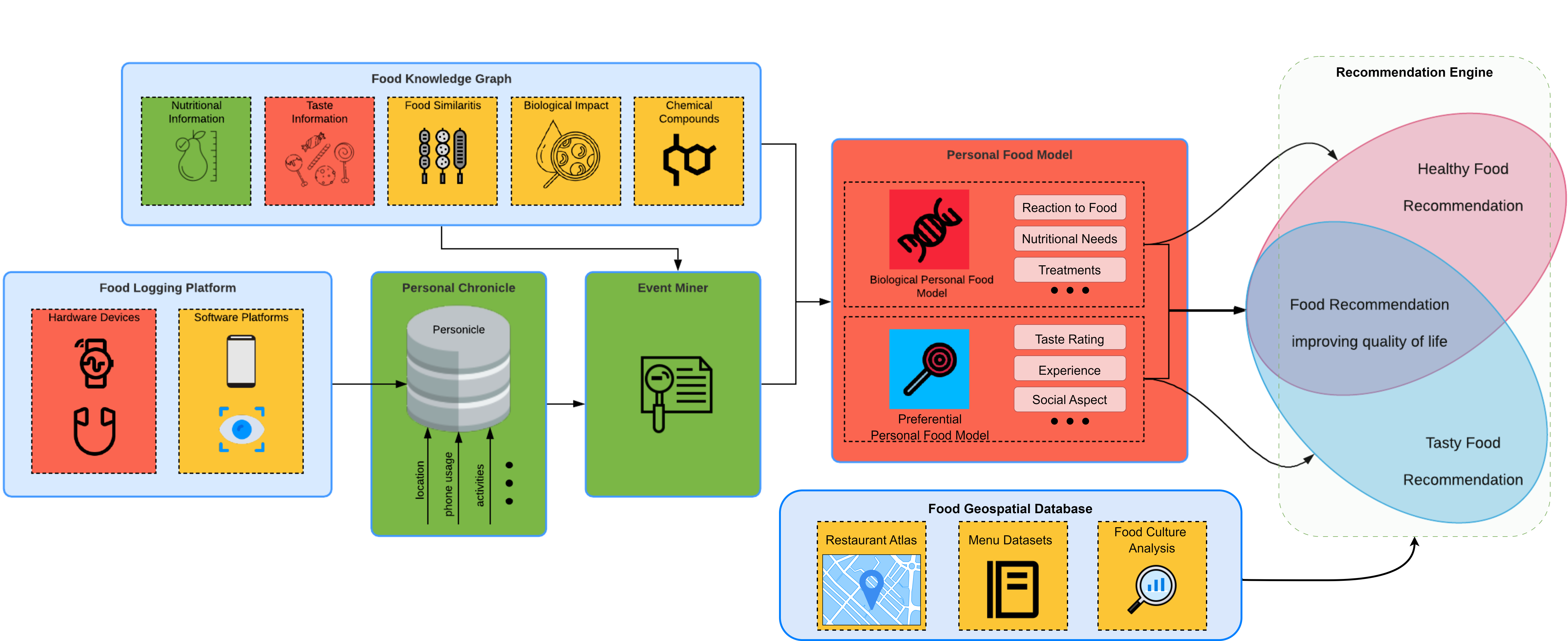}
    \caption{Overview of the Contextual Personal Food Recommendation Components. The red, orange, and green color indicating how established are the research field of each of these components in an individual level in the field of computer science. Showing the complexity of the problem even as individual components let alone as an integrated framework. Red: Requires intensive improvements, orange: fairly established, green: advanced level of establishment.}
    \label{fig:components}
\end{figure*}

Section 2.1 of this chapter provides an overview of the components that make up food recommendation systems. As seen in \ref{fig:components} these components include the multimodal food logger, personal chronicle, personal model, food knowledge graph, recommendation engine, and food atlas. The food recommendation system is a complex system that requires the integration of various components to provide personalized food recommendations. In this section, we will briefly delve into each of these components and discuss their importance and how they are used in the personalized food recommendation process. We will review how each component contributes to the accuracy and effectiveness of the system, and we will explore the challenges that come with implementing each component in more detail through out the remaining chapters.

\subsubsection{Multimodal Food Logger}

Traditional food logging methods, relying on manual diaries, are cumbersome and prone to inaccuracies. While advancements in computer vision have enhanced food recognition capabilities, challenges remain in comprehensive international food identification, ingredient analysis, and accurate volume estimation. To address these limitations, a novel multimedia food logging platform is proposed. This platform leverages complementary information from diverse sources, including visual recognition, speech-based systems, payment data, barcodes, and sensors. Upon food item recognition and volume estimation, the system accesses nutritional databases to obtain detailed nutritional information. Additionally, the platform incorporates user feedback, capturing both enjoyment and physiological responses through sensors (e.g., heart rate, glucose levels). This comprehensive, multimedia approach facilitates the collection of more accurate and nuanced data, which is essential for building highly personalized and effective food recommendation systems.

\begin{figure}[!ht]
  \centering
  
  \includegraphics[width=1.0\linewidth]%
    {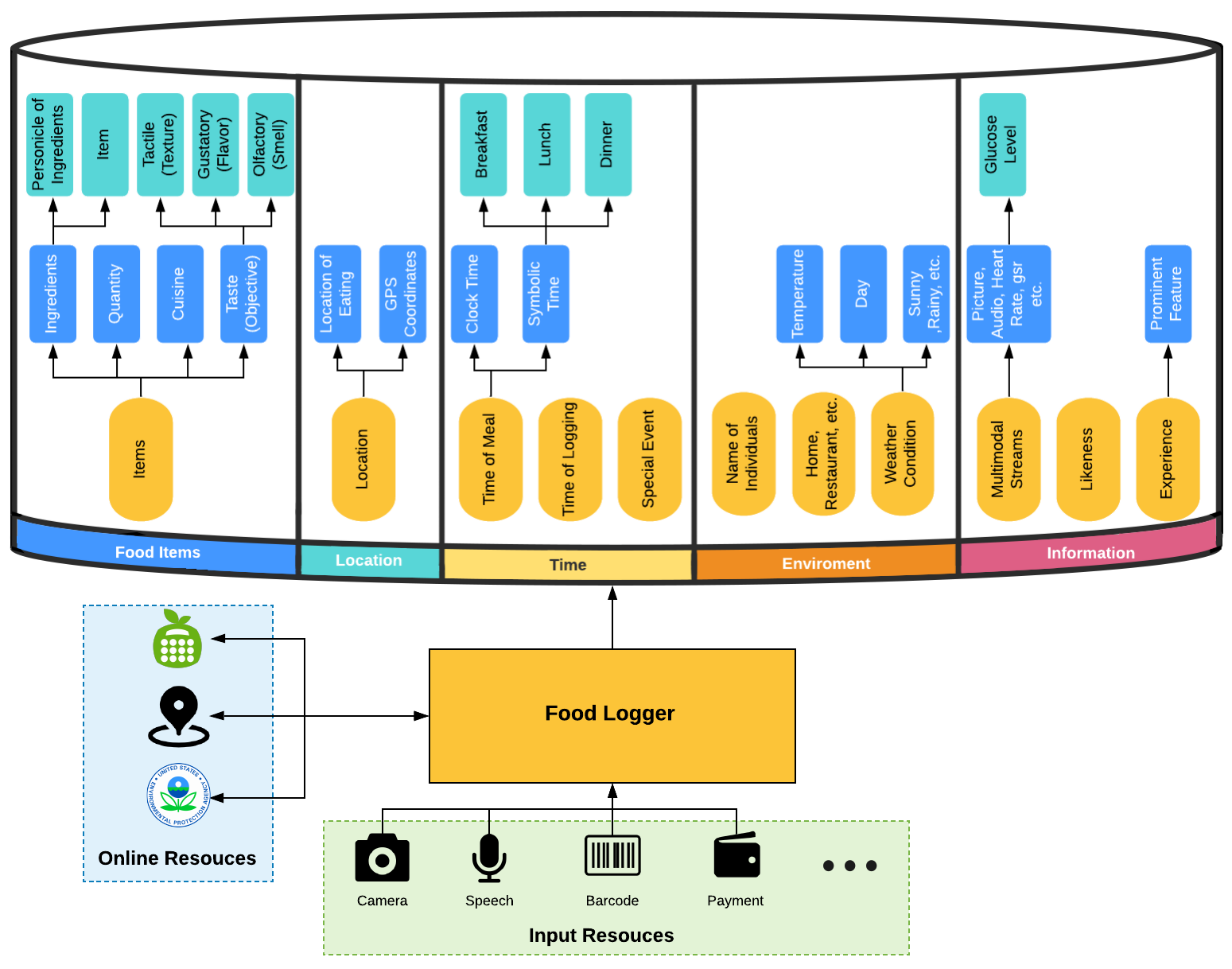}
    \caption{Food logging will use multimedia input sources and complement information from online databases to log each meal and all metadata related to the meal. It captures information about the food (dish name, ingredients, quantity), location (place of eating), time (eating and logging), social context (companions), causal aspects (nutritional and flavor information), and multimedia and experiential information about the food.}
    \label{fig:foodlogger}
\end{figure}

\subsubsection{Personal Chronicle}
A personal chronicle dataset, meticulously aggregating diverse user data streams from multiple sources, is crucial for advanced personalized recommendation systems. This dataset enables the construction of personalized models that capture individual behaviors and preferences.  It integrates data from wearables, mobile devices, social media, and IoT sensors, encompassing dietary patterns, physical activity, location, sleep, and other relevant health metrics. Systems like Personicle provide platforms for collecting, storing, and analyzing this data using machine learning algorithms.  Within personalized food recommendation, personal chronicles supply the foundation for models that predict food choices and generate tailored recommendations, representing an active area of research with significant potential.



\subsubsection{Personal Food Model (PFM)}
Personal modeling is a foundational technique in personalized recommendation systems, enabling the creation of user profiles based on preferences, behaviors, and context. Particularly in food recommendation systems, it captures users' dietary preferences and habits to tailor recommendations. Personal food modeling, a subset of personal modeling, constructs a user-specific food model incorporating preferential and biological aspects. Preferential modeling assesses users' taste preferences using data such as consumption history and ratings, while biological modeling examines physiological responses to food through sensor data. This dual approach underpins personalized food recommendations, aiming to align with users' culinary preferences and health objectives. Future research will further refine these models to enhance food recommendation systems' accuracy and user relevance.
\begin{figure}[!ht]
  \centering
  
  \includegraphics[width=0.4\linewidth]%
    {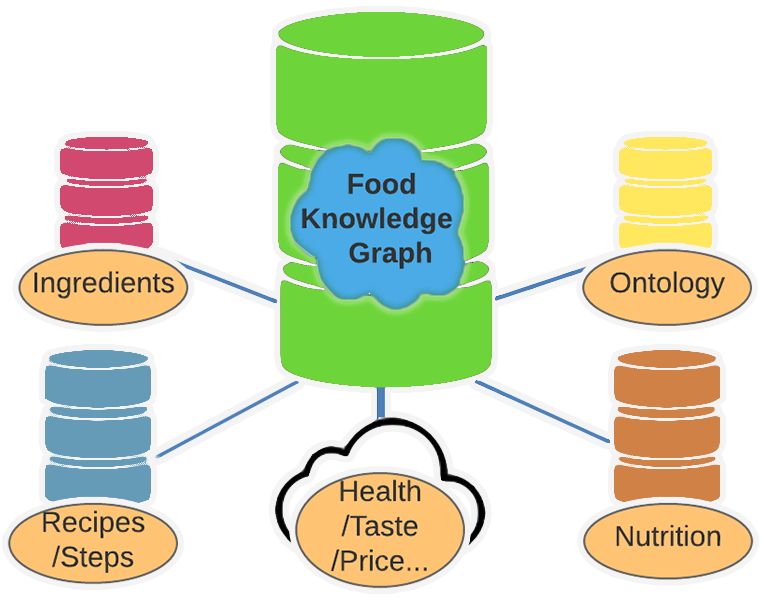}
    \caption{Food Knowledge Graph aggregating and unifying food-related information from multiple offline and online sources.}
    \label{fig:FKGflow}
\end{figure}
\subsubsection{Food Knowledge Graph (FKG)}
An intricate network detailing the multifaceted nature of food through interconnected nodes and hierarchical structures derived from food ontologies. This graph distinguishes between concrete and abstract ingredient nodes, recipe connections, and entity nodes with geographical and nutritional data, enabling comprehensive insights into food relationships and nutritional content. The creation of the FKG involves integrating diverse data sources to represent global food knowledge, facilitating expansion through community contributions. Despite its potential, constructing the FKG's base poses challenges across various research domains such as knowledge graph optimization and natural language processing. Exploration of Food Knowledge Graph is a deep topic on its own and is beyond the scope of this thesis. As an we pick a recipe dataset as our food knowledge graph to demonstrate the basic functionalities of this component within the framework which could be expanded into much more complex datastructre in future works.

\subsubsection{World Food Atlas (WFA)}
A significant hurdle in advancing personal food recommendation systems lies in the development of an electronic World Food Atlas (WFA), a comprehensive database encompassing a global spectrum of dishes and food items available to individuals regardless of location. The task of compiling such an atlas is formidable, considering the vast diversity of consumable foods, along with the myriad ingredients and recipes that exist worldwide. However, constructing a WFA is imperative not only for improving individual quality of life but also for sustaining global environmental health. This endeavor represents an initial stride towards realizing this lofty objective. Our aim is to catalyze the formation of an international consortium comprising scientists, technologists, and culinary experts dedicated to the assembly of the WFA. This document outlines the preliminary efforts in this direction, marking the beginning of a collaborative journey towards creating a universally accessible food information resource.

\subsubsection{Recommendation Engine}

In personalized food recommendation systems, the recommendation engine is pivotal, leveraging data from sources like personal chronicles and food knowledge graphs to craft personalized suggestions. While traditional engines include collaborative filtering, content-based filtering, and hybrid methods combining the two, each has its nuances. Collaborative filtering generates recommendations based on the preferences of similar users, content-based filtering focuses on the attributes of food items, and hybrid approaches aim to mitigate the limitations of the former two by blending their strategies.

Recent advancements have introduced deep learning to enhance recommendation engines, with notable studies proposing sophisticated models that integrate user feedback and food item features to refine food recommendations. Despite the chosen algorithm, factors such as item diversity, recommendation novelty, and serendipity crucially influence user satisfaction.

Given the array of existing recommendation engines, this thesis will focus on adopting a Large Language Model (LLM)-based recommendation engine, exploring its potential to advance personalized food recommendation systems. This decision aligns with the evolving landscape of recommendation technologies and the continuous search for methods that yield more accurate and user-aligned suggestions.




\subsubsection{Integration with Other Technologies}
Integration with other technologies is one of the key challenges in developing personalized food recommendation systems. There are various technologies and platforms that can be integrated to enhance the performance and effectiveness of these systems. For instance, integrating social media platforms such as Facebook, Twitter, and Instagram can provide additional user data that can be used to enhance the recommendation algorithms. In addition, integrating wearable devices such as fitness trackers can provide real-time information about the user's physical activities, which can be used to improve the accuracy of the recommendations. Additionally, integrating voice assistants such as Amazon Alexa and Google Assistant can provide a convenient and seamless user experience.

However, integrating these technologies also poses several challenges. For instance, integrating social media platforms and wearable devices requires access to user data, which raises privacy concerns. Moreover, integrating different technologies can also lead to compatibility issues, which can affect the performance of the system. Therefore, it is important to carefully consider the benefits and challenges of integrating different technologies before implementing them in personalized food recommendation systems.

Overall, integrating other technologies is an important direction for the development of personalized food recommendation systems, but it requires careful consideration of the benefits and challenges associated with each technology.

\begin{figure}[!ht]
  \centering
  
  \includegraphics[width=1.0\linewidth]
    {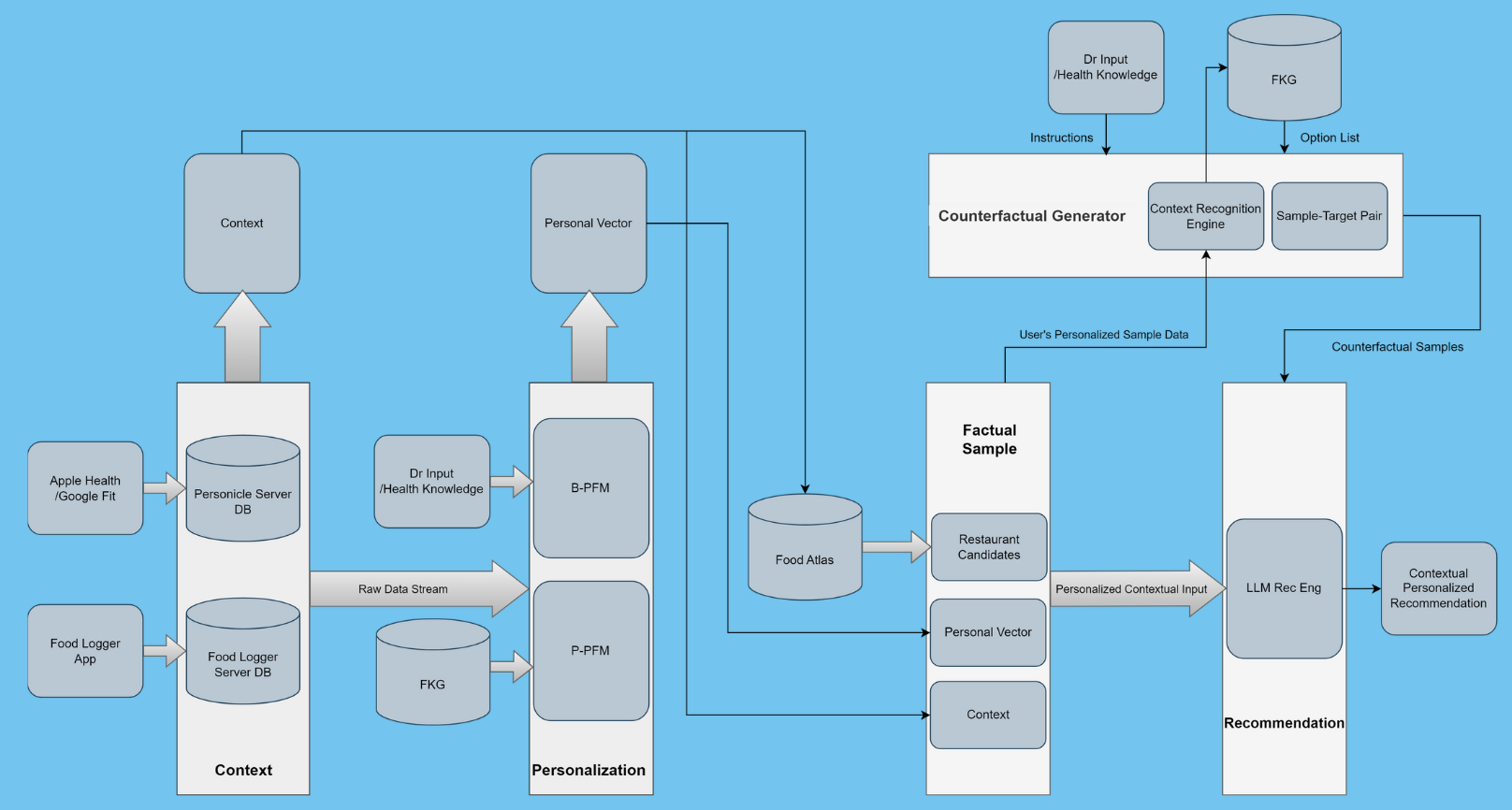}
    \caption{In what follows in this thesis we explore this proposed holistic architecture. Each of the key components in this diagram will have a dedicated chapter which explores the component more deeply and then in Chapter 7 we integrate the entire system together and showcase a prove of concept of the entire architecture.}
    \label{fig:over_arch}
\end{figure}

\section{Summary}
This chapter has provided a comprehensive overview of the foundation, evolution, and current challenges within the field of personalized food recommendation systems. Beginning with a detailed exploration of personal food modeling and the various types of recommendation engines, including collaborative filtering, content-based filtering, hybrid methods, and the emerging deep learning approaches, it sets the stage for understanding the complexities of tailoring food suggestions to individual preferences, behaviors, and dietary needs.

The discussion highlighted the critical role of the recommendation engine, leveraging data from personal chronicles and food knowledge graphs, and underscored the potential of Large Language Model (LLM)-based engines in transcending the limitations of traditional models. By examining the gaps in existing research—particularly the absence of a comprehensive, LLM-based framework tailored for the unique demands of food recommendation—the chapter elucidates the necessity for innovations that integrate detailed component designs, such as the Multimedia Food Logger and the World Food Atlas, within a holistic system.

The culmination of this exploration is the proposition of an integrated framework as presented in \ref{fig:over_arch} which explores the key components in the following chapters which is an important contribution of this thesis, in addition proposing an LLM-based recommendation engine, signifying a paradigm shift towards a more integrated, contextually aware approach to personal food recommendations. This thesis aims to bridge the identified gaps through the development of a novel framework that not only incorporates domain-specific elements essential for precise food recommendations but also harnesses the power of LLMs to manage the extensive variability of food categories effectively.

As we conclude this section, the stage is set to delve into the subsequent chapters, which will detail the design, implementation, and evaluation of an integrated contextual personal LLM-based food recommendation system. This transition marks a critical juncture in the thesis, moving from the theoretical and empirical groundwork laid in the current chapter to the practical application and validation of the proposed innovations in personalized food recommendation.
\chapter{Personal Food Model (PFM)}
\section{Background}
Personal Food Model (PFM), begins by acknowledging the pivotal role of food in determining the quality of human life \cite{rostami2020_personal_food_model}. Food not only fuels our bodies with necessary energy and nutrients but also serves as a cornerstone of pleasure and social interaction. Yet, the pursuit of culinary enjoyment often clashes with the nutritional requirements vital for maintaining optimal health, contributing to a surge in diet-related ailments such as obesity, diabetes, and hypertension. This raises a critical inquiry: Why is food enjoyment so compelling, and how can it coexist with the need for healthful eating?

The sensory experience of food, deeply rooted in multimedia and multimodal interactions, involves the engagement of all senses, heavily influenced by past encounters. This rich, multisensory experience significantly affects our physiological and biochemical responses to food, illustrating the complex relationship between dietary choices and health outcomes. The integration of various sensors for monitoring health indicators such as blood glucose levels and heart rate further highlights the dynamic interplay between food consumption and individual health.

Despite food's integral role in our lives and its profound impact on personal and social well-being, a unified computational examination of its multifaceted nature remains elusive, unlike other life domains such as social media or sports. Recent work by \cite{Min2019FoodChallenges} has tried to address different challenges in food data collection. Defining food computing as the study of all food-related data will encourage more research in different dimensions of how people think about food, how much they eat, and how that affects society and health.

This chapter aims to explore the concept of a PFM within this broader context of food computing, proposing a systematic approach to reconcile the dual desires for gastronomic pleasure and nutritional health. By delving into the complexities of food as a multimodal experience, this section sets the stage for a detailed examination of how personalized food models can bridge the gap between enjoyment and well-being, offering insights into crafting dietary recommendations that satisfy both taste and health requirements.

Personalized modeling serves as an integral component in the development of tailored recommendation systems, enabling the creation of detailed user profiles by evaluating personal preferences, activities, and situational factors. This approach is applied across various industries such as e-commerce, healthcare, and entertainment, significantly contributing to the refinement of recommendation accuracy. In the sphere of food recommendations, personalized modeling is critical for customizing suggestions that are congruent with an individual's dietary preferences and habits.

Within the specialized field of personal food modeling, this methodology concentrates on collecting and analyzing food-related data to construct a model centered around the user. This model integrates the user's food preferences, eating habits, and biological responses to food, laying the groundwork for customized dietary advice. Personal food modeling is categorized into preferential and biological segments, each targeting specific dimensions of the user's interaction with food.

Preferential modeling focuses on capturing the user's food likes and dislikes based on historical consumption, feedback, and online interactions, aiming to build a model reflective of the user's taste preferences and the food attributes that appeal to them. Advanced methods, including collaborative filtering and deep learning, are utilized for this purpose, facilitating the discernment of preferred food qualities.

On the other hand, biological modeling examines the physical impact of food intake, utilizing sensor data to monitor the user's physiological reactions to different dietary choices. The objective is to establish a biological food model that accurately depicts the user's responses, guiding the formulation of recommendations that align with health goals and nutritional needs.

Personal food modeling thus stands as a pivotal element within personalized food recommendation systems, merging preferential and biological data to generate recommendations that align with the user's dietary preferences and health requirements. Future research endeavors will focus on refining personal food modeling techniques and enhancing the accuracy and effectiveness of personalized food recommendation systems.

This chapter's primary contribution lies in the integration of a personal model that blends the culinary multimedia experience with aspects of biological health. This model is employed within an intricate recommendation system that views food items through the lens of features enhancing both pleasure and nutritional value, aiming to improve specific health metrics like sleep quality. By examining an individual's diverse food experiences and the nuanced contextual elements tied to various food items and dishes, the system endeavors to balance factors of enjoyment and nutritional benefit, thereby facilitating the selection of appropriate food choices within specific contexts.

\begin{figure*}[h]
  \includegraphics[width=1\textwidth]{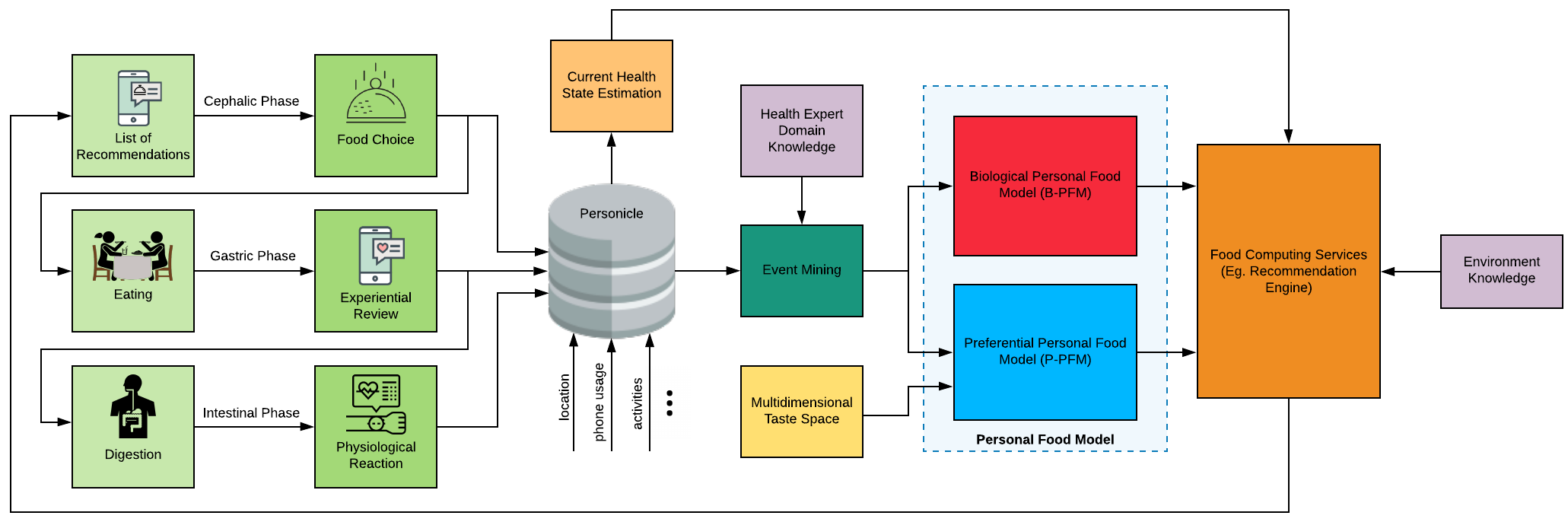}
  \caption{Food Recommendation Architecture: Data from the 3 digestion phases are being collected alongside other data-streams to create the PFM: the heart of Food Recommendation.}
  \label{fig:PFMSystemsOverview}
\end{figure*}

\section{PFM}
The Personal Food Model (PFM) represents a digital encapsulation of an individual's food-related attributes, serving as a cornerstone in food recommendation systems to offer dietary suggestions aimed at enhancing life quality. Despite the myriad factors complicating dietary choices, existing models lack a holistic framework that encompasses food as a multimedia experience, integrating aspects such as taste, visual appeal, social influence, and experiential factors. This section demonstrates PFM's capability to forecast an individual's multifaceted food preferences across various contexts, achieved by leveraging diverse data streams, including location history, vital signs, and documented food intake through text, voice, and images. Future endeavors will seek to broaden the data sources informing the personal model, incorporating additional streams like calendar entries, social media activity, and transaction records.

PFM's complexity arises from its multidimensional nature. The biological dimension of the model addresses how different food items can meet specific nutritional objectives, such as weight management or athletic performance enhancement. Moreover, a nuanced, contextual understanding of user requirements is essential for accurately computing real-time dietary needs. The influence of biological and life events on food choices is an indirect yet critical component that should be integrated into the model.

Figure \ref{fig:PFMSystemsOverview} illustrates the process through which the personicle aggregates various data streams over extended periods. Data from the personicle informs the PFM, which bifurcates into biological and preferential segments. The Biological PFM documents the body's response to diverse food items, including allergies and nutritional demands, playing a vital role in food-related decision-making. However, biological considerations represent just one facet of the decision process. The Preferential Personal Food Model, constituting the user's taste profile, captures prior food experiences and potential interest in untried dishes, offering a comprehensive view of the user's gastronomic preferences.

\begin{figure}[!ht]
  \centering
  
  \includegraphics[width=0.7\linewidth]%
    {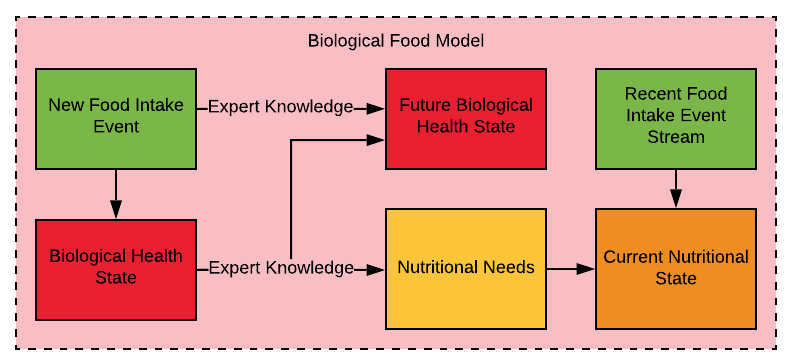}
    \caption{The interactions in the biological food model}
    \label{fig:biological-model}
\end{figure}

\subsection{Biological food models}
Constructing a Biological Personal Food Model (B-PFM) exclusively through data-driven approaches presents significant challenges. Applications such as MyFitnessPal, which track dietary intake and physical activity, are narrowly tailored towards fitness objectives and do not encompass the breadth required for a comprehensive biological model. To address this, we advocate for a hybrid methodology that marries user data with domain-specific knowledge to establish a foundational rule-based population model, which is progressively personalized with accruing data. These rules elucidate the influence of dietary habits on biological metrics, for instance, the correlation between pre-sleep heavy meals and diminished sleep quality. By collating such insights from authoritative sources, we assess the relevance and accuracy of these patterns across varying contexts, thus crafting the B-PFM through context-sensitive regulations.

Moreover, it's crucial to consider the multifaceted effects of dietary choices and events on an individual's health status, acknowledging that users may have divergent health objectives potentially leading to conflicting dietary recommendations. Maintaining equilibrium among various biological aims, while factoring in immutable personal attributes like allergies, intolerances, and genetic predispositions, is essential. Figure \ref{fig:biological-model} delineates the influence of nutritional intake on the user's current health state, tailored to their specific biological requirements. Through the process of event mining, discussed subsequently, we transform domain expertise into operational rules. These rules are then verified against user data to forecast future biological states, ensuring a comprehensive and personalized biological food model.

\begin{figure}[!ht]
  \centering
  
  \includegraphics[width=0.9\linewidth]%
    {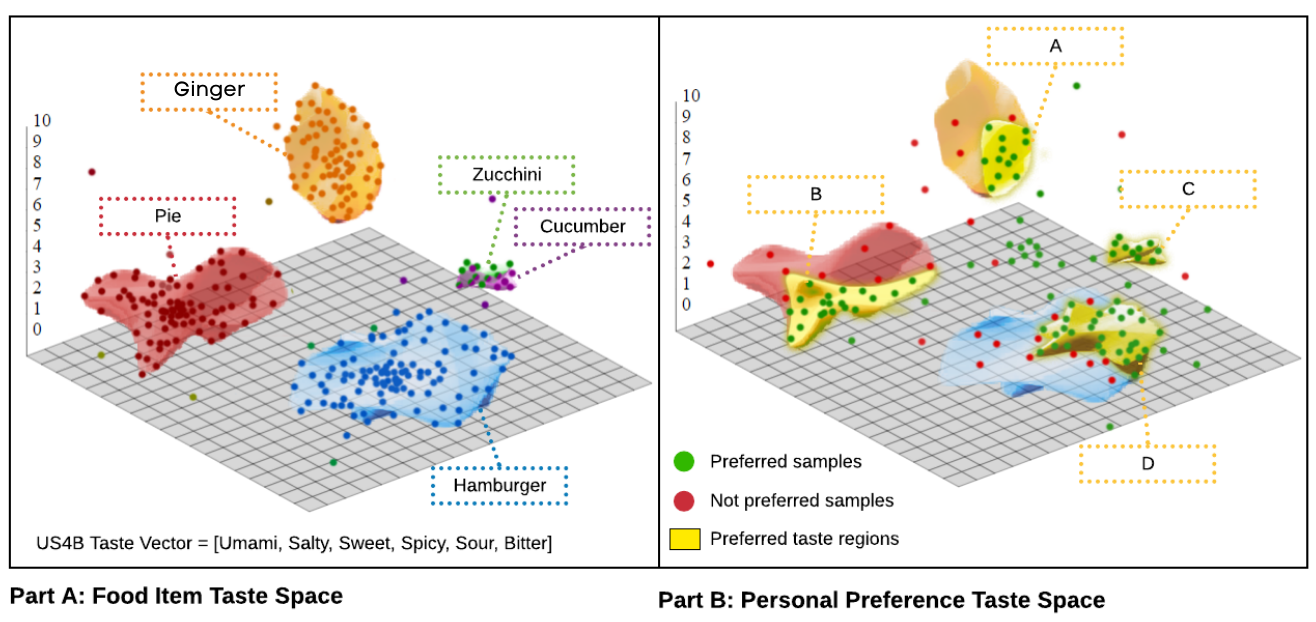}
    \caption{Visualization of the US4B Taste Space. Part A: The collection of all taste samples from all users determine the hypervolume of the taste range of food items. Part B: Past taste sample values and  ratings from the user determine user's preferred taste region within the food item taste range hypervolume.}
    \label{fig:hd-taste-space}
\end{figure}

\subsection{Preferential food models}

In this section of the thesis, we introduce an innovative methodology for delineating and quantifying human taste perception, laying the groundwork for the development of the Preferential Personal Food Model (P-PFM). Traditional food recommendation systems often overlook nuanced preference models, opting instead to prioritize health-centric advice or superficially gauge user preferences through extensive lists of ingredients and dishes, without delving into the underlying reasons for those preferences. Our approach transcends this limitation by conceptualizing a taste space defined by six primary tastes, encapsulated within the US4B taste model. This model, representing a multidimensional and additive framework, incorporates umami, salty, sweet, spicy, sour, and bitter tastes (USSSSB or US4B) to capture the complexity of human taste experiences.

Drawing parallels to the RGB color space's pivotal role in multimedia advancements, the US4B taste space is posited as a foundational element for future food-related technological innovations. Within this taste space, each food item is assigned a specific value across the US4B dimensions, facilitating the creation of a Hyperdimensional Taste Space (HD-Taste-Space). This space allows for the differentiation of food items based on their taste profiles, such as distinguishing between an unripe Brazilian mango and a ripe Indian mango, or identifying the shared taste region of zucchini and cucumber.

By analyzing recipe databases, we estimate the taste profile hypervolume for dishes within this hyperdimensional space, a novel approach that extends beyond ingredient-based recipe correlations. This technique enables the construction of the P-PFM by mapping a user's food log to the HD-Taste-Space, identifying hypervolumes that represent the user's taste preferences. This critical aspect of the P-PFM informs on both familiar and potentially likable yet untried foods for the user, facilitating the identification of healthier alternatives within their taste preference range. For instance, diet soda serves as a prime example, mirroring the sensory attributes of regular soda but differing in biological impact. By situating food items within the US4B taste space and pinpointing the user's taste preference regions, we can tailor food recommendations to align with individual taste profiles and dietary needs, marking a significant advancement in personalized food recommendation systems.

\section{Collecting Data: Food Chronicle}
A pivotal element in the architecture of sophisticated personalized recommendation systems is the integration of a personal chronicle dataset. This dataset serves as an exhaustive repository, aggregating diverse streams of user data from myriad sources and cataloging them as discrete events. Such personal chronicles offer an in-depth narrative of an individual’s behaviors, preferences, and activities, laying the groundwork for crafting highly tailored models specific to the individual. These models, whether aimed at health-related insights or preference delineation, necessitate the assimilation of data spanning various domains to be effectively constructed.

Models are built using data. Most successful search engines, social media, and e-commerce systems utilize personal models to provide people with the right information at the right time in the right context, usually even before a user articulates his need \cite{AdomaviciusTowardExtensions}. Personal food models play the same role in food recommendation systems \cite{Min2019FoodChallenges}. We need to log food consumed by a person and all the relevant metadata over a long period for the user. While initial food logging efforts required cumbersome manual food diaries, smartphones and cameras can drastically improve the quality and ease of logging. Aizawa \cite{FoodLog:Applications} was the first multimedia researcher to champion the idea of logging food using a smartphone camera and remains a very active researcher. Applications for camera-based food-logging have been developed in many other countries \cite{Chen2016Deep-basedRetrieval}, \cite{Bossard2014Food-101Forests}. 
Multimedia and computer vision research communities have been actively exploring food-logging systems. These systems use computer vision techniques to recognize items, their ingredients, and even the volume consumed by the user \cite{Chen2016Deep-basedRetrieval}, \cite{Oh2018MultimodalJournaling}. Unfortunately, there is no generalized logger for international food, and identifying ingredients and volume remains a challenge. A useful review of many visual approaches and descriptions of databases used for training is included in \cite{Min2019AComputing}.

Conversational voice interfaces are becoming quite popular, making rapid progress.  Systems like Alexa and Siri are available at home, in phones, and watches. People can report what they eat, volume, and reaction to food using a simple sentence. Many packaged food and processed ready to prepare food items have barcodes.  Since barcode readers are now omnipresent even in smartphones, one can get all food information from these.  Some sensors measure muscle activity and try to infer food items from that.  These are placed on the chest, near the ear, or neck \cite{Chu2019RespirationSensors}.  These have shown some progress in recognizing eating events but have not gone much beyond that yet.

We propose a multimedia food logging platform, shown in Figure \ref{fig:foodlogger}, that could use many relevant sources to log food items and find all information that may be needed to build a PFM. Multimedia uses complementary and correlated information and provides more comprehensive and precise information than any one medium. Moreover, we will keep adding new sensors and technologies to keep the logger useful.

\begin{figure}[!ht]
  \centering
  
  \includegraphics[width=\linewidth]%
    {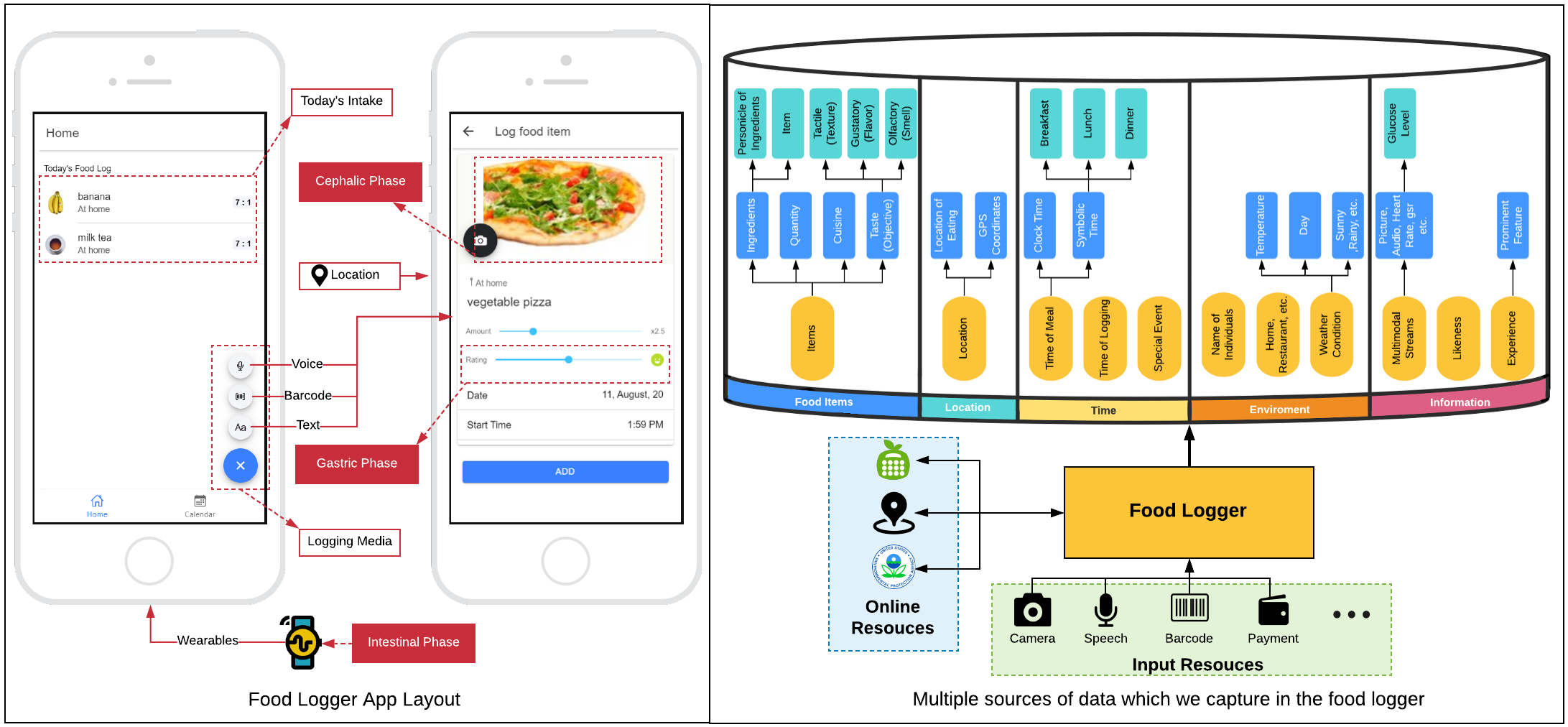}
\caption{Food logging will use multimedia input sources and complement information from online databases to log each meal and all metadata related to the meal. It captures information about the food (dish name, ingredients, quantity), location (place of eating), time (eating and logging), social context (companions), causal aspects (nutritional and flavor information), and multimedia and experiential information about the food.}
    \label{fig:foodlogger}
\end{figure}

This is the beginning of building a robust multimedia solution to the problem of logging.  There are three important aspects to this platform:
\begin{itemize}
    \item We can design a multimedia platform that uses visual food recognition, speech-based systems, payment-based options, barcodes, sensors to determine food chewing and the content of the food, and several similar emerging approaches.
    \item Once a dish or food item is recognized and the amount consumed is known, systems must find the nutritional data using governmental or commercial databases.  Similarly, weather information, social context, and other metadata related to food required by the PFM may come from other sources. 
    \item The log must contain the user's reaction, both in terms of enjoyment and bodily reaction.  The enjoyment information may come from asking the user, and the bodily reactions may come from sensors such as heart rate, glucose measurement, and respiration rate.
\end{itemize}

\subsection{Utilizing Other Knowledge and Data sources}
We should refer to personicle here and associate the food log with other personicle measures (VP).
We enrich the food events with associated nutritional, culinary, and contextual information using databases from different public and private organizations. These include nutrition (NutritionIX, USDA food database), weather, air quality (now provided by the EPA), and place (Google Places, Yelp).
\newline
We may also want to capture some biomarkers characterizing the health of the person. These parameters may be continuously recorded and could be used to identify physiological responses to food items \cite{Oh2018MultimodalJournaling}. A personicle-like system \cite{Oh2017FromChronicles} can capture this information, and the time-indexed nature of the data and events makes them readily available for associating and analyzing with the food events.

\subsection{Data Model for the Foodlog}
Food logs are collected for 
\begin{itemize}
    \item Building PFM to understand the nutritional requirements and taste preferences of the user.
\item Understanding the health state of the user.
\end{itemize}

These two goals may require different information from the food log. Building PFM requires as much longitudinal data as is available, while health state estimation requires PFM and recent lifestyle and biological data. We need to keep these goals in mind while designing the food log.
We have followed the HW5 (how, what, when, why, where, what) model as described in \cite{XieEventStreams}, \cite{WestermannTowardApplications} to identify what information can fully describe a food event and maximize its utility for a variety of applications. The different aspects and associated information are detailed in figure \ref{fig:foodlogger}.
There can be three types of data collected:
\begin{enumerate}
    \item Observed data: directly captured using a sensor.
    \item Derived data: We can derive some data and information using sensors and knowledge sources.  This information will depend on the algorithms and data sources used. 
    \item Subjective data: The system may prompt the user or some other human source to get specific information.  This data is prone to errors as it depends on human perception.
\end{enumerate}
We should utilize the different types of measurements in different manners to minimize the error in our analyses and predictions.

\subsection{Current Status}

In this chapter, we describe the data and knowledge needed to build a PFM directed at improving sleep quality. We considered that sleep quality is affected by stress, activity, and food \cite{Azimi2019PersonalizedStudy}.  We are implementing a food logging platform.  We decided to focus on data collection using voice, text, and barcodes for the current version.  We will include visual recognition approaches soon.

We add food metadata to the log using weather and reverse-geo databases.  The foodlogger asks the user about their reaction to each item entered.  We used the NutrionIX platform to get information about calories and nutrients in each food item.  The current food logger has information about how much a user likes a dish to build the preferred personal food model. However, information about the taste and flavor of a food item is not readily available from any source.  We are working towards deriving such information about food items from different sources.  This is an excellent open opportunity for the multimedia community to take the lead in solving this critical problem.
Our multimedia platform is progressing well.
\begin{figure}[!ht]
  \centering
  
  \includegraphics[width=0.9\linewidth]%
    {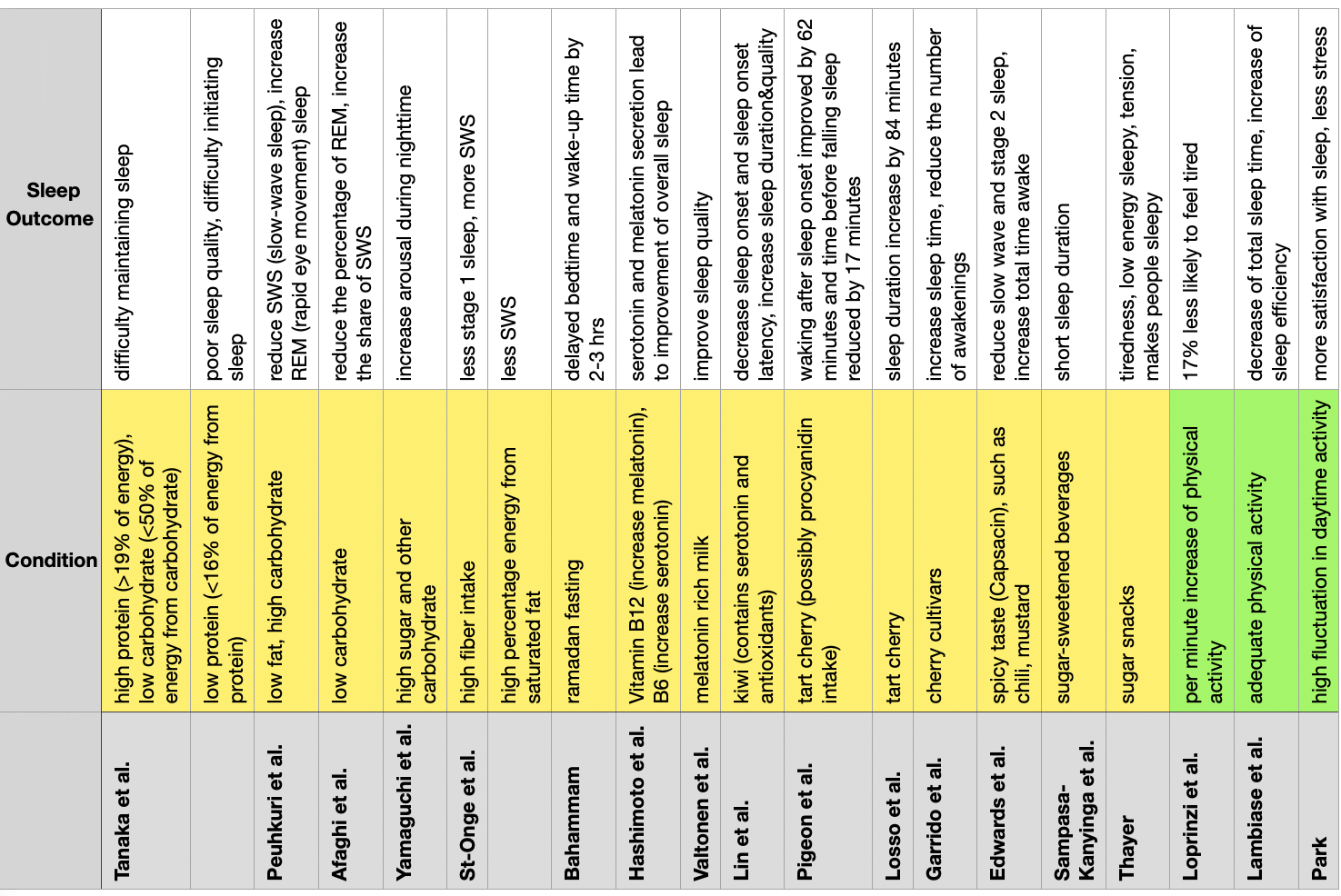}
\caption{Attributes of food events that impact sleep quality. These relationships form the basis for the Biological Personal Food Model.}
    \label{fig:relatedresearch}
\end{figure}
\section{Knowledge Collection}
Knowledge collection is not just the extraction of information from published papers; it comes from multiple sources.

As stated in the previous sections, the personal model should be able to incorporate existing knowledge sources. We have surveyed papers that explore the relationship between dietary inputs and sleep outcomes.
and used their results to initialize the relationships in the personal food model. 
We summarize our findings in figure \ref{fig:relatedresearch}.
We found that macronutrients have a great impact on sleep outcomes \cite{Tanaka2013AssociationsWorkers}, \cite{Peuhkuri2012DietaryMelatonin}, \cite{Peuhkuri2012DietQuality}, \cite{Yamaguchi2013RelationshipRegularity}, \cite{Afaghi2008AcuteIndices}, \cite{St-Onge2016FiberSleep}. Some micronutrients contribute to melatonin secretion and hence can have a significant impact on sleep quality \cite{Hashimoto1996VitaminHumans}, \cite{Valtonen2005EffectsSubjects}. Additionally, there are some studies that explore the effect of specific food items such as kiwi fruit \cite{Lin2011EffectProblems} and cherries \cite{Pigeon2010EffectsStudy}, \cite{Losso2018PilotMechanisms}, \cite{Garrido2013AAging.} on sleep. Some chemicals responsible for specific tastes, such as capsaicin \cite{Edwards1992SpicyThermoregulation} and sugar \cite{Sampasa-Kanyinga2018SleepAdolescents}, \cite{ThayerEnergyExercise}, can also impact sleep. Fasting contributes to the change of bedtime \cite{BaHammam2013TheAssessment} as well.
\newline
We have also included some studies about the impact of exercise and physical activity on sleep \cite{Loprinzi2011Association2005-2006}, \cite{Loprinzi2012TheWomen}, \cite{Lambiase2013TemporalWomen}, \cite{Park2014AssociationsAdolescents} as it is an important confounding variable that impacts both nutritional needs and sleep quality.

\begin{figure}[!ht]
  \includegraphics[width=\linewidth]{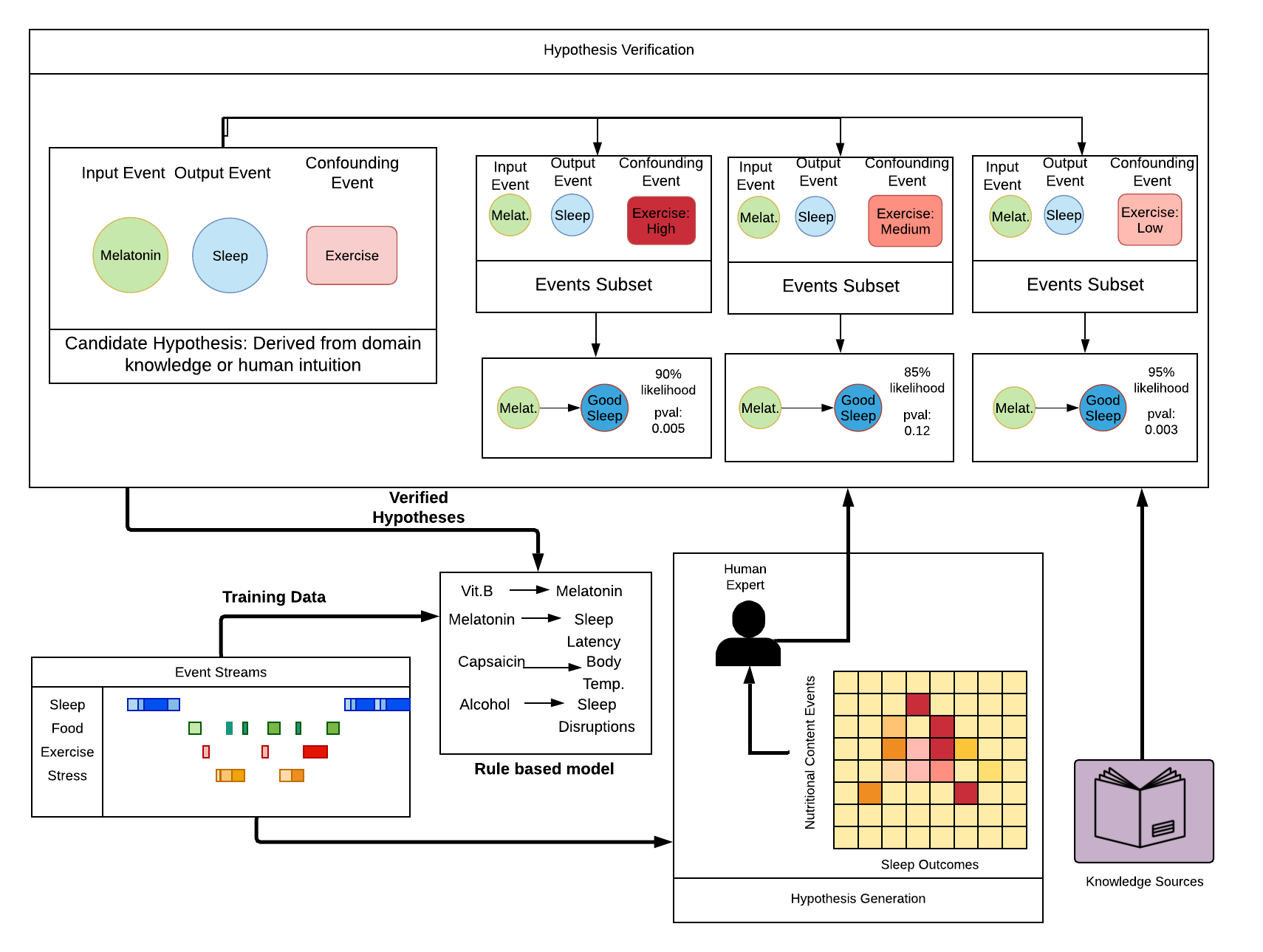}
\caption{Event Mining workflow: Hypothesis generation operators allow us to find frequently occurring sequences of events. These can be converted to hypotheses by including confounding variables, and they can then be tested in the presence of these confounding factors using a hypothesis verification operator. These verified hypotheses serve as a rule-based model for the user's behavior.}
  \label{fig:SleepModel}
\end{figure}

\section{Event Mining}
Multimedia research in event mining has focused on event recognition and situation understanding (eg., sports and surveillance videos). There has not been much research on how we can utilize event mining to run n-of-1 experiments using a person's events and data streams and derive rules that describe their behavior in different situations.
Event mining allows us to find patterns and relationships between different events in our daily lives. We can find relationships between different events in a person's lifelog data and derive an explainable personal model \cite{PandeyUbiquitousHealth}. 

Event mining results in rules of the form $Event_i \xrightarrow{C}Event_o$, where $Event_i$ is the input event, and we want to find out its effect in the occurrences of the outcome, $Event_o$. $C$ defines the set of confounding variables and temporal conditions that might affect this relationship. For Biological-PFM, the input events are lifestyle events that have a causal impact on some observable biological outcome \cite{Pandey2020ContinuousRetrieval}. While, in the Preferential-PFM, the input events capture the contextual situations that affect the user's culinary preferences.
This view of events and their impacts is in line with the potential outcomes framework for causal inference (provided the required assumptions, eg., SUTVA are valid) \cite{Rubin2005CausalDecisions} and are explored in detail later in this section.
\newline
We perform a two-step analysis with a human expert acting as an intermediary to select non-spurious relationships. The event patterns language described in \cite{Jalali2016InteractiveStreams} allows us to describe the relationships as temporal patterns of events. \textbf{Hypothesis generation} is used as a preliminary investigation tool that allows a human expert to identify any behavioral patterns in the form of events co-occurrences and \textbf{Hypothesis verification} tells us whether the relationship is causally significant in the presence of the confounding variables.

\subsection{Hypothesis generation: Discovering new behavioral patterns}

Users' event logs contain all of their daily habits and biological responses to different events. Hypothesis generation operators allow us to discover these habits and patterns that can be tested and used for prediction.
This step of the analysis is mostly data-driven and starts with a human expert specifying the event streams that they believe to be correlated. The output is a heat map with different combinations of events occupying different positions (Figure \ref{fig:SleepModel}).
The patterns with relatively higher frequency may represent a significant relationship and selected for hypothesis verification.
\newline
The frequent patterns would then need to be converted to candidate hypotheses. The user would need to specify the cause and effect events along with any confounding factors. This hypothesis can then be verified using the hypothesis verification operator.


\subsection{Hypothesis verification: Verifying patterns under different contexts}
Users can also verify their beliefs by encoding those as patterns of events and specifying the variables that define the contextual situation. 
We defined these patterns using scientific literature, as described in the previous section.
Each occurrence of the pattern represents an instance of the input event (treatment), and we want to measure its impact on the outcome. Thus, each occurrence of the pattern becomes a single unit in the potential outcomes framework \cite{Rubin2005CausalDecisions}, and we can compare different units while matching them by the confounding factors, to estimate the causal effect of the treatment.
\newline
Once we have found all the pattern occurrences and the confounding variables, we follow a two-step process to find the validity of the rule. 

\begin{enumerate}
    \item \textbf{Find similar situations based on confounding variables (Contextual Matching)}. The confounding variables define the situation in which the input event (treatment) occurs, and can affect the event relationship we want to analyze. Therefore, we want to compare the events that occur in similar contexts and compare the impact of the input event on the outcome in an unbiased manner. We can do it by either clustering the values of confounding variables or converting the confounding variables to events and find matching confounding event patterns. 
    \item \textbf{Find the validity of the relationship for each situation.} Once we have performed the contextual matching, for each contextual group, we can find the effect of the treatment on the outcome using an appropriate statistical test. We can compare the difference in the outcome for different input events, and this would tell us the relative causal effect of the different input events. 
\end{enumerate}
This two-step hypothesis verification allows us to simulate an N-of-1 experiment on the user's event log while also incorporating the existing scientific knowledge in the form of candidate hypotheses and identifying confounding variables. 

\subsection{Deriving Personal Food Model using Event mining}
We need to analyze the food log in conjunction with other events from the personicle to create an explainable and personalized food model for every individual. The model would predict the impact of food events on other aspects of a person's life; and how different lifestyle and biological factors impact our food choices. In this chapter, we are exploring the relationship between food events and sleep outcomes; therefore, we will include behavioral factors that would impact these two events, such as physical activity (exercise, step count).

We can identify different behavioral habits of the user using hypothesis generation operators. We can also visualize the relationship between various nutritional factors and different sleep outcomes to find if these are worth exploring further. Once we have identified such relationships, we can start verifying these hypotheses.
We derive the hypotheses from data or existing biomedical literature. These relationships have been detailed in the previous section and are also depicted in figure \ref{fig:relatedresearch}.
\newline
Figure \ref{fig:SleepModel} shows the complete event mining process for the personal food model. The verified hypothesis contain event relationships that hold true in the specified contextual situation.
These relationships form a set of rules with varying degrees of accuracy in different contextual situations. For example, if we have verified that cow's milk has a positive impact on sleep latency, then the relationship would be quantified in the form of minutes reduced in latency. It will have a different value for different contextual situations described by physical activity, day's meals, and last night's sleep. These rules could thus be used to identify the potential outcome of different foods and recommend items with the desired sleep outcome.

\section{Going Forward: Multimedia for Personal Models}

Though this chapter focuses primarily on personal food models, it is really about building personal health models using disparate data and information sources.  A personal health model's importance is apparent in these days of a pandemic that has disrupted lives globally.  In this section, we discuss interesting challenges that we need to address.  We believe that multimedia computing offers concepts, techniques, and practical experiences related to key areas mentioned in the paper.

\begin{enumerate}
    \item User Privacy: User privacy and data protection are integral to developing a multimedia personal model. Without adequate security measures the model is unlikely to be widely adopted, regardless of the performance or utility. 
    This is an important challenge for multimedia, artificial intelligence, and privacy and security research groups and we are actively looking for collaborations in this area. 
    There are learning techniques such as federated learning \cite{Yang2019FederatedApplications} that allow us to build models and share insights without taking users' data from their device. We need to incorporate such methods in our platforms so that the users have complete ownership of their data. 
    \item Taste Space: Taste and flavor of food are very complex.  Food taste space depends on the ingredients and recipe as well as visual presentations.  On the other hand, each person has their own preferred taste space that must be determined by observations over a long time. We are exploring 6-dimensional taste space. This is less than the tip of the proverbial iceberg.  Such representations will result in labeling food items better so that people can select what they will enjoy eating and will be healthy.
    \item Multimedia Logging Platform: Multimedia community has focused on food logging using only visual recognition approaches and has been limited only to dish and ingredient recognition.  Food logging is not just recognizing dishes from pictures, but identifying all characteristics of an eating event.  We need to build a multimedia logging platform to collect all food-related information relevant to building PFM. Such logs could be used for studying population for health as well as for business reasons.
    \item Multimodal event detection: The health state of a person is usually estimated by combining multimedia (audio-visual) and multimodal (heart rate, EEG, respiration rate, Glucose content) signals.  Estimation of health state is a great challenge for researchers that will also help ALL humans. 
    \item Multimodal Knowledge Collection: Much of the diagnosis and prescription related to health is multimodal and will require extending traditional knowledge graph \cite{Zulaika2018EnhancingGraphs}\cite{HaussmannFoodKG:Recommendation} techniques.
    \item Event Mining: Mining multiple sequences of event streams detected from disparate data streams is essential for both building models such as PFM as well as for health state estimation.  Event mining may offer more challenging problems in predictive and preventive approaches in several application areas, including health using novel forms of machine learning than object recognition offered in computer vision.  We have already started building a platform for this.
    \item Recommendation System to motivate behavioral change: In context of eating habits, a recommendation system which always promotes the healthiest option is not necessarily the best one. A good recommendation must consider personal food preferences and healthiness together to suggest not just healthy but correct amount of 'healthy and tasty' food. Correct recommendation should be given at the correct place and the appropriate time to motivate behavioral change \cite{Patel2015WearableChange}, \cite{Motivate:Publication}. PFM is the first step towards context-aware recommendation in food domain but this is just the beginning of a long journey.
\end{enumerate}
\chapter{Data Collection}
\section{Background}
Data collection plays a pivotal role in the development of personalized food recommendation systems, as it serves as the basis for constructing personal food models. These models rely on accurate and diverse data to understand users' preferences, dietary habits, and nutritional needs. In this introductory section, we will discuss the various types of data collected from different sources and the challenges associated with gathering personal food intake records and other food-related information. We will also explore the potential benefits of integrating additional data sources, such as smartwatch data and well-being metrics, to create more comprehensive personal food models.

Personal food recommendation systems require a diverse array of data to generate tailored recommendations that promote healthier eating habits and improved overall well-being. This data can be broadly divided into two categories: personal data and food-related data. Personal data encompasses information about the individual, such as food intake records, biological data from sensors and lab tests, and lifestyle data from smartwatches. Food-related data, on the other hand, comprises spatial food data (stored in the World Food Atlas) and non-spatial food data (stored in the Food Knowledge Graph), which provide context for understanding food items and their properties. Spatial Data stored in the World Food Atlas is discussed in the following chapters of this thesis, as we require a much more detailed look into it.

Collecting personal food intake records in daily life poses several challenges, such as the reliance on user input, potential inaccuracies in self-reported data, and the inherent complexity of food items and their properties. Researchers have been exploring various methods to facilitate the logging of food intake, including text search, voice command, and image input. Smart applications have been developed to estimate the nutritional values of food items based on user inputs, eliminating the need for users to know the exact nutritional breakdown of the food they consume.

In addition to personal food intake records, collecting biological and lifestyle data from sensors, smartwatches, and lab tests can provide a more holistic understanding of users' lifestyles and health conditions. By integrating these additional data sources, personalized food recommendation systems can generate more context-aware recommendations that better align with users' unique needs and circumstances.

The collection of food-related data, both spatial and non-spatial, is crucial for understanding the food context and providing intelligent recommendations. Spatial food data, stored in the World Food Atlas, includes information about the geographic distribution of food items and their properties, which will be discussed in Chapter 6. Non-spatial food data, including personal food logs and other data, stored in the personal chronicle database and Food Knowledge Graph, encompasses information about food items, their properties, and their relationships, which are the main focus of this chapter. It may be trivial to represent numerical data, but how do we really log food?

In this chapter, we will focus on the collection of personal data, such as food intake records and biological records, for individual users. We will discuss various methods and tools for gathering this data, as well as the challenges and opportunities associated with integrating diverse data sources to create comprehensive personal food models. The subsequent chapters will delve into the World Food Atlas, the Food Knowledge Graph, and other food-related data collection and processing techniques.

\section{Multimedia Food logger}
In this section, we will explore the importance of multimedia food loggers and their applications in capturing essential food-related information. We will discuss the various multimedia input options available to log food intake and the benefits they provide to different stakeholders in the food ecosystem. Additionally, we will examine the integration of food logging data with other physiological data streams to enhance personalized food recommendations.

Food logging serves as a pivotal element within food recommendation systems, necessitating the documentation of an individual's dietary intake and associated metadata over extended periods. Initial efforts in food logging relied on labor-intensive manual diaries, yet advancements in smartphone and camera technologies have significantly enhanced the efficiency and fidelity of these logs. The multimedia and computer vision research communities have made strides in developing food logging systems employing computer vision techniques to identify food items, their constituents, and the quantities ingested. Despite these advancements, the absence of a universal logger capable of accommodating international cuisine and accurately determining ingredient composition and consumption volume persists as a notable challenge.

In response to these limitations, this chapter introduces a comprehensive multimedia food logging platform that leverages a myriad of data sources, including visual recognition, speech recognition systems, payment information, barcode scanning, and sensor data, to facilitate the logging process. Upon identifying a food item and quantifying its consumption, the platform utilizes governmental or commercial databases to retrieve the corresponding nutritional information. Crucially, the development of a Personal Food Model (PFM) necessitates the inclusion of user feedback regarding both the gastronomic experience and physiological responses, which can be derived from various sensors monitoring heart rate, glucose levels, and respiration rate. This multifaceted approach to data collection offers a robust and detailed mechanism for capturing dietary information, essential for the formulation of precise food recommendation systems. The proposed multimedia food logging platform, depicted in Figure \ref{fig:foodlogger}, represents a dynamic tool that integrates diverse, complementary sources of information, ensuring comprehensive and accurate data collection. Furthermore, the platform is designed to evolve by incorporating emerging technologies, thereby maintaining its relevance and utility in enhancing food logging practices \cite{Rostami2020}.

\subsection{Motivation and Importance}
Models are built using data. Most successful search engines, social media, and e-commerce systems utilize personal models to provide people with the right information, at the right time, in the right context, usually even before a user articulates his need \cite{AdomaviciusTowardExtensions}. Personal food model plays the same role in food recommendation systems \cite{Min2019FoodChallenges}. We need to log food consumed by a person and all the relevant metadata over a long period for the user. While initial food logging efforts required cumbersome manual food diaries, smartphones and cameras can drastically improve the quality and ease of logging. Aizawa \cite{FoodLog:Applications} was the first multimedia researcher to champion the idea of logging food using a smartphone camera and remains a very active researcher. Applications for camera-based food-logging have been developed in many other countries \cite{Chen2016Deep-basedRetrieval}, \cite{Bossard2014Food-101Forests}. 
Multimedia and computer vision research communities have been actively exploring food-logging systems. These systems use computer vision techniques to recognize items, their ingredients, and even the volume consumed by the user \cite{Chen2016Deep-basedRetrieval}, \cite{Oh2018MultimodalJournaling}. Unfortunately, there is no generalized logger for international food, and identifying ingredients and volume remains a challenge. A useful review of many visual approaches and descriptions of databases used for training is included in \cite{Min2019AComputing}.

Conversational voice interfaces are becoming quite popular, making rapid progress.  Systems like Alexa and Siri are available at home, in phones, and watches. People can report what they eat, volume, and reaction to food using a simple sentence. Many packaged foods and processed ready-to-eat foods should have barcodes.  Since barcode readers are now omnipresent even in smartphones, one can get all food information from these.  Some sensors measure muscle activity and try to infer food items from that.  These are placed on the chest, near the ear, or neck \cite{Chu2019RespirationSensors}.  These have shown some progress in recognizing eating events but have not gone much beyond that yet.

To overcome these, we propose a multimedia food logging platform, shown in Figure \ref{fig:foodlogger}, that could use many relevant sources to log food items and find all information that may be needed to build a PFM. Multimedia uses complementary and correlated information and provides more comprehensive and precise information than any one medium. Moreover, we will keep adding new technology as it becomes available to keep the logger useful.
Multimedia food loggers are crucial for capturing food intake records in context, with detailed metadata that can be used in learning, analytics, and recommendations for both users and other stakeholders. By understanding the effects of food on various aspects of our lives, such as biology, mood, and overall well-being, multimedia food loggers can serve as the foundation for numerous food-related studies and applications.

\subsection{Food Logger App Layout}

The main purpose of this app is to log food. The app is designed so that anyone without any technical background can use this app in their everyday life to log their food intake in an efficient way, without spending too much time but reasonably precise and accurate by capturing the most important aspects of a food event. This includes basic information about the person, the food item, the nutrients, the time, the location, and the environmental information. The following subsections discuss the layout of the food logger app which will be presented at the demo. 

\subsubsection{Home Page}

The homepage displays a simple representation of today's food intakes and the corresponding times of each today's food events. The first screenshot from the right in figure 2 illustrates the homepage. There is a button at the bottom right of the page which takes the user to the logging page and another button at the bottom left which takes the user to the food journal history. These two parts will be discussed in the next subsections. There is also an optional study-specific region in the middle of the homepage which could visualize the current nutrient intake state to the user. This component is not a main component of the food logger since many studies might not intend to trigger any intervention. Many studies just want to capture the original food intake without any intervention so in those studies this component could be removed. 

\subsubsection{Food Journal History}

The app has a section which displays the food intake journal of previous days as shown in figure \ref{fig:foodlogger}. The user can edit the history and add or delete food intakes for any previous day. The app provides cues such as images and locations to remind the user of that specific day so it can assist the user in remembering the accurate food event better.

\subsubsection{Food Logging Page}
The logging part is the most important part of the app. As it is shown in part B of figure \ref{fig:foodlogger}, we demonstrate how the user can use different media, and how we convert image, voice and text to capture all the important information associated with the food intake by using nutrition databases and other online resources.

\subsection{Multimedia Inputs and Outputs}

\begin{figure}[!ht]
  \centering
  \includegraphics[width=\linewidth]%
    {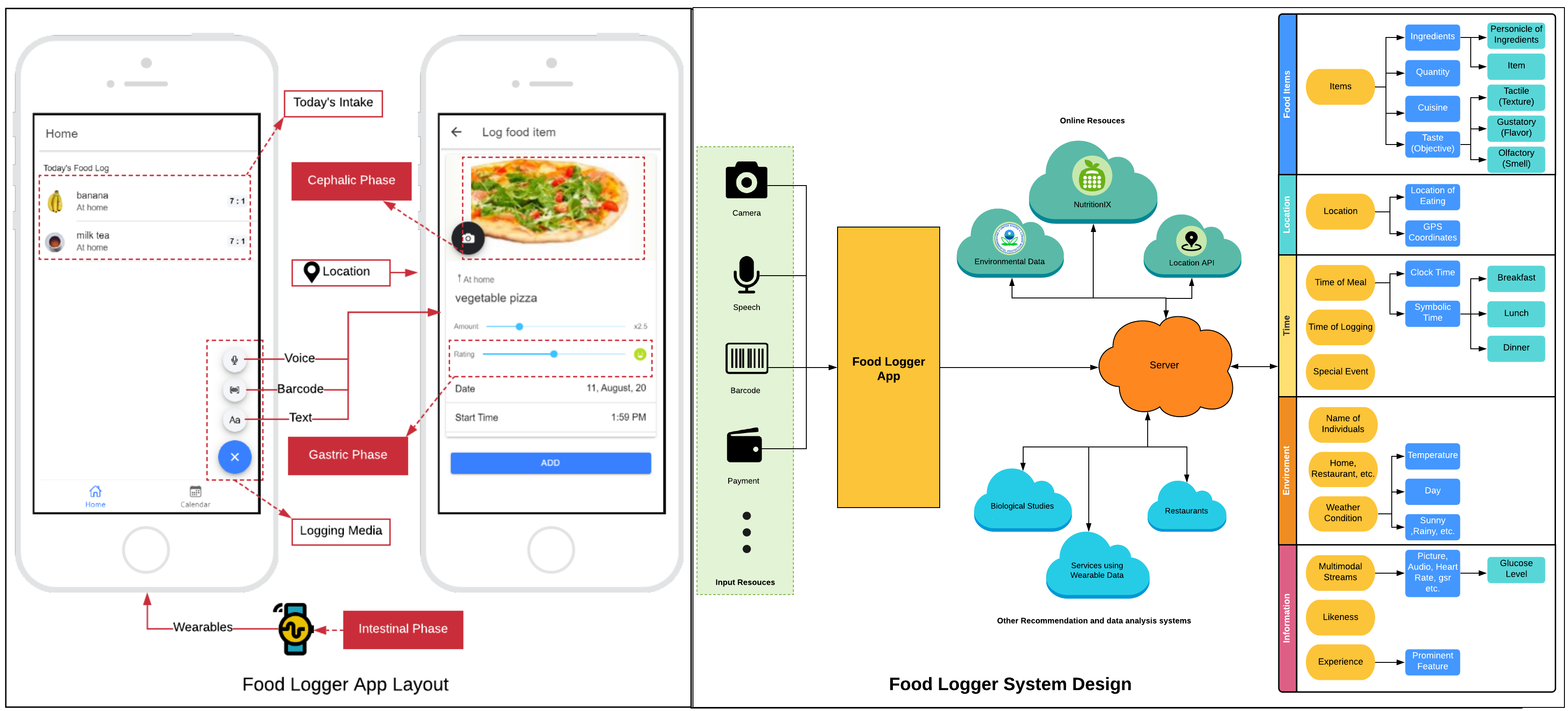}
    \caption{Food logger uses multimedia input sources and uses online sources to add additional information to create a complete food log and provides these integrated information to different external platforms for various applications \cite{FoodLog:Applications}.}
    \label{fig:highlevellog}
\end{figure}

We will demonstrate what goes into the food log and what will be stored for each meal which is logged by the food logger. This information is shown in the right part of the figure \ref{fig:highlevellog}. Some of these are logged directly from the app such as the food item, the image, time, location, the amount and the user's rating of the food \cite{Azimi2019PersonalizedStudy}, \cite{WestermannTowardApplications}. And we also show that some additional information will be collected using other online sources such as NutritioniX, geolocation APIs and environmental databases to pull information such as nutrients, location type and the overall context and integrate all these different information in one place to record a meal event. To signify the multimedia nature that the food logger inherits \cite{Spence2015MultisensoryPerception}, we need to record food in multiple ways and illustrate each approach.

Furthermore, we illustrate how this integrated information is useful. Any application can use our system to obtain the food log of its users and our system will provide them with the integrated information all in one place. We use multiple media types to capture the food log as shown in Figure \ref{fig:foodlogger}. We demonstrate how the data collection is designed to work using voice, text, barcode, visual image recognition. The audience can try logging their food using a variation of these media types. However we acknowledge that visual recognition is not as satisfactory as we would like in this version of the app, but all the other approaches are completely available for the audience to test and experience.

Multimedia food loggers can utilize various media types, such as voice, text, barcode, and visual image recognition, to capture food intake records. Each input method has its unique advantages, making the process of logging food practical and efficient in different circumstances.

Food logging will use multimedia input sources and complement information from online databases to log each meal and all metadata related to the meal. It captures information about the food (dish name, ingredients, quantity), location (place of eating), time (eating and logging), social context (companions), causal aspects (nutritional and flavor information), and multimedia and experiential information about the food.

The different multimedia input options of our app make the process of logging food practically in different circumstances yet much useful information as shown in Figure \ref{fig:foodlogger} is captured for each meal. This information can be used in different data analysis applications \cite{FoodLog:Applications}. Various fields can benefit from the output of the food logger: Better advertisement can be tailored for the taste of individuals, restaurants gain access to the knowledge that what people like to eat in a certain area, and the agriculture industry can predict what percentage should be dedicated to each type of crops to be grown for the next season. Moreover, we discuss that in addition to the food preference analysis and finding tasty foods using the food logger \cite{NewtonAndersonEveryoneCulture}, the data captured by the food logger can be combined with other physiological data streams \cite{Nag2019SynchronizingMonitoring} such as step count, heart rate or any other data streams to determine the healthiness and effects of different food groups on varying groups of people \cite{Oh2018MultimodalJournaling}, \cite{Oh2017FromChronicles}.  

\subsection{Food Journal History}
The food logger app is designed to be user-friendly, with a simple interface that displays daily food intake records and corresponding times. The app also provides access to food journal history and a logging page where users can input new food intake events. An optional study-specific region can visualize the current nutrient intake state for users if desired by researchers.
\begin{figure}[!ht]
  \centering
  \includegraphics[width=1\linewidth]%
    {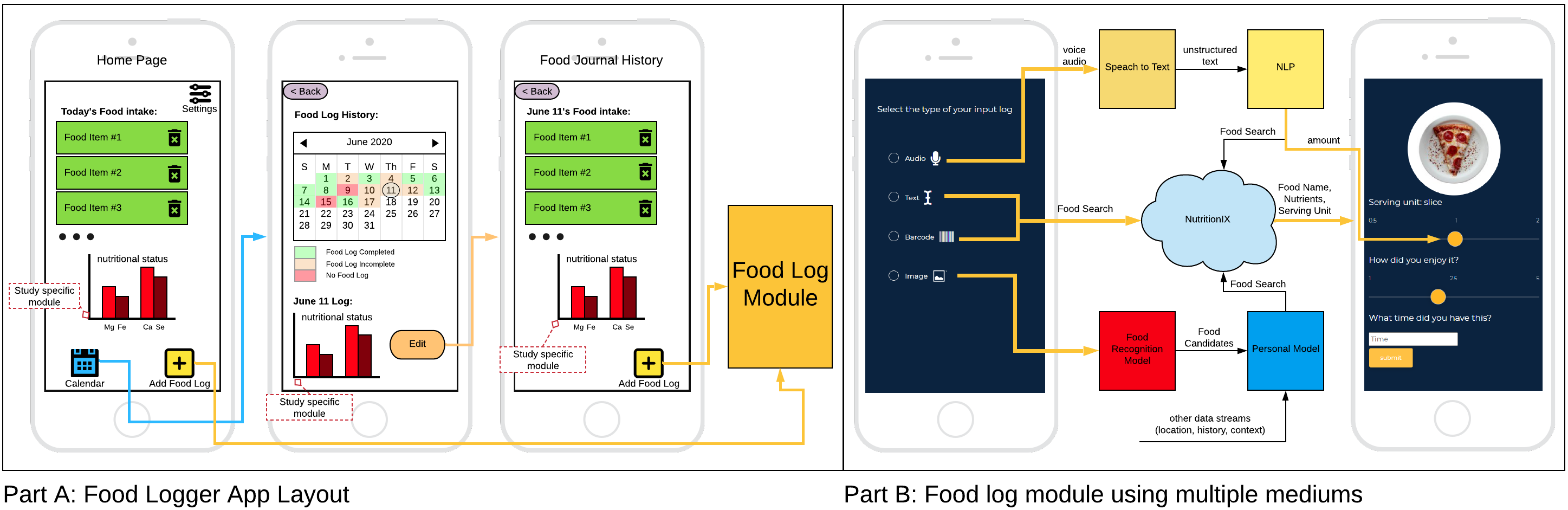}
    \caption{Food logger app architecture.}
    \label{fig:AppArchitecture}
\end{figure}

The app has a section which displays the food intake journal of previous days as shown in figure \ref{fig:AppArchitecture}. The user can edit the history and add or delete food intakes for any previous day. The app provides cues such as images and locations to remind the user of that specific day so it can assist the user in remembering the accurate food event better.

\subsection{Logging Food Intake}
The logging component of the app enables users to input food intake events using various multimedia options. By leveraging nutrition databases and other online resources, the app can capture essential information associated with the food intake, such as food items, nutrients, time, location, and environmental context.

The logging part is the most important part of the app. As it is shown in part B of figure \ref{fig:AppArchitecture}, we demonstrate how the user can use different media, and how we convert image, voice and text to capture all the important information associated with the food intake by using nutrition databases and other online resources.

\subsection{Benefits and Applications}
Multimedia food loggers offer numerous benefits and applications across different fields. They can be used to tailor better advertisements for individual tastes, help restaurants understand local food preferences, and inform the agriculture industry about crop demands. Furthermore, when combined with other physiological data streams, food logging data can be used to determine the healthiness and effects of different food groups on various populations.
This is the beginning towards building a robust multimedia solution to the problem of logging.  There are three important aspects to this platform:
\begin{itemize}
    \item We introduce a multimedia platform that uses speech-based systems, barcodes, and connects to Apple Health and Android Fit to collect biological data. In addition, it is designed in such a way that it could easily be integrated with online services such as Google Lens, which provide visual recognition to log food only by capturing a photo. In the future, many other modalities could be added, such as payment activity analysis, sensors to determine food intake and eating activity and recognize the content of the food, and several similar emerging approaches.
    \item Once a dish or food item is recognized and the amount consumed is known, systems must find the nutritional data using governmental or commercial databases.  Similarly, weather information, social context, and other metadata related to food required by the PFM may come from other sources. 
    \item Finally, for building PFM, the log must contain the user's reaction both in terms of enjoyment and bodily reaction.  The enjoyment information may come from asking the user, and the bodily reactions may come from sensors such as heart rate, glucose measurement, and respiration rate.
\end{itemize}
\subsection{Integration with Physiological Data Streams}
By integrating food logging data with other physiological data streams, such as step count and heart rate, multimedia food loggers can provide more comprehensive insights into the effects of food intake on individuals' health and well-being. This integration can enhance personalized food recommendations and contribute to a better understanding of the relationships between food intake, health, and lifestyle factors.

We enrich the food events with associated nutritional, culinary, and contextual information using databases from different public and private organizations. These include nutrition (NutritionIX, USDA food database), weather, air-quality (airnow provided by EPA), and place (Google Places, Yelp).
\newline
We may also want to capture some biomarkers characterizing the health of the person. These parameters may be continuously recorded and could be used to identify physiological responses to food items \cite{Oh2018MultimodalJournaling}. A personicle like system \cite{Oh2017FromChronicles} can capture this information, and the time-indexed nature of the data and events makes it readily available associating and analyzing with the food events.

These two goals may require different information from the food log. In general, building PFM requires as much longitudinal data as available, while health state estimation requires PFM and recent lifestyle and biological data. We need to keep these goals in mind while designing the food log.
We have followed the HW5 (how, what, when, why, where, what) model as described in \cite{XieEventStreams}, \cite{WestermannTowardApplications} to identify what information can fully describe a food event and maximize its utility for a variety of applications. The different aspects and associated information we capture are detailed in figure \ref{fig:AppArchitecture}.
There can be three types of data collected:
\begin{enumerate}
    \item Observed data: Directly captured using a sensor.
    \item Derived data: We can derive some data and information using sensors and knowledge sources.  This information will depend on the algorithms and data sources used. 
    \item Subjective data: The system may prompt the user or some other human source to get specific information.  This data is prone to errors as it depends on human perception.
\end{enumerate}
The different types of measurements are used in different manners to minimize the error in our analyses and predictions.

\subsection{User Feedback}
In personalized food recommendation systems, user feedback is crucial for improving the accuracy and relevance of recommendations. User feedback can be in the form of explicit feedback, such as ratings or reviews, or implicit feedback, such as click-through rates or purchase history. The integration of user feedback into the recommendation system can provide valuable insights into the user's preferences, tastes, and dietary restrictions, which can be used to enhance the personalization of recommendations.

One of the challenges in incorporating user feedback is the sparsity of data. Users may not always provide feedback, and when they do, it may be incomplete or biased. Therefore, techniques such as matrix factorization and collaborative filtering have been developed to address these issues and improve the quality of recommendations.

Another important aspect of user feedback is the timeliness and relevance of the feedback. In personalized food recommendation systems, users may provide feedback immediately after a meal or days later, which can affect the accuracy and relevance of recommendations. Therefore, incorporating temporal and contextual information into the recommendation system can enhance the quality of recommendations and provide a more personalized experience for the user.

Overall, user feedback is an essential component of personalized food recommendation systems, and its integration can significantly improve the accuracy and relevance of recommendations. However, the challenge remains in addressing the sparsity and timeliness of user feedback and developing effective techniques for incorporating it into the recommendation process.

\subsection{Privacy and Security}
Personalized food recommendation systems rely heavily on collecting and processing user data, which may raise concerns about privacy and security. Users may be reluctant to share their personal data with these systems, especially sensitive information such as health data, without adequate privacy protection measures in place. Furthermore, the use of location-based data for personalized recommendations may reveal users' movements and habits, which could lead to privacy violations.

To address these concerns, researchers have proposed various privacy-preserving techniques for personalized recommendation systems, such as differential privacy, homomorphic encryption, and federated learning. These techniques aim to protect users' data while still allowing for personalized recommendations to be made. Additionally, regulations such as the European Union's General Data Protection Regulation (GDPR) and the California Consumer Privacy Act (CCPA) have been enacted to protect users' privacy rights.

\section{Food Knowledge Graph}
\begin{figure*}[h]
  \includegraphics[width=0.8\textwidth]{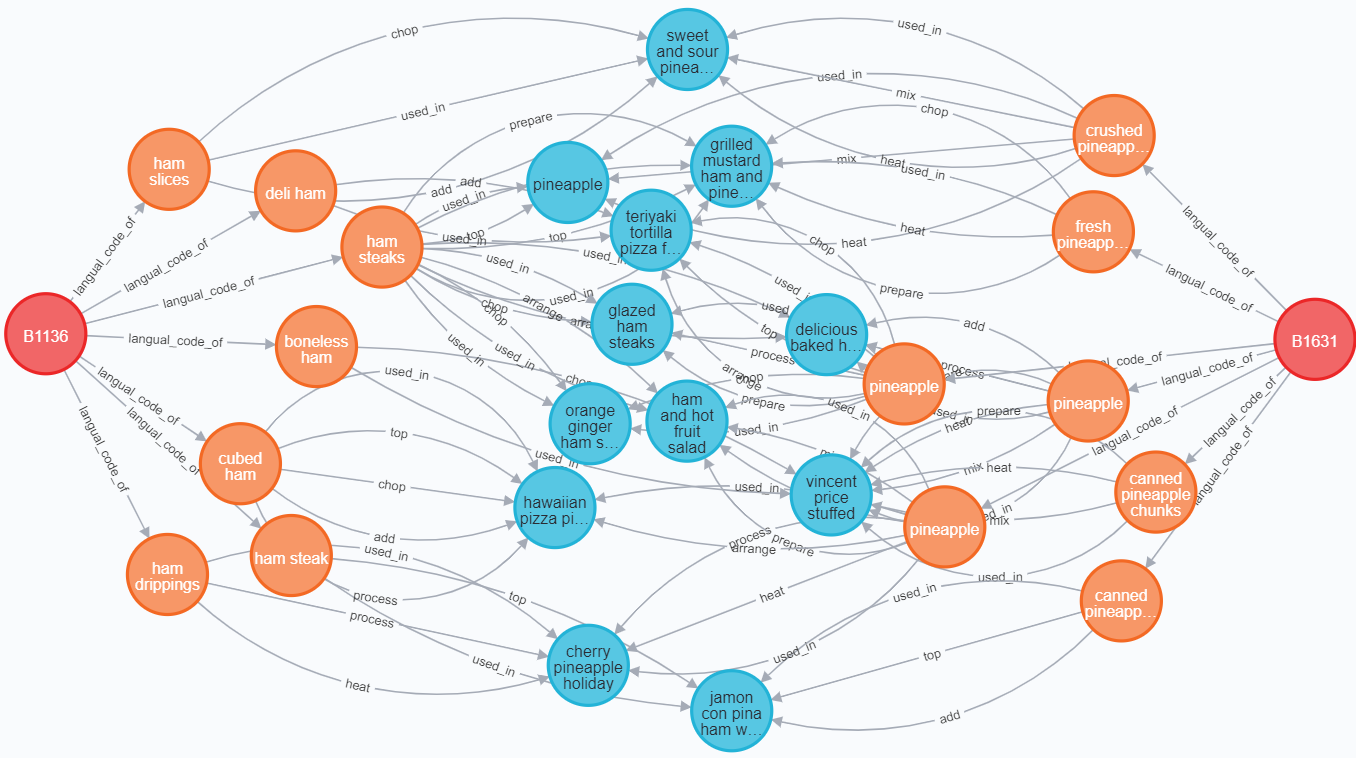}
  \caption{Food Knowledge Graph representation in the Neo4j environment. B1136 is the general pork node and B1631 is the general pineapple node. This image shows the result of the query looking for food items which have both pork and pineapple as ingredients.}
  \label{fig:FKG_snap}
\end{figure*}
FKG is basically the collection of interlinked descriptive information about different aspects of food. FKG contains different dimensions of connectivity and different node types. The ingredient nodes are interconnected by a semantically hierarchical structure obtained from the food ontology database \cite{Dooley2018FoodIntegration} which provides a cue about the similarity between different ingredients and acts as the backbone of the formal semantics of the FKG. Some nodes in the ingredient category are actual ingredient nodes such as cow's milk or chicken, and some other ingredient nodes are abstract nodes which represent a class of ingredients such as dairy or meat. The hierarchical connections between the ingredient nodes makes sure that ingredients within the same category always have a shorter path to each other than to nodes outside of the category. For example cow's milk should have a shorter distance to Greek yogurt than to chicken since cow's milk and Greek yogurt are both a subset of the dairy node. More complex nodes are the recipe nodes, in another dimension of connectivity, the recipe nodes are connected to the corresponding ingredient nodes. The edge connecting a recipe node to its corresponding ingredients contains information about the portion and also information about the step in which the ingredient is involved in, such as frying or mixing. Entity nodes must be able to contain geographical information such as geographical availability, popularity and origin of dishes and some ingredients. Furthermore the detailed nutritional content of ingredients are associate to ingredient nodes which enables the derivation of the nutritional contents of the dish nodes since they are connected to the ingredient nodes. The entity descriptions contribute to one another by providing context for further interpretations which is the main promise of the FKG. In order to create the FKG data from multiple sources need to be brought together to form a foundation which represents the food knowledge as shown in Figure \ref{fig:FKGflow}. Once the FKG is designed with some initial nodes allowing both people and computers to process these formal semantics an efficient and unambiguous manner, expansion standards can enable public contribution from parties with various backgrounds in a semi-supervised manner. However creating the initial backbone of the FKG is in facet a challenging task since it involves problems which are currently hot research topics in different fields such as multimedia processing, knowledge graph formation and optimization, food entity resolution, query refinement and natural language processing. Figure \ref{fig:FKG_snap} shows a sub-graph of the FKG which is the result which was generated for a query asking for all the food items which contain both pork and pineapple.

\chapter{Context-Aware Personal Food Modelling}
\section{Background}
Food serves many functions at an individual level. It provides us with the energy and building blocks to sustain our lives while also serving as a source of personal fulfillment and social glue. Our taste and sensory preferences are a significant causal factor behind our food decisions and affect our health. For this reason, there is a rapidly growing need for personalized food services that guide users towards a healthier lifestyle while also ensuring the food's enjoyment. With the advancement of technology, especially in the recommendation and sensing fields, it is possible to guide users towards a healthier lifestyle by understanding their underlying taste profile and their daily lifestyles to provide healthier recommendations that still appeal to the user's tastes \cite{Rostami21PPFM}. Food is an essential part of our lives, and advancements in applications such as food logging platforms and recipe recommendations can help us identify and improve our eating behavior.
\begin{figure}[!ht]
  \centering
  
  \includegraphics[width=0.7\linewidth]%
    {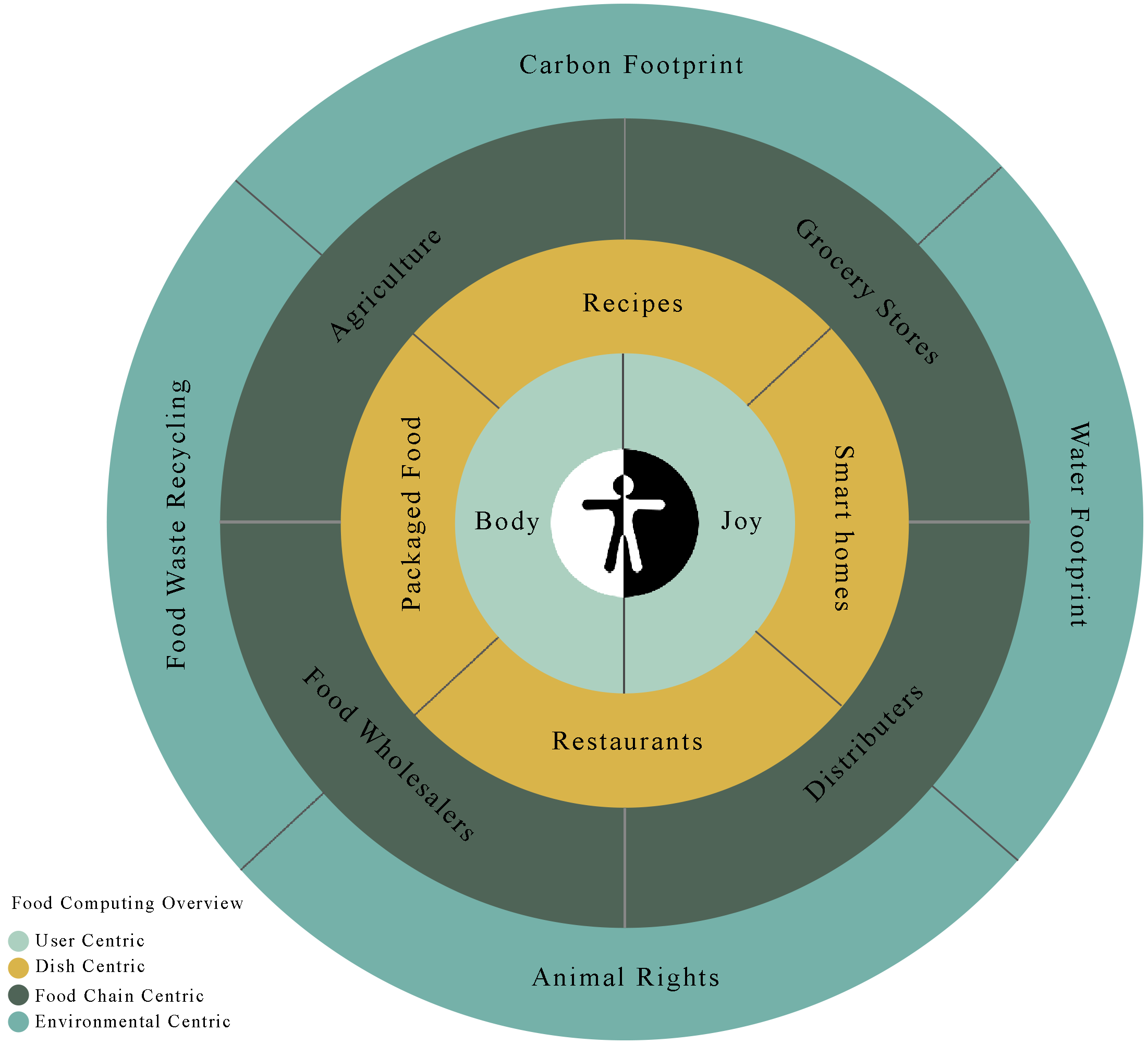}
    \caption{Personal food computing overview and relevant proposed layers.}
    \label{fig:personcentric}
\end{figure}
At the heart of these personalized food services lies the Personal Food Model\cite{Rostami2020PersonalModelb}. As shown in figure \ref{fig:personcentric}, it (PFM) has two main components: 1) the biological component and 2) the preferential component. The biological component determines how different food items interact with our biology and health. In contrast, the preferential component captures how different contextual and environmental factors impact our food preferences and, in turn, affect our choices. 
There has been a lot of work done on context-aware preference modeling and recommendations. However, the current approaches are still far from truly personalizing these recommendations.
Especially for health and food-related applications, the contextual factors can be captured by different multi-modal devices and applications. Typically, these applications store individual data in their silos, which do not interact with other applications. 
We propose a comprehensive food event model that could provide a mechanism for these applications to cross-utilize each other's data. We also demonstrate how we could utilize the data collected using such applications to create a user preference model and how it varies with different contextual factors such as stress and temperature. We use event mining principles to model the contextual relationships in an unsupervised manner.

\section{Related Work and Motivation}
We first take a look at current food models and services. Diet and chronic diseases are closely correlated \cite{Shivappa2019DietInflammation}. A Western diet characterized by a high intake of energy-dense and processed food is a risk factor for many chronic diseases, including diabetes, obesity, and cardiovascular diseases \cite{Shi2019GutDiseases}. A significant hurdle towards maintaining a healthy diet is the draw towards pleasure (rather than health) from eating. Tasty and palatable food is often linked to chronic health risks. According to some interesting experiments by \cite{Raghunathan2006TheProducts}, the more unhealthy dish items are perceived to have better taste and are enjoyed more during actual consumption. Thus, future meals have a greater preference for it, especially for tasks when a hedonic goal is more salient. However, this is not necessarily true. Many other instances show that a healthy option can be tasty.
A good example is discussed in \cite{Apaolaza2018EatWellbeing}, where the result of the study shows that not only organic food has a great positive impact on well-being but can also be pleasurable. Furthermore, the Mediterranean diet, which appears to be tasty to many, is linked to the prevention of Chronic Diseases \cite{Romagnolo2017MediterraneanDiseases}. It may seem that delicious foods must be unhealthy and healthy foods must be dull, but this is not the case. Since the taste preference of different people varies considerably based on many factors such as environmental, cultural, and genetics \cite{Risso2017ACulture}, it is impossible to manually come up with a healthy diet that would appear tasty to everyone. People have their personal taste profile, which could be quite complex and changes with the context. Personalized food services need to create the PFM of the user, which understands how the food affects the user's body and the user's taste profile and what kind of food the user enjoys \cite{Rostami2020PersonalModelb}. But there aren't any concrete methods to create the user's taste profile since taste digitization and quantification are challenging. \cite{Rostami2020PersonalModelb} presents a novel US4B taste model as a unified model to capture taste. While this model is very promising, there is still no precise method to obtain the taste value of different food items. That's why most food models and recommendation services completely ignore the pleasure aspect of the food and purely focus on health and nutrition like \cite{Drescher2007AMeasures}, \cite{Abhari2019AAspects} and \cite{Namgung2019MenuChildren}. 
\newline
Some other works have recognized preference as an important factor and have tried to create a simple preference model. For example, \cite{Li2018ApplicationRestaurants} considers the spiciness of the food as the cue for its taste and as a reference point, which is an interesting improvement. Still, the spiciness alone is not adequate to model the taste profile. \cite{Harvey2013YouPrediction} proposes a recipe recommendation method based on similar ingredients in the dishes with a high rating from the user. Ingredient-similarity-based methods like \cite{Nirmal2018OptimizationApproach} are a significant step towards finding healthier options that the user can enjoy. However, they are still limited to replacing one or two ingredients, and the gap to understanding the taste profile of the food remains. 

There are many methods for providing foods with suitable nutritional content, as reviewed in \cite{TrangTran2018AnDomain}. However, food is more than just refueling energy; it is an experience. Therefore it is essential to capture all the aspects of the experience to have enough information to create the personal food model. Different aspects of a food event can change the impact on the body and affect the enjoyment aspect. \cite{Adam2007StressSystemb} brings good intuition that stress has a strong correlation with food. Time of eating is also an important factor in how the food affects the body \cite{Chaix2019Time-RestrictedDiseases}. \cite{Rostami2020PersonalModelb} uses food events to capture many different aspects of a meal such as location, time, and amount, but a standard food event model doesn't exist. 
The causal aspect of events poses an interesting problem as compared to the other aspects \cite{Westermann2007TowardApplications}. It requires events from other sources as well and finding the causal contextual factors. But the personalized causal aspect of the food events have not been studied to date. Some population studies have explored the causal aspect of specific diets and their relation with age and weight like \cite{vanMeer2016FoodAge}, but these insights do not generalize to every individual.

Some studies have explored the causal aspect modelling in other health related applications such as health state estimation and context-driven nutrition recommendation. The eventual goal of causal aspect modelling, especially for food events, is to enable context-driven health state navigation that would help individuals achieve their health-related goals.

There are multiple statistical techniques available for causal and context-driven modelling. There has been an increase in embedding causal relationships within recommendation systems \cite{Bonner2018CausalRecommendation}. We have adopted an event driven approach and utilize event mining \cite{Jalali2016InteractiveStreams} for finding causal relationships between different lifestyle events \cite{Pandey2020PersonalizedSleep}. This approach utilizes principles from Pearl's \textit{do-calculus} \cite{Pearl2009CausalOverview} and tests for causal relationships in different bins of confounding factors. This approach results in relationships between events, and such insights are inherently interpretable, which is a desirable quality for personal models\cite{Schafer2017TowardsSystems}. This approach would also be able to fully leverage the benefits of a comprehensive event model.
\begin{figure}[!ht]
  \centering
  
  \includegraphics[width=0.7\linewidth]%
    {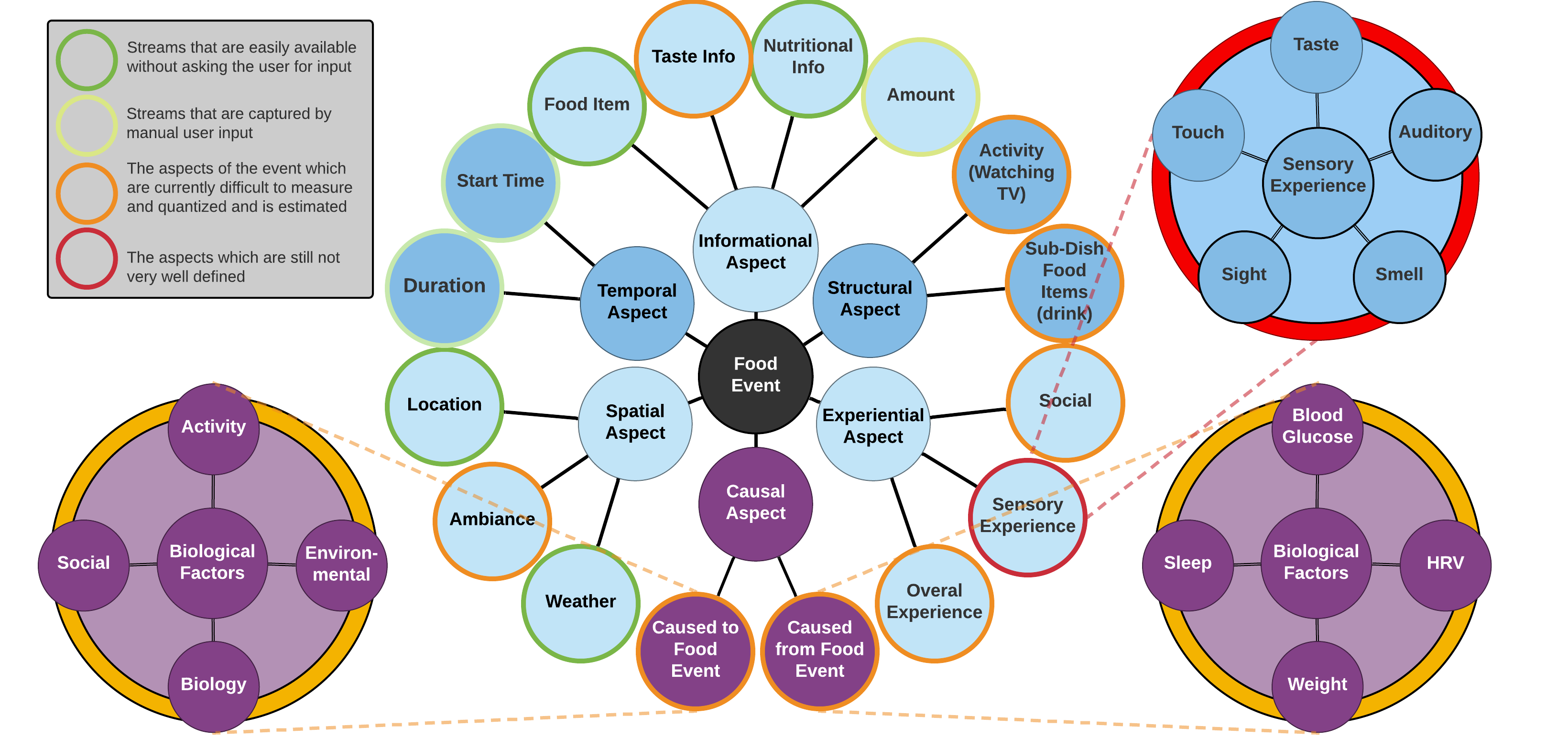}
    \caption{Food event causal analysis.}
    \label{fig:food-event}
\end{figure}
\section{Food Event Model}
A single food event has multiple facets captured by different applications. The food being eaten, the time of the day, the amount of food, the location type, the ambience, the person eating the food, and other people involved in that event are some examples. If we capture sufficient contextual information about the food event, it will be possible to find what caused the food event. Furthermore, collecting information about the body's response to the food event opens the door to understanding the biological responses to the food event. We propose a comprehensive food event model that includes all the elements defining the food event, as depicted in figure \ref{fig:food-event} which draws inspiration from Westermann and Jain's multimedia event model described in \cite{Westermann2007TowardApplications}. The food event model consists of 6 main aspects: \textbf{spatial}, \textbf{experiential}, \textbf{informational}, \textbf{structural}, \textbf{temporal} and \textbf{causal} aspect. Each of these aspects has sub-components, as illustrated in figure \ref{fig:food-event}. Some of these aspects have been studied extensively and are widely captured, such as location and time. We can capture the biological aspects in free-living conditions thanks to the advances in Internet-of-Things (IoT) and wearable technologies such as sleep monitoring \cite{AsgariMehrabadi2020SleepPreprint}. Physiological data-streams such as Electrocardiography (ECG) and Photoplethysmography (PPG) are non-invasive and low-cost techniques and enable continuous health and well-being data collection \cite{Naeini2019AInternet-of-Things}, \cite{KasaeyanNaeini2019AnMonitoring}. However, some other sub-components are challenging to collect as they are not understood very well, such as the sensory experience. Auditory and visual information are the only exceptions and are well understood in the multimedia field; however, no such model exists for the sense of taste and smell. 
The taste information of a food event is crucial for building the preferential side of the personal food model, but to the best of our knowledge, there is currently no method to map food items to ingredients to taste information. 
In this chapter, we propose a novel approach to capture information about the taste experience of a dish driven by the informational aspect discussed in detail in the Experiment Design section.

\section{Causal Analysis}

\subsection{The Causal Aspect}
Identifying an event's cause(s) is not an easy question to answer even in trivial cases. Numerous factors could affect a food event, such as physical activity, social gathering, weather, or the time of the day. Modeling this aspect of a food event is extremely challenging using current methods; however, it is critical for building a Personal Model. As shown in figure \ref{fig:food-event}, the causal aspect has two sub-components: events that caused the food event and events that the food event has caused. Events caused by a food event are mostly reflected in the biological impact of food. This includes changes in heart rate variability, sleep quality, and other effects on the individual's body and health.
On the other hand, the events that caused a food event could be more complicated, as external factors could also influence and initiate a food event. These include social events, environmental factors, and weather conditions. Many environmental and biological factors have been known to affect a food event. Some biological factors may be easier to model, for example, age and weight, which are shown to affect the food decision making process \cite{vanMeer2016FoodAge}. Psychological aspects may be more challenging to measure, like mental stress; however, many studies have shown that it can be measured using wearable devices and even social media usage \cite{Saha2017ModelingCampuses}\cite{Saha2019AUse}. \cite{Adam2007StressSystemb} brings excellent intuition on how stress can have a strong influence on food choice. A food event also depends on environmental factors such as the weather \cite{Ito2018ModelingProcess}. Complex environmental causal factors are often missed, which can bring substantial help in the context reconstruction. For example, a pandemic, such as the COVID-19, can drastically affect the food habits of populations. \cite{Mehrabadi2020TheAnalysis} shows that for geographical regions with higher numbers of daily COVID-19 cases, the historical trends in search queries related to bars and restaurants are strongly correlated with re-openings happening in those areas. Therefore the environmental knowledge is an essential factor in the causal aspect of the food. 
In this chapter, we picked two important causal factors: stress and weather, to demonstrate how the context can change the user's food preference profile and analyze how they affect the dietary choices, which will be discussed in the following sections.

\begin{figure}[!ht]
  \centering
  
  \includegraphics[width=1\linewidth]%
    {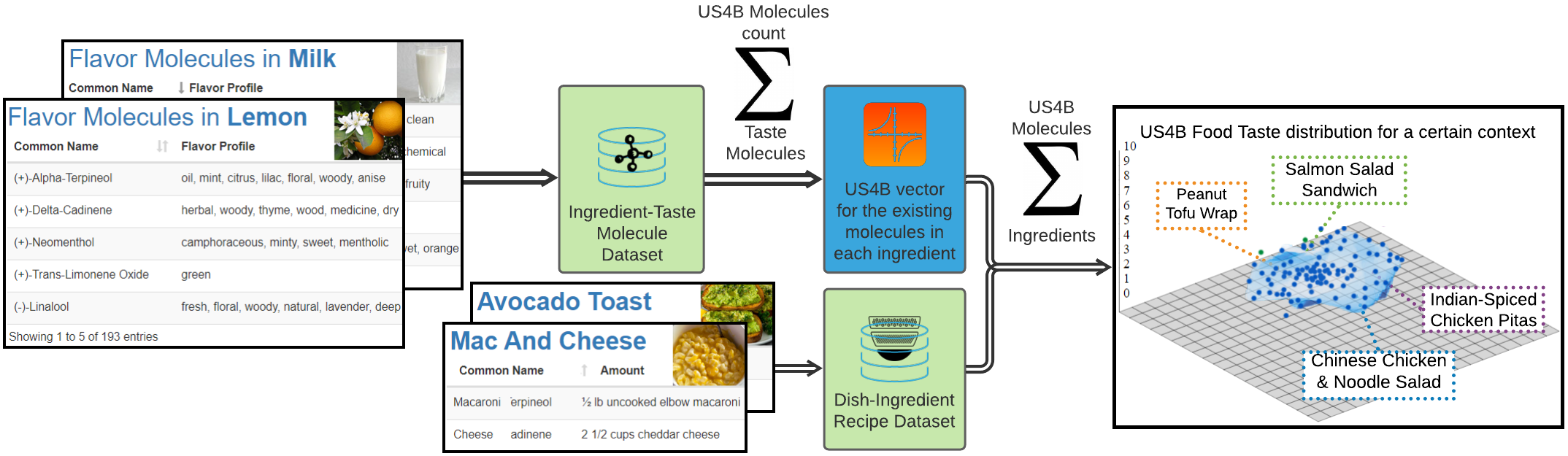}
    \caption{Taste Space Generation: This figure illustrates our approach to map the food items to a corresponding US4B vector in the taste space using the taste information associated with the present molecules in each ingredient. }
    \label{fig:taste-space}
\end{figure}
\section{Food Preference Space: Taste space}
The taste of food items can be very complex. The taste sensory aspect has not been modeled as robustly as the visual and auditory senses, which have standard models such as the RGB color space model. In \cite{Rostami2020PersonalModelb}, the authors demonstrated how a unified and robust taste space model is required to create a preferential personal food model. They presented the US4B taste space, which includes six dimensions: umami, salty, sweet, sour, spicy, and bitter. However, that work was the initial step and did not provide any actual taste dataset or concrete approach to build such a dataset. There is currently no available method that could approximate the US4B values of a set of dishes using available resources. We provide a novel approach that uses the taste molecules to estimate a dish's US4B taste space using its ingredients. 
FlavorDB \cite{Garg2018FlavorDB:Molecules} is the only existing publicly available extensive data set on food taste that contains the list of taste molecules associated with each food item, and a list of taste and smell attributes to each molecule. However, we could not derive the US4B values for dish items directly from flavorDB as there are a few limitations to this data set. This dataset only has the information for ingredients and lacks the information for dish items and recipes. The dataset does not have any information about the intensity of the taste for different molecules. As shown in figure \ref{fig:taste-space}, we start by counting the taste molecules associated with each element of the US4B taste model and create a taste vector for each ingredient item. Then we use a recipe data set containing a list of ingredient items for each dish and use the calculated US4B vector for ingredients in the previous step to calculate a US4B value for each dish based on adding the taste vectors of the ingredients. The taste values for dishes create the personal food model based on the food items they consume in different situations. 
We used this approach to create a taste profile of 60 different dishes. We picked 20 dish recipes for each meal type: breakfast, lunch, dinner. Each meal type has ten dishes for a heavy meal option and 10 dishes for a light meal option. This data is fed to a randomized markov-chain event generator, described in the following section, to create a randomized events log, including food events.
\begin{figure}[!ht]
  \centering
  
  \includegraphics[width=1\linewidth]%
    {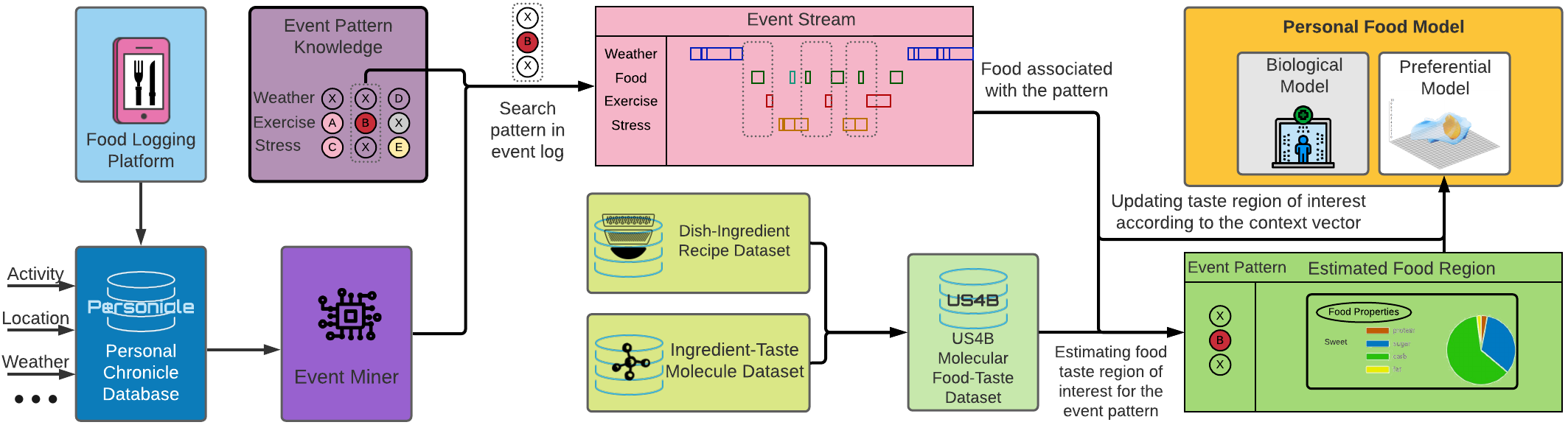}
    \caption{Causal Preferential Model Architecture: The food logging platform captures the different aspects of the food events. We use event mining to find contextual patterns and build a taste profile for each pattern and update the preferential subsection of the personal food model.}
    \label{fig:causal-preference-model}
\end{figure}
\section{Location Data Sources for Personalized Food Recommendation}
Personalized food recommendation systems rely heavily on location data to provide relevant recommendations to users. The accuracy and reliability of these systems depends on the quality and diversity of the location data sources used. In this section, we will explore some of the common location data sources used in personalized food recommendation systems.


Global Positioning System (GPS) data is the most common location data source used in personalized food recommendation systems. GPS data provides accurate location information to the system, which can be used to recommend nearby restaurants, food trucks, or grocery stores. GPS data can also be used to track user movement patterns, which can help the system learn more about their preferences and habits. However, GPS data has some limitations, such as battery consumption, accuracy in urban areas, and signal loss in indoor environments.


User check-ins are another common location data source used in personalized food recommendation systems. User check-ins provide information about the user's current location and the food or restaurant they are visiting. This data can be used to recommend similar restaurants or food items to the user. User check-ins also provide valuable social data that can help the system understand the user's preferences and interests. However, the accuracy and reliability of user check-ins can vary, as some users may forget to check-in or intentionally provide false information.


Geotagged social media posts can also be used as a location data source in personalized food recommendation systems. These posts provide information about the user's current location and the food they are consuming. This data can be used to recommend similar food items or restaurants to the user. Geotagged social media posts also provide valuable social data that can help the system understand the user's preferences and interests. However, the accuracy and reliability of geotagged social media posts can be limited, as not all users geotag their posts and some may intentionally provide false information.


Wi-Fi access points can also be used as a location data source in personalized food recommendation systems. Wi-Fi access points provide information about the user's location based on the Wi-Fi networks they are connected to. This data can be used to recommend nearby restaurants or food trucks. Wi-Fi access points also provide valuable indoor location data, which can be used to recommend food items or restaurants within a specific building or complex. However, the accuracy and reliability of Wi-Fi access points can be limited, as not all locations have Wi-Fi access points and some may be incorrectly labeled.

Overall, location data sources are a critical component of personalized food recommendation systems. By using a combination of different location data sources, these systems can provide accurate and relevant recommendations to users. However, each data source has its limitations and challenges, which must be carefully considered when designing and implementing these systems.

\section{Experimental Design}
We present a novel food preference model that considers causal factors to estimate the taste preferences in a particular context. The food model captures the user's preferred taste region, which could change with context. Figure \ref{fig:causal-preference-model} illustrates the overall architecture of the preferential food model. Food logger \cite{Rostami2020MultimediaLogger} collects information about the food event and stores in the Personicle. The Personicle is a database containing different data streams about the user over a long time in one place\cite{Oh2017FromChronicles}. We apply event mining operators on the user's personicle \cite{Pandey2018UbiquitousHealth}\cite{Pandey2020PersonalizedSleep} to identify contextual factors that impact food preferences. We create event patterns relating different contextual factors with the meal events and find all occurrences of these contextual event patterns in the event streams. This allows us to find food items consumed in different contexts. We can then aggregate the corresponding taste vectors to find the contextual taste preference 
\begin{figure}[!ht]
  \centering
  
  \includegraphics[width=\linewidth]%
    {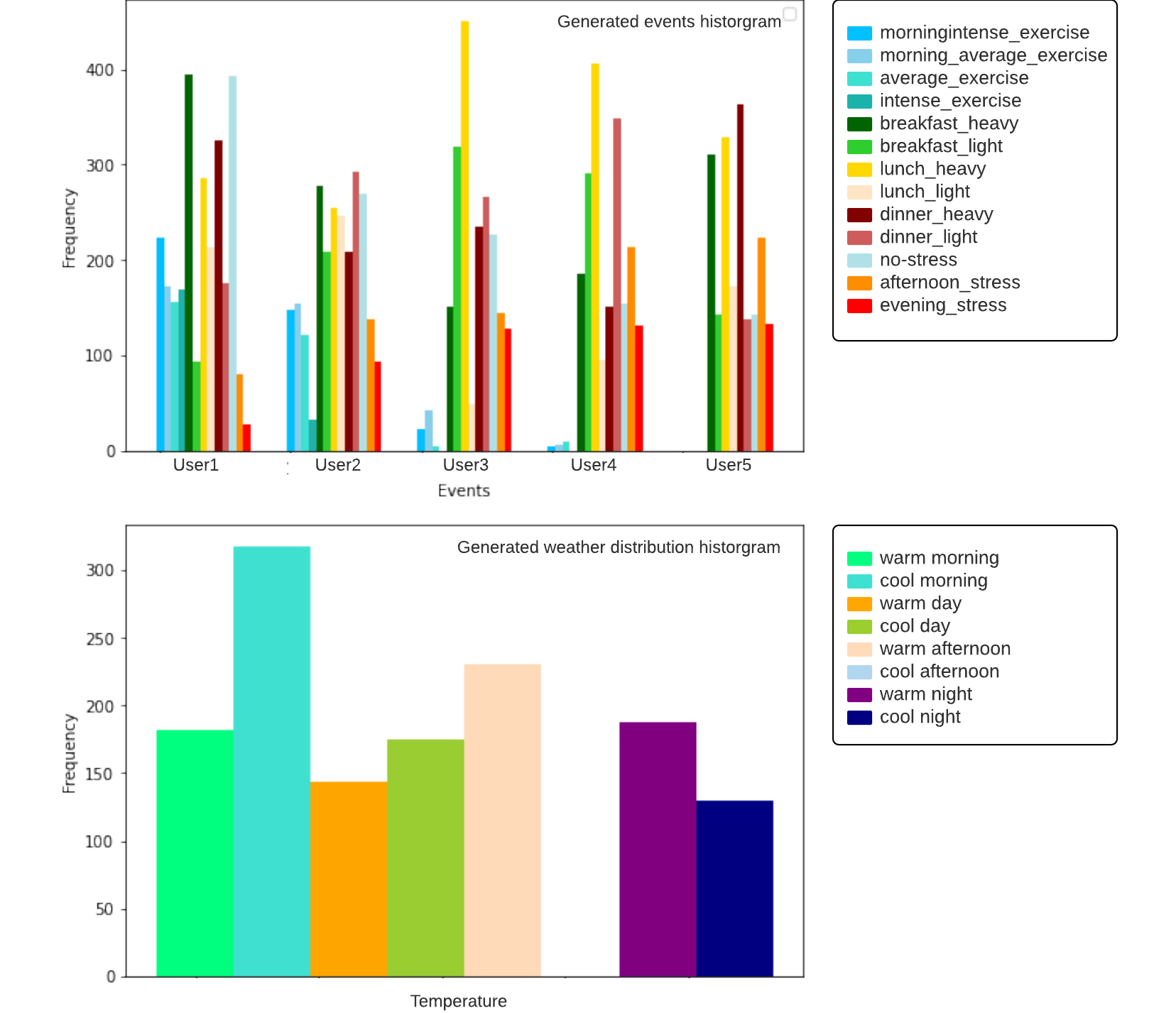}
    \caption{Dataset Summary: This figure displays a summary of our events dataset. This includes the frequency distribution for the different events that are present in the events log for the five people in our dataset. The event relationships were encoded as probabilistic transitions in a Markov-chain model. Concurrent and past contextual events also affect the parameters of the lifestyle events.}
    \label{fig:data-distribution}
\end{figure}


We opted to utilize synthesized data for the experiments because we can use the ground truth of contextual factors' impact on taste to validate the model, just like in many other works such as \cite{Barnard2002AData}. By using synthesized training data, the dataset size can be significantly increased with little human labor \cite{Chen2018SampleAhead:Data}. \cite{Patki2016TheVault} introduces a system that automatically creates synthetic data to enable data scientists. \cite{Patki2016TheVault} suggests that synthetic data can successfully replace original data for data science if it meets two requirements: First, it must somewhat resemble the original data statistically to ensure realism and keep problems engaging for data scientists. Second, it must also formally and structurally resemble the original data so that any software written on top of it can be reused.

We designed a rich event stream database, and the occurrences and parameters of these events depend on the contextual factors such as time of the day, temperature, and stress. We use the novel US4B taste estimation method to estimate the taste-related molecules' quantity in a dish as the taste cue for the personal model. We showcase multiple experiments on our rich event stream data set generated using a randomized Markov-chain based event generator.
\newline
We created five different lifestyle profiles, which would determine the generated events for five different people over a period of 500 days with approximately three food events a day for a total 7373 food events (Figure \ref{fig:data-distribution}). The lifestyle profiles consist of the parameters needed for the Markov-chain model to generate the event streams. These parameters include the probability distribution of each event occurrence based on the previous event, designed to imitate a natural event stream. These events include food events that are controlled by contextual variables such as stress and weather. Research shows that stress correlates with eating more palatable and delicious food \cite{Adam2007StressSystemb}. Even though the relationship could be both ways, either overeating or not eating as much depending on the person. We designed the parameters associated with the stress-related causal aspect of a food choice based on the available findings such that if a person had a stressful day, it would impact their food choice towards more palatable foods for some subjects and towards less appetite for others. Accordingly, some of our synthetic subjects have a higher probability of having a stressful day than the others to achieve a greater variety within the dataset. The causal relation of weather context with food choice has also been studied. \cite{Ito2018ModelingProcess} shows that combining weather context in food profile modeling yields better results. We also have distribution parameters regarding the weather condition and parameters that affect each subject's food choice based on the weather context. 
We then use event mining operators to find causal relationships between contextual factors and meal events in the generated data. The underlying causal relationships in the synthetic data were hidden from the event mining system and the person doing the analysis.

\begin{figure}[!ht]
  \centering
  
  \includegraphics[width=1\linewidth]%
    {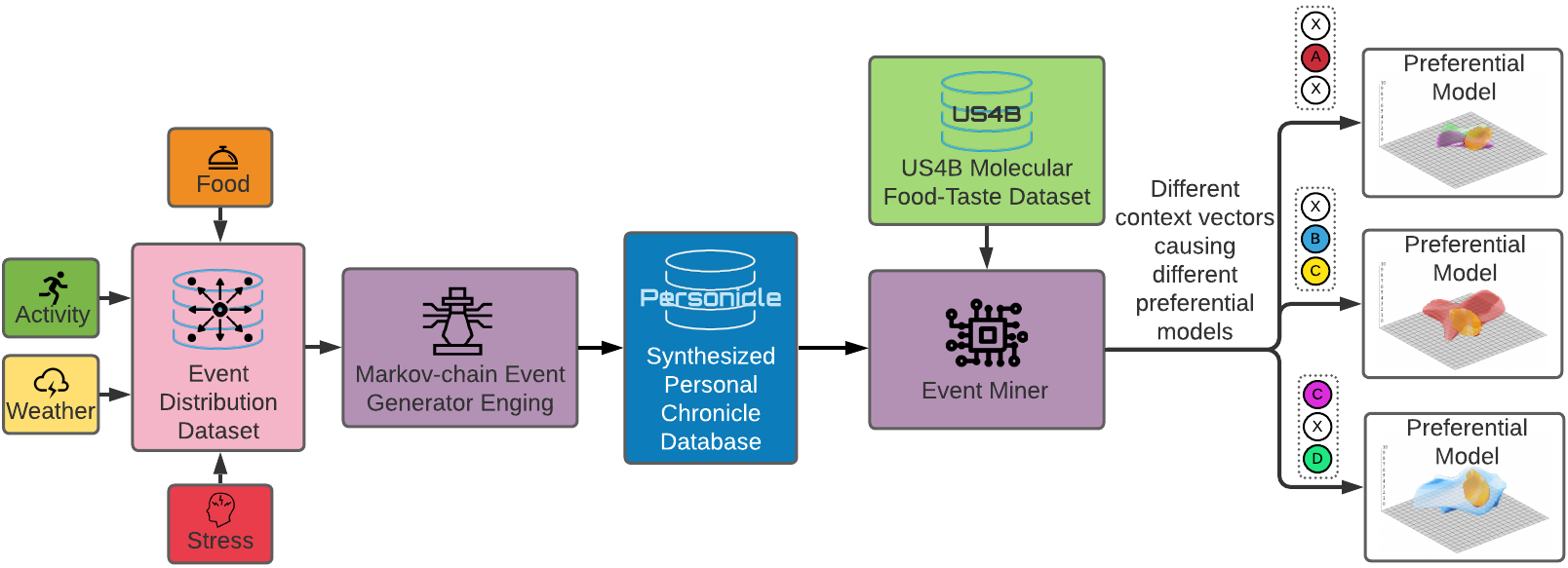}
    \caption{Experimental Design: This figure illustrates how we perform hypothesis testing using synthetic data. The events dataset contains the list of event types which are to be generated such as food and activity. The events must resemble the real data statistically so the parameters are carefully selected and are fed to the Markov-chain event generator engine to create the synthesized dataset. Then we use event mining to apply our model to the dataset and test its viability in action. }
    \label{fig:architecture}
\end{figure}

\section{Results}

We attempted to answer three research questions (RQ) in our experiments:
\begin{enumerate}
    \item How does the individual taste preference vector change with changes in contextual parameters?
    \item How does adding context-awareness change the predictive performance of the preference model?
    \item How much data is needed to create a stable model?
\end{enumerate}

\subsection{RQ1: Contextual variation in taste profile}
\begin{figure}[!ht]
  \centering
  
  \includegraphics[width=1\linewidth]%
    {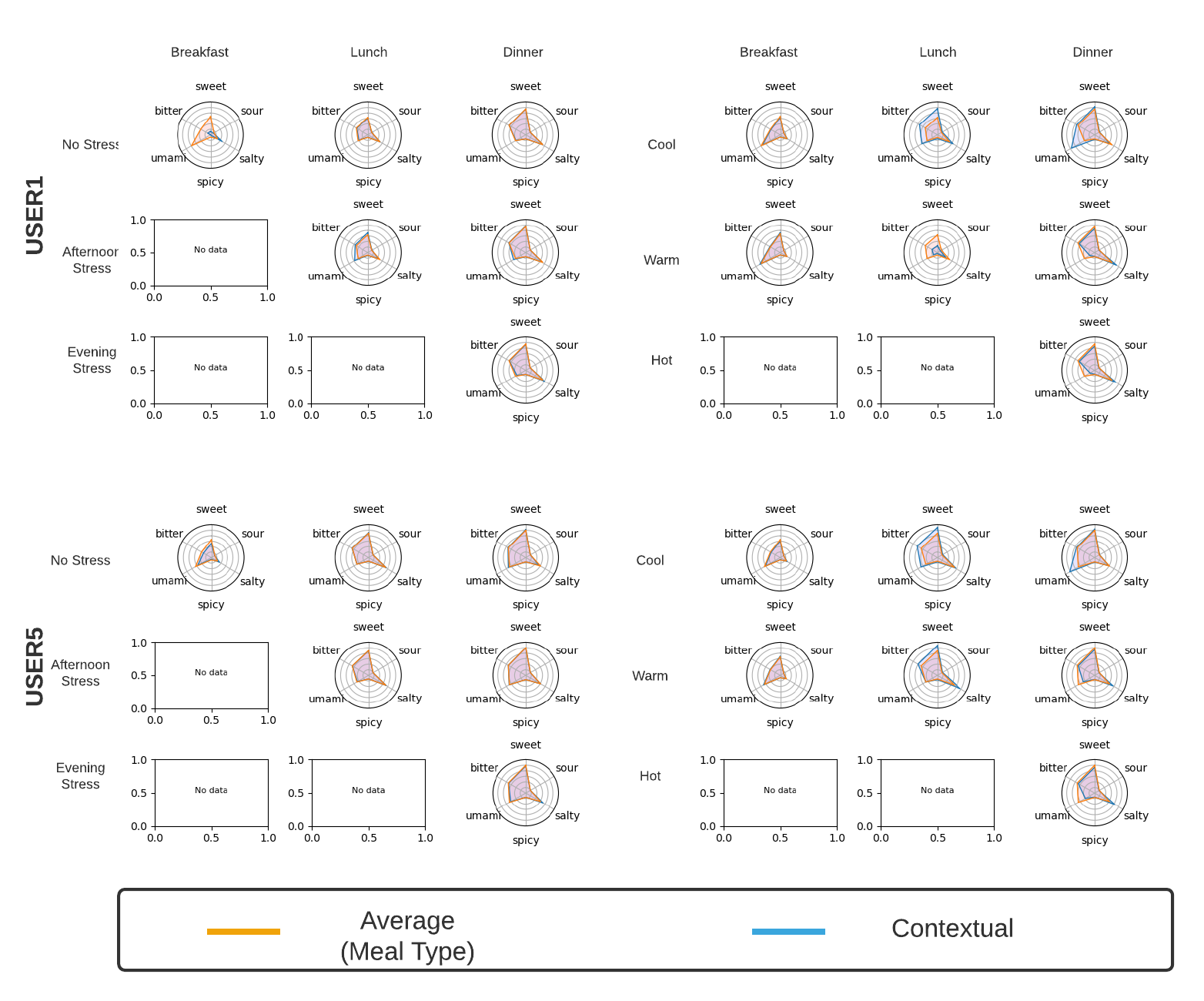}
    \caption{Variation in taste preferences with context. This figure shows how the preferences for different taste aspects change with context. We can see that for User1, the preference for sweet, bitter, and umami flavors during lunch goes down with increase in temperature. }
    \label{fig:contextual-result}
\end{figure}

Figure \ref{fig:contextual-result} shows the variation in the preferred taste profile with different contextual variables for the five individuals in our dataset. We created the individuals' contextual taste profile by averaging the US4B taste vectors for meals consumed in different contextual situations. Thus, every individual has nine contextual preference vectors (3 temperature levels * 3 stress levels) for every meal (breakfast, lunch, and dinner). The contextual preference vectors are compared against the average preference vector for the three meals in the radar-plots in fig. \ref{fig:contextual-result}. We have included the radar plots for two users. 
We can see that \textbf{user5} has an increased preference for umami flavored food for dinner when it's cool outside, but that preference goes down with an increase in temperature, and \textbf{user1}'s preference for sweet, bitter, and umami flavors during lunch goes down with increase in temperature.

\begin{figure}[ht]
  \centering
  
  \includegraphics[width=\linewidth]%
    {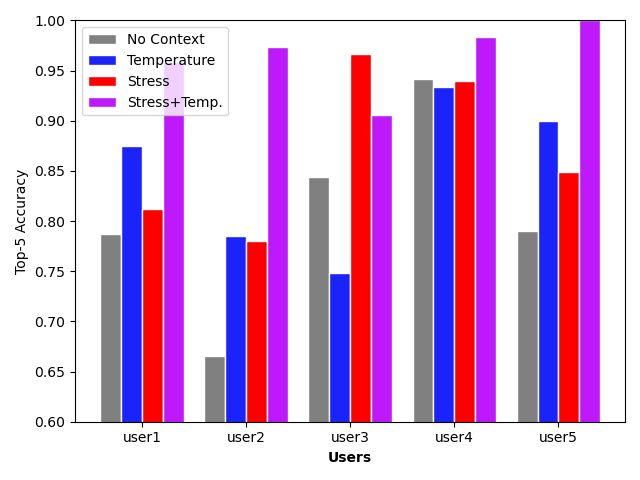}
    \caption{Model performance using Top-5 predictions accuracy. We can see that for all users adding all contextual factors (Stress+Temperature) leads to a better model than no contextual information. }
    \label{fig:prediction-accuracy}
\end{figure}
\begin{figure}[ht]
  \centering
  
  \includegraphics[width=0.8\linewidth]%
    {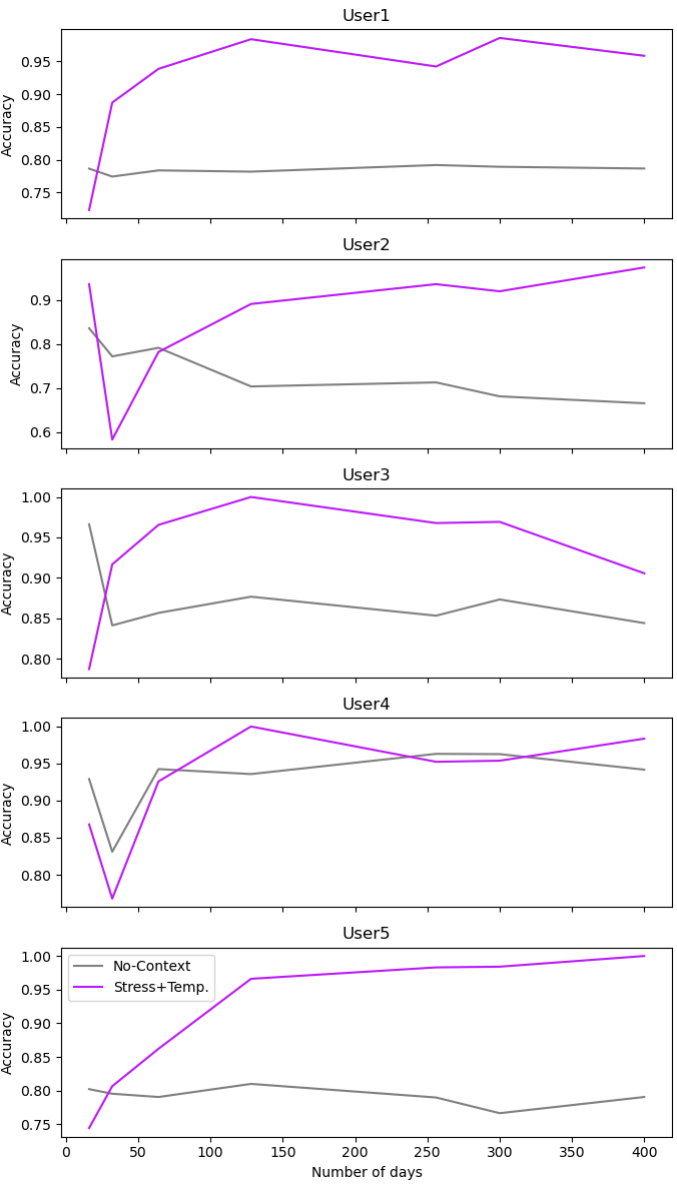}
    \caption{Model accuracy vs training data volume. With adequate volume, the model gains higher accuracy.}
    \label{fig:accuracy-vs-volume}
\end{figure}
\subsection{RQ2: Comparison of prediction performance}

We compared the context-aware preference model's predictive performance against the "No-Context" preference model using Top-5 accuracy as our performance measure. We used an 80-20 train-test split while maintaining the chronological order (test set samples were from a period after the training set samples) and report the models' performance on the test set.
We used a nearest-neighbor approach to match an individual's preference vector with the available food items using cosine similarity. We predicted the five most likely food items for every meal event in the test set and compared the predictions against the meal event's actual dish.
\newline
Figure \ref{fig:prediction-accuracy} compares the performance of models with different levels of contextual information. As we expected, adding contextual information leads to a better performance than the "No-Context" model for all five individuals in our dataset.

\subsection{RQ3: Model accuracy with training data volume}
We also performed experiments to find how the model accuracy varies with the amount of training data. We used a fixed test set containing events data for 100 days. The training set size was varied on a logarithmic scale from 4 to 400 (with a factor of 2). We used the top-5 accuracy metric, and the results are reported in fig. \ref{fig:accuracy-vs-volume} for all users. We observe that initially, the non-contextual model outperforms the context-aware model. This could be due to a lack of data in different contextual situations. This explanation is supported by the observation that as the size of the training dataset increases, the context-aware model outperforms the non-contextual model. The accuracy graph starts flattening at 128 days; thus, we believe that we would need to collect about 100 days of events to train and use this model effectively.

\section{Context in the Integrated Feasibility Study}
This section elucidates the role and composition of context within our integrated holistic feasibility study, particularly addressing the significance of location as both an input and output within the contextual framework.

Context in our study is multifaceted, incorporating various factors and subcomponents derived from numerous sources. Location serves as a prime example of such data, initially captured by the food logger before being fed into the Personicle. The Personicle, in turn, processes this and other aggregated data to construct a personal chronicle—a time-ordered array of events, most of which include location attributes. Consequently, the aggregated data is further processed to delineate the individual's context, encapsulating location among its elements.

It's important to clarify why location appears at both the beginning and end of this process. While initially a raw data point collected by the food logger, location gains contextual significance once integrated into the personal chronicle and analyzed in conjunction with other data streams. This dual presence of location highlights its pivotal role in constructing a meaningful context for personalized recommendations.

Moreover, contextual factors such as a person's health state embody a higher level of complexity, often resulting from the amalgamation of diverse data streams and external knowledge bases. The intricacy of context in relation to food is substantial, yet for the scope of this feasibility study, we narrow our focus to location alone. This streamlined approach aims to assess the robustness of the entire pipeline and the potential of this methodological framework, setting the stage for future research to delve into the broader dimensions of contextual factors.

Location emerges as a crucial contextual factor in food recommendation systems. A person's available food options are significantly influenced by their geographic location, underscoring the essence of incorporating location data to achieve truly contextualized food recommendations. By prioritizing location in this preliminary study, we seek to establish a foundational understanding of context's role in enhancing the relevance and precision of food recommendations, paving the way for more comprehensive explorations in subsequent research endeavors.

\section{Moving Forward}
We presented a novel approach which extends a new door of possibilities in multimedia research. Our dataset is available open-source to help with further investigation of this topic. The results demonstrate the importance of the causal aspect of food events and the estimation of the taste preference model by regarding food as media and more advanced mapping of the food items to the US4B taste space. This observation suggests much stronger future work on the study of the food event's causality by studying more complex factors involved in the causal aspect. Furthermore, studies with rich data streams from real users, including individuals differing in biological status and living-lab environments, will be invaluable to unravel the causal factors that shape food decisions.


\chapter{World Food Atlas Processing and Analysis}

\section{background}
Humans have been scrutinizing food from diverse perspectives.
Societal evolution and the associated technological revolution have given rise to a data-driven approach toward food.
Data related to food is diverse, rich, and multidimensional and interacts with multiple components \cite{min2019survey}.
Food has a direct impact on people's health \cite{lederer2022relation}.
Positive changes to food habits and dietary practices lead to better health outcomes.
Receiving the correct food recommendation at the appropriate time has the potential to reduce symptoms of chronic diseases, enable people to make healthier lifestyle choices, and increase overall happiness.
For a food navigation system to be effective and not result in a hedonic treadmill, the recommendations should be based on a person's food model rather than being derived from a population-level hypothesis.
We discuss the components and requirements of contextual food navigation systems with a focus on health and personal preference to formulate the food recommendation.

In most cases, the search query is about the food to be consumed \cite{Abhari2019_review_nutrition_recSys}\cite{kumar2016survey}.
We currently lack an effective food navigator that provides accurate food recommendations due to a disconnect between the data sources and the limited consideration of an individual's food profile, demography, and location.
We build the World Food Atlas to unify food data across the globe in a standardized manner with a special emphasis on location-based data to answer complex food-related queries at the right time.
Food has a wide-ranging impact across multiple sectors in the fields of healthcare, agriculture, public policy, and the environment.

Recent work \cite{Rostami2021WorldProject} provides an overview about a world food atlas to connect different aspects of food.
However, there is no clear architecture presented to achieve this goal.
Furthermore, while \cite{Haussmann2019_food_kg} mentions the need to build interconnected food networks, a working knowledge graph is yet off the table.
A knowledge graph contains rich information about its entities and acts as a tool that can be leveraged to build future food navigation systems.
The World Food Atlas is a platform that enables food navigation by extracting data from multiple sources needed from different locations. 

\begin{figure}
    \centering
    \includegraphics[width=0.47\textwidth]{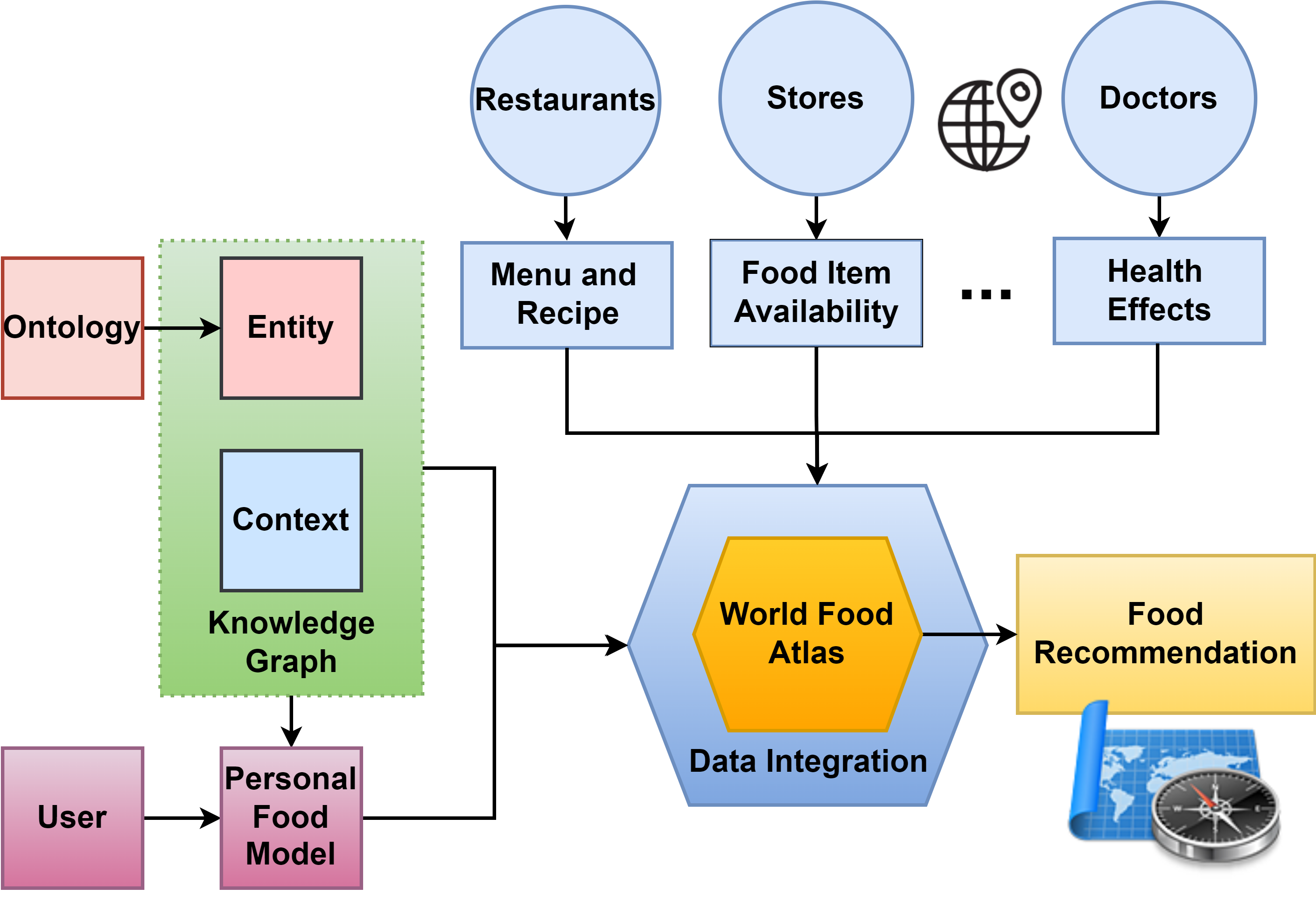}
    \caption{World Food Atlas for Food Recommendation}
    \label{fig:wfa_fr_flow}
\end{figure}

In this chapter, we make the following contributions.  We discuss what food navigation systems are and how every food-related problem is a recommendation problem. Later, we discuss various queries from stakeholders in the food ecosystem. We then present a world food atlas architecture to answer food queries for food recommendation and navigation as shown in Figure ~\ref{fig:wfa_fr_flow}. Finally, we showcase an experiment where we interview physicians, nutritionists, and doctors from Stanford University to find out what kind of queries they would ask a WFA and we design a WFA schema to demonstrate a full platform to solve real-world food-related problems \cite{Rostami2021_}, \cite{Rostami2022WFA}.  

\section{World Food Atlas}

The World Food Atlas (WFA) emerges as an indispensable element for any effectively operational personal Food Recommendation System (Food-RecSys). This section underscores the critical need for immediate, geolocational food inquiries and introduces a novel framework and architectural design for the WFA.

A food recommendation system lacking the capability to execute geospatial queries for identifying locally available food options fundamentally compromises its utility on a global scale. Such functionality necessitates a geospatial database adept at processing location-based queries to furnish real-time information on available food selections, menu items, eateries, and other pertinent locational data. Regrettably, this crucial component has historically been overlooked, a gap this thesis endeavors to bridge by underlining its significance.

The challenge of creating an electronic WFA, a database that catalogues a worldwide array of dishes and food items accessible to users irrespective of their geographical location, is substantial. The diversity of global cuisine, inclusive of countless ingredients and recipes, poses a significant obstacle. Nevertheless, the establishment of a WFA is essential, not merely for the enhancement of personal well-being but also for the preservation of global environmental health. This initiative seeks to spearhead the formation of a global coalition of scientists, technologists, and food specialists aimed at developing the WFA. This manuscript delineates the initial steps towards this ambitious goal, signaling the commencement of a cooperative venture to develop a food information repository with universal accessibility.



\subsection{Related Work and Motivation}


The idea of a world food atlas is first proposed in \cite{Rostami2021WorldProject} captures information about food across the world by integrating food logs about athletes and recipe logs \cite{Aizawa2019FoodLog:Application}. This chapter outlines a food knowledge graph (FKG) capturing information about restaurants, ingredients, recipes, nutritional content, and ontological \cite{dooley2018_FoodOn_Ontology} information about food products. However, apart from providing a preliminary design, detailed descriptions of different systems components and relationships between them are missing.

A collection of research articles, data sources, and identification of common themes of food knowledge graphs and their applications in the industry is presented in \cite {Min2021_foodKGReview}. Food knowledge graphs have different meanings depending on the context and application. However, synthesis and critical review of different sources are missing. 
The value of constructing food knowledge graphs for personalized food recommendation systems is discussed in \cite{chen2021_pFoodReQ}. However, the recommendation system works on a limited dataset and provides limited details about the design process involved in food knowledge graphs. 
Though the semantics and design of the food knowledge graphs for food recommendation are presented in \cite{Haussmann2019_food_kg}, it lacks a crowdsourcing mechanism for expanding data and components for the food knowledge graph.

The food environment atlas \cite{breneman2013food} is a platform to capture data-driven information about food choices and diet quality to identify patterns that can be used to formulate policy interventions for the community's health. The atlas provides information about food insecurity, food assistance programs, physical activity, and the socio-economic status of the population across the United States. The atlas is among the few that have detailed information about the data collection process, associated documentation, and more importantly downloadable datasets for further use. Though the maps can be customized, the user interface is not intuitive. The tool is academic-oriented and is used by researchers for policy-making, data analysis, and predictions. There is a need to make the environment atlas more user-friendly by improving the user interface, and design.

The food research access atlas \cite{rhone2017ers} is another atlas developed by the USDA to capture information about access to supermarkets, especially by low-income populations. It provides downloadable census tract data that can be used for creating an accessible food network and for population-level planning. Like the food environment atlas, though the food access research atlas allows customize-able visualization and download of food access data for different populations and sub-populations, there exists scope for improvement to the user interface to allow use beyond the research community.

Vermont's Food System Atlas \cite{vermontfoodatlas} is an interactive online tool that provides information about resources across the food supply chain ranging from production to marketing. With options to download and export data, the atlas serves as a valuable tool to understand the food landscape.
Wisconsin's Farm Fresh Atlas \cite{farmfreshatlaswisconsin} is another online searchable atlas that provides information about farmer's markets, restaurants, grocery stores, and businesses within the region. 
The Maine Food Atlas \cite{mainefoodatlas} is a comprehensive atlas that encompasses people, businesses, and organizations with Maine's food system. Researchers, scientists, and businesses can request raw data to improve the food security landscape in the region.
Another popular food atlas is the taste atlas \cite{tasteatlas}. The atlas consists of an online world map that details common food items, their recipes, and their historical and cultural background. However, there is a need to improve the robustness of the tool by increasing the number of dishes from various cuisines and improving the search functionality of nearby restaurants serving dishes captured on the map.

The food waste atlas \cite{foodwasteatlas} is a comprehensive tool that allows users to query food loss and food waste data across multiple stages in the food supply chain and across different sources. The database is however updated only till 2013 and there is a need to integrate the latest data from the United Nations. Further, a web interface for customized visualization of trends across years, for different countries and filtering based on data sources will be valuable in understanding historical trends.
We face the double burden of food waste and hunger across the world.
The Hunger Map \cite{worldhungermap} is a live interactive tool developed by the World Food Programme to monitor, predict and display information about food insecurity across populations throughout the world. 
A social food atlas \cite{socialfoodatlas} based in Italy to increase awareness about socially relevant food projects thereby creating a community geared towards sustainable food practices. Vienna's food atlas \cite{viennafoodatlas} is an effort to capture the historical and cultural significance of common foods in the region increasing awareness about food and spurring conversations about creating sustainable ecosystems.

The current food atlases display data from a limited number of sources and do not provide contextualized information.
They operate in silos and cater only to specific population groups such as researchers and businesses.
However, individual contributors and the larger community of data generators and contributors are neglected.
There is no standardized process to submit data which discourages users from submitting data.
We overcome these pitfalls by designing a world food atlas that acts as a food navigation system for stakeholders across the food ecosystem. 
Our unique WFA platform architecture collects and integrates data from multiple sources which serves as the precursor to food recommendation systems.
To enable stakeholders to contribute data positively, we develop a novel WFA schema that gives structure to the data input and enables incorporating data from multiple avenues.  

\begin{figure*}[t]
    \centering
    \includegraphics[width=1\textwidth]{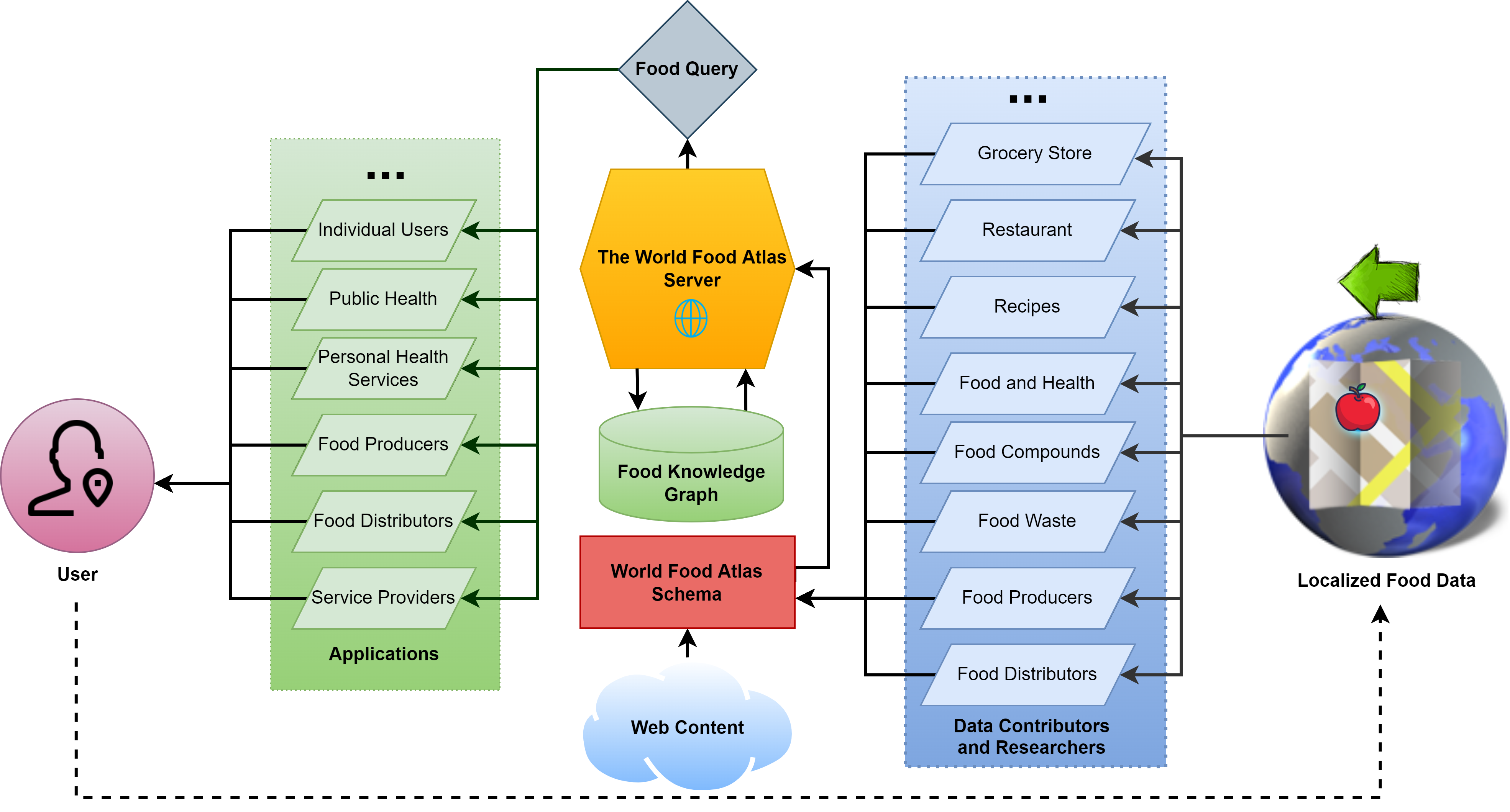}
    \caption{World Food Atlas Architecture: The localized food data is posted onto WFA via the schema and is available for various food-related applications.}
    \label{fig:wfa_architecture}
\end{figure*}

\subsection{Querying the World Food Atlas}
Food data is multidisciplinary, multidimensional, and multimodal and plays a key role in every aspect of society. 
It is generated and used by a variety of stakeholders in the healthcare, agriculture, cosmetics, and entertainment industry.
Each stakeholder may have questions related to personalized food recommendations, health effects of food, and lifestyle management.
Food navigation platforms service queries while collecting food- and health-related data to build a user's personal food model (PFM) \cite{rostami2020_personal_food_model}.
When combined with the location, time, and relationship with other entities, the queries serve to answer the food recommendation problem.

\begin{figure}[!ht]
  \centering
  \includegraphics[width=\linewidth]%
    {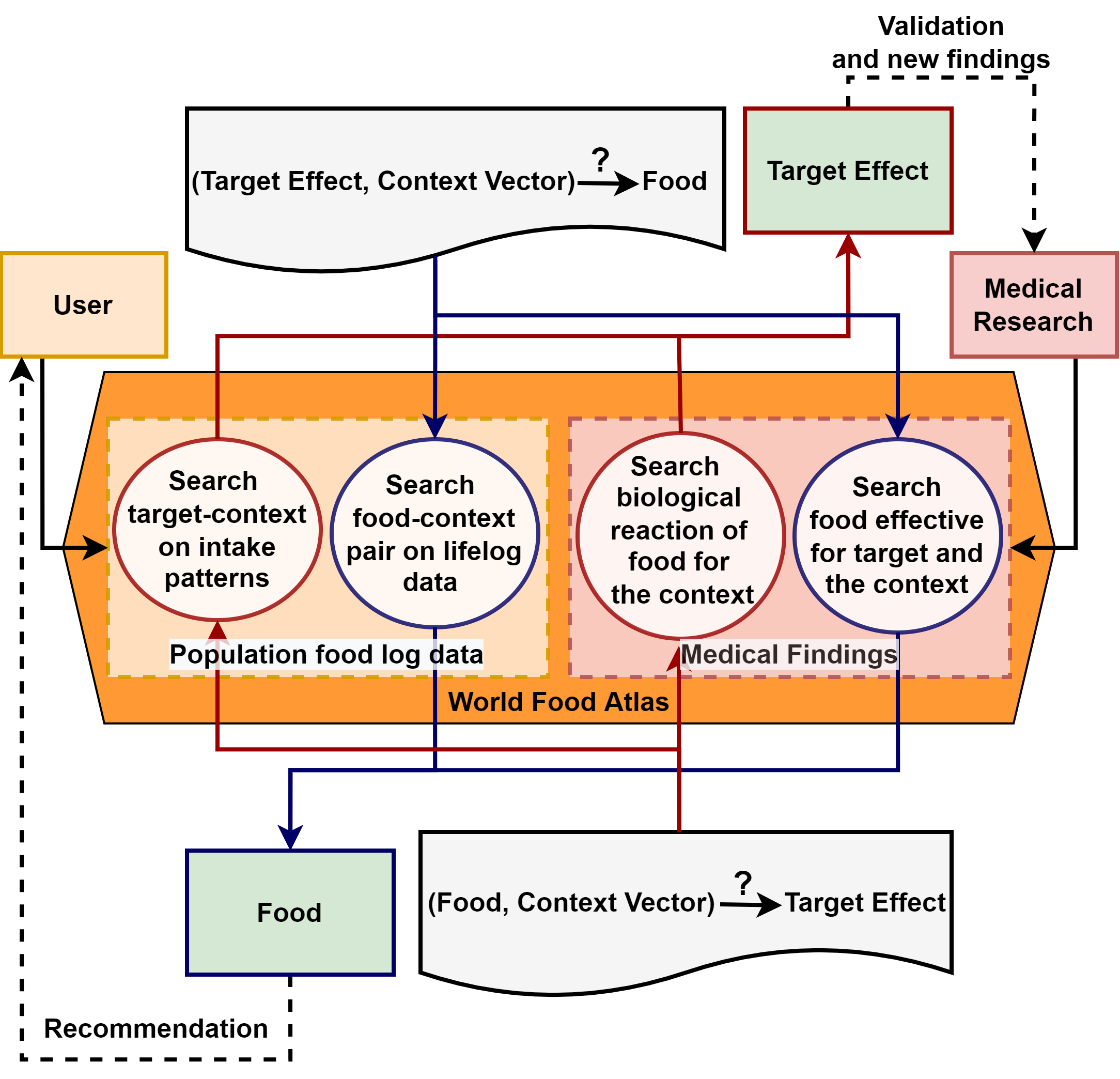}
    \caption{Health-related queries encompass questions that affect the user's biology. We generalize complex questions by categorizing them into specific query classes. We ask about the impact of certain types of food on the desired health state and measure the health effects of consuming certain kinds of food, both within a context. We aim to answer these questions through the World Food Atlas through an event mining approach and by leveraging medical research findings.}
    \label{fig:healthqueries}
\end{figure}

Using the world food atlas, we process person-entity-context queries which enable food recommendation platforms to move towards contextual and personalized data-driven approaches to recommend the right kind of food at the right time.
We list certain queries that are related to food and that have an impact on health which can be answered using the World Food Atlas.
We show the relationship between food, health, context, and user with the world food atlas while answering health-related queries in Figure ~\ref{fig:healthqueries}.
We detail specific query examples by potential users and outline general query classes that could be generated by members of the community.

Every query has a question-answer pair where the input is a question and the output is the answer from the WFA. 
A common query type includes searching for ingredients or food items useful to attain health goals.
A few examples of health class questions include - What food items help relieve stress? Which snack can contribute towards improving my sleep quality and costs less than 10 USD? 
A food item has a target health effect within a given context and can be represented as follows.
\begin{algorithm}
\caption{Contextual Query to Search Food Based on Effect}
$$ ( Effect_t, Context_v ) \longrightarrow Food $$
The $Effect_t$ represents the target effect and the $Context_v$ represents the context vector.
\end{algorithm}

On the contrary, users may ask questions about the impact certain kinds of food can have on their health within a particular context. For example, what impact will drinking three cups of coffee have on me as a 50-year-old African-American woman?
The relationship is represented as follows.
\begin{algorithm}
\caption{Contextual Query to Search Effect Based on Food}
$$ ( Food, Context_v ) \longrightarrow Effect_t $$
\end{algorithm}

Questions of this nature vary based on an individual's political, religious, and geographical backgrounds and may have answers that contradict medical findings.
Additionally, a person's food model \cite{rostami2020_personal_food_model} leads to different health effects for the same food consumption patterns and quantity. 

As shown in Figure ~\ref{fig:healthqueries}, using the WFA, we obtain responses to these queries using real-world data.
We examine current medical findings to find research-backed food recommendations by the medical community for the desired health effect for a specific context.
Through the WFA, we also examine population-level life log data of people across the world to recommend food items that help achieve health goals in a specific context.
Conversely, if users search for the effect of eating a certain type of food on their health, the WFA scrutinizes lifelogs from multiple stakeholders and medical findings across the research community to find potential health effects of food items.

In some instances such as dietary restrictions for users, contextual information bears prominence for food and health-related queries. Existing work \cite{majumder-etal-2019-generating} generates recipes based on user's preference history using limited data about food. Such platforms consist of contextual preferential queries on food. 
A few examples are - What dishes can I make to prevent soy allergies? What recipes can I use to make lamb curry that is low in cholesterol and saturated fat? And that can take into consideration all the limitations that a person might have for cooking at home or for ordering the food.
\begin{algorithm}
\caption{Query to Search Recipe Based on Requirements}
$$ ( Requirement_v ) \longrightarrow Recipes $$
The $Requirement_v$ represents the requirements vector.
\end{algorithm}
What restaurants can I go near me which has vegan food? 

\begin{algorithm}
\caption{Contextual Query to Search Restaurant Based on requirements}
$$ ( Requirement_v, Context_v ) \longrightarrow Restaurant $$
\end{algorithm}

Data generated by users through lifestyle navigation apps are used to build personal models \cite{rostami2020_personal_food_model} around food, emotion, sleep, stress, and physical activity.
Food recommendation systems have aimed to model the preference of the users \cite{Pandey2021EventModelling} to answer some questions such as:
What do I eat when I am stressed? In what context do I prefer drinking tea in the evening? 
When do I eat spicy food? When do I crave sweets? What foods cause sleep discomfort at night? How does my current diet affect acid reflux?
These queries can be represented using the following class.
\begin{algorithm}
\caption{Contextual Query to Search Food Based on Personal Log}
$$ ( Personal Log, Context_v ) \longrightarrow Food $$
The $Personal Log$ represents the personal lifelog and food log data of the person. 
\end{algorithm}

Today with various wearable sensors, logging real-time biological signals is possible for a wide range of users across the world \cite{greene2021electromagnetic}.
Compact sensors play a critical role in the acquisition of physical, chemical, or biological data \cite{nikzamir2022highly}. 
Food storage spaces can be monitored using sensors like temperature or humidity sensors \cite{law2010sub} and nutrition can be assessed accurately using nutrition detection sensors \cite{khansili2018label}. 
The WFA aggregates a large connection of data from various contributors around the globe to process such queries as will be discussed in the following sections.
We, therefore, design a World Food Atlas that takes a holistic data-driven approach for answering all food-related queries.

\begin{figure*}[t]
    \centering
    \includegraphics[width=\textwidth]{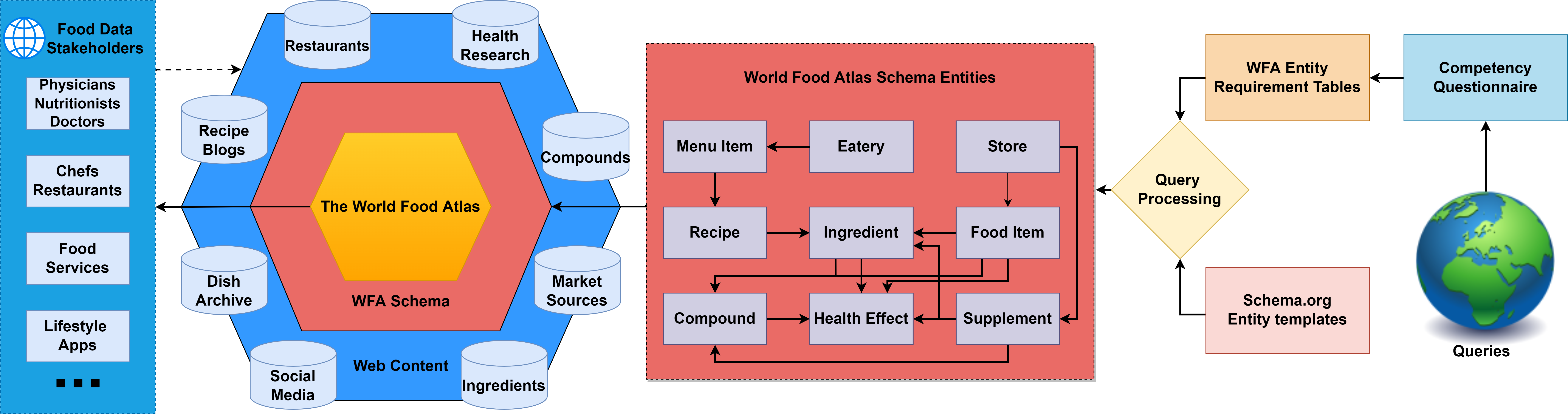}
    \caption{World Food Atlas schema Design for processing real-world queries}
    \label{fig:wfa_schema}
\end{figure*}
\section{World Food Atlas Architecture}

Stakeholders across the food supply chain such as individual users, food producers, food distributors, personal health service providers, and public health authorities are contributors and consumers of food-related data.
The sources of data, their format, size, frequency, collection methodology, and applications vary\cite{min2019survey}.
The geographical context and the temporal features of data play a crucial role in the way data is collected, processed, and analyzed for use in food recommendation applications\cite{rostami2020_personal_food_model}.
However, there is a lack of a unified platform for congregating all food-related information to provide lifestyle navigation.
To this end, we construct the World Food Atlas which provides an overarching framework for data collection, data representation, and food-related information retrieval.

The detailed architecture of the World Food Atlas platform is shown in Figure ~\ref{fig:wfa_architecture}.
There are multiple contributors to food-related data as shown by the blocks in the blue column. 
Congregating data will help explore similarities and differences between content.
Most data is generated and made available by grocery stores, restaurants, food manufacturers, distributors, individual users, researchers, government agencies, research institutions, and independent agencies at the state, national, and international levels.
Data from these sources is multidimensional and in varying formats.
Web content also serves as an input to the world food atlas platform. Methods such as \cite{kiddon2016globally} use models to reconstruct the recipe information from available text such as title or ingredients.

The most challenging part is connecting the data from multiple sources and across multiple platforms in a systematic manner. 
We create a world food atlas schema that allows consolidating data into a unified framework.
The schema forms the backbone of the world food atlas.
We use the schema to standardize data sources and bring them into the world food atlas.
The schema fits well into a database and can be stored on a server.
When a user poses a query based on their food needs, information can be extracted from the server based on the integrated mega atlas that maps relationships between multiple sources.
The data is then presented in a downloadable format.
The encrypted data can be sent to individual contributors for updating their database and can be used for future use. For example, the data can be fed into a machine learning model for training and can be used for food recommendations.
On the other hand, the output data from the food queries can be released publicly with appropriate citations for use by individuals, public health providers, personal health service providers, food manufacturers, food distributors, etc.
This data can be integrated with food logging platforms \cite{Rostami2020MultimediaLogger}, life logging applications, and population-level data by the United States Department of Agriculture, United Nations Food and Agricultural Organization, and the World Food Program.

The different stakeholders that can benefit from the world food atlas are highlighted in the green column through responses to application-specific queries.
For instance, service providers can improve their food quality, food distributors can benefit from better logistics management, and food producers can leverage the WFA for designing better products.
Individual users can directly use the open-source platform and public health organizations can leverage localized data output based on inputs from multiple sources.
All these applications benefit from the automation capabilities offered by the world food atlas both at an individual level and at the population level. In turn, they contribute to the world food atlas through food logging applications, life logging applications, and population-level studies.
There exists a feedback loop that allows sending data back to contributors. 

We are among the first ones to create a simple and standardized data collection schema and architecture platform to collect food-related data thereby creating incentives for stakeholders to contribute data.
We ensure that each contributor gets credit for their data by creating an identity profile for each data source which increases awareness about the data source and allows users to cite resources appropriately when used.
In this way, our world food atlas architecture allows dynamic data creation, storage, and retrieval which is advanced compared to the existing atlases.

\subsection{The Schema Experiment}

The World Food Atlas is designed to become a global data hub with a location-based and time-based focus on food. 
Large-scale data generated by multiple stakeholders should be seamlessly imported into the platform.
Doing so requires formulating multiple data classes, detailing relationships between classes, and a comprehensive data collection schema.
We design a novel world food atlas schema that allows users to contribute data to the platform without having detailed knowledge of the underlying food knowledge graph. 
The users share data under the corresponding schema since we handle data integration internally.
The World Food Atlas entity classes, their relations, and schema form the backbone of the world food atlas architecture.

The schema provides structure and consistency to the incoming data of a certain class.
We show the design process and framework of the world food atlas schema in Figure ~\ref{fig:wfa_schema}.
The different food data stakeholders include physicians, nutritionists, doctors, chefs, and individual users through data from lifestyle navigation applications such as Apple Health Kit, Google Fitbit, etc, represented by the blue rectangular column.
The hexagon indicates the mechanism through which data from multiple sources is aggregated and fed into the World Food Atlas. 

The stakeholders can either directly contribute to the atlas through the WFA schema or via content generated by them on the web.
For example, if a restaurant wants to upload its data into the atlas, it can use the 'Eatery' schema template.
Similarly, food bloggers can use the 'Recipe' schema and 'Ingredient' schema templates to upload data about their favorite recipes. Research groups \cite{marin2019recipe1m+} have put valuable effort to gather a dataset of different recipes with many images associated to each recipe and has demonstrated the great potential of such data. An effort like this requires a substantial amount of resources. While very valuable, it lacks an automated method to expand the dataset, whereas, WFA schema is a hub for large-scale recipe datasets and is constantly growing.
There is a two-way street between stakeholders/contributors of food data such as doctors, chefs, food services, lifestyle navigation apps, and the world food atlas. They provide data to the atlas through the schema and they receive data based on their queries.

\begin{table*}[t]
    \centering
    \scalebox{0.45}
    {
    \begin{tabular}{|p{0.20\linewidth}|p{0.20\linewidth}|p{0.20\linewidth}|p{0.15\linewidth}|p{0.15\linewidth}|p{0.20\linewidth}|p{0.15\linewidth}|p{0.20\linewidth}|p{0.27\linewidth}|p{0.20\linewidth}|p{0.20\linewidth}}
    \textbf{Eatery}	&	\textbf{Stores}	&	\textbf{Recipe}	&	\textbf{Menu Item}	&	\textbf{Food Item}	&	\textbf{Dietary Supplement}	&	\textbf{Ingredient}	&	\textbf{Compound}	&	\textbf{Health Effect}	&	\textbf{Nutrition}	\\
    \hline
\textbf{Menu Item}	&	\textbf{Dietary Supplement}	&	\textbf{Ingredient}	&	\textbf{Ingredient}	&	\textbf{Ingredient}	&	\textbf{Ingredient}	&	\textbf{Compound}	&	\textbf{Health Effect}	&	Health Benefits	&	Calories	\\
Address	&	\textbf{Food Item}	&	\textbf{Nutrition}	&	\textbf{Health Effect}	&	\textbf{Health Effect}	&	\textbf{Health Effect}	&	\textbf{Health Effect}	&	BioChem Interaction	&	Misconceptions	&	Carbohydrate	\\
Cuisine	&	Address	&	Category	&	\textbf{Nutrition}	&	\textbf{Nutrition}	&	\textbf{Nutrition}	&	\textbf{Nutrition}	&	BioChem Similarity	&	Symptoms	&	Cholestorol	\\
Description	&	Description	&	Cook Time	&	Category	&	Category	&	Active Ingredients	&	Category	&	Chemical Role	&	\cellcolor[HTML]{dae8fc} Allergen Effect	&	Fat	\\
Drive Thru	&	Drive Thru	&	Cooking Method	&	Description	&	Description	&	Category	&	Cost	&	Composition	&	\cellcolor[HTML]{dae8fc} Effect on blood glucose level	&	Fiber	\\
Email	&	Email	&	Copyright Holder	&	Name	&	Distributor	&	Cost	&	Description	&	Description	&	\cellcolor[HTML]{dae8fc} Effect on cardio-metabolic health	&	Protein	\\
Item Price	&	Item Price	&	Copyright Notice	&	Serving Size	&	Name	&	Description	&	Name	&	Genetic Expression	&	\cellcolor[HTML]{dae8fc} Effect on Pregnancy	&	Saturated Fat	\\
Latitude and Longitude	&	Latitude and Longitude	&	Copyright Year	&	Suitable for Diet	&	Net Weight	&	Guidelines	&	Suitable for Diet	&	inChI	&	\cellcolor[HTML]{dae8fc} Effect on triglycerides	&	Serving Size	\\
Logo	&	Logo	&	Cuisine	&	\cellcolor[HTML]{cccccc} Deals/Offers	&	Serving Size	&	Legal Status	&	\cellcolor[HTML]{cccccc} Deals/Offers	&	inChIKey	&	\cellcolor[HTML]{dae8fc} Fatty acid profile	&	Sugar	\\
Name	&	Name	&	Date Created	&	\cellcolor[HTML]{dae8fc} Allergens	&	Suitable for Diet	&	Manufacturer	&	\cellcolor[HTML]{dae8fc} Allergens	&	IUPAC Name	&	\cellcolor[HTML]{dae8fc} Pre-, pro-, post- biotic content	&	Trans Fat	\\
Open Hours	&	Open Hours	&	Date Modified	&	\cellcolor[HTML]{dae8fc} Cost	&	\cellcolor[HTML]{dae8fc} Taste	&	Maximum Intake	&	\cellcolor[HTML]{dae8fc} Glycemic Index	&	Molecular Formula	&	\cellcolor[HTML]{dae8fc} Side Effects	&	Unsaturated Fat	\\
Payment Method	&	Payment Method	&	Description	&	\cellcolor[HTML]{dae8fc} Taste	&	\cellcolor[HTML]{cccccc} Deals/Offers	&	Mechanism of Action	&	\cellcolor[HTML]{dae8fc} Insulin Index	&	Molecular Function	&		&	\cellcolor[HTML]{dae8fc} \% Daily Value	\\
Phone	&	Phone	&	Instructions	&		&	\cellcolor[HTML]{cccccc} Growing Location	&	Medicine System	&	\cellcolor[HTML]{dae8fc} Shelf Life	&	Molecular Weight	&		&	\cellcolor[HTML]{dae8fc} Electrolytes	\\
Photo	&	Photo	&	Name	&		&	\cellcolor[HTML]{cccccc} Origin City	&	Name	&	\cellcolor[HTML]{dae8fc} Taste	&	Name	&		&	\cellcolor[HTML]{dae8fc} Minerals	\\
Price Range	&	Price Range	&	Photos	&		&	\cellcolor[HTML]{cccccc} Origin Country	&	Proprietary Name	&		&	Potential Use	&		&	\cellcolor[HTML]{dae8fc} Vitamins	\\
Reservation	&	Review	&	Prep Time	&		&	\cellcolor[HTML]{dae8fc} Allergens	&	Recognizing Authority	&		&	Proprietary Name	&		&		\\
Review	&	Star Rating	&	Reviews	&		&	\cellcolor[HTML]{dae8fc} Cost	&	Recommended Intake	&		&	Smiles	&		&		\\
Star Rating	&	Website	&	Star Rating	&		&	\cellcolor[HTML]{dae8fc} Shelf Life	&	Safety Consideration	&		&	Subcellular Location	&		&		\\
Website	&	\cellcolor[HTML]{cccccc} Curbside Pickup	&	Steps	&		&		&	Suitable for Diet	&		&	Taxonomic Range	&		&		\\
\cellcolor[HTML]{cccccc} Curbside Pickup	&	\cellcolor[HTML]{cccccc} Deals/Offers	&	Suitable for Diet	&		&		&	Target Population	&		&	\cellcolor[HTML]{dae8fc} Taste	&		&		\\
\cellcolor[HTML]{cccccc} Deals/Offers	&	\cellcolor[HTML]{cccccc} Dine-in	&	Total Time	&		&		&	\cellcolor[HTML]{cccccc} Deals/Offers	&		&		&		&		\\
\cellcolor[HTML]{cccccc} Dine-in	&	\cellcolor[HTML]{cccccc} Events	&	Website	&		&		&	\cellcolor[HTML]{dae8fc} Allergens	&		&		&		&		\\
\cellcolor[HTML]{cccccc} Events	&	\cellcolor[HTML]{cccccc} Non-Contact Delivery	&	Yield	&		&		&	\cellcolor[HTML]{dae8fc} Shelf Life	&		&		&		&		\\
\cellcolor[HTML]{cccccc} Non-Contact Delivery	&	\cellcolor[HTML]{cccccc} Online Order	&	\cellcolor[HTML]{cccccc} Videos	&		&		&	\cellcolor[HTML]{dae8fc} Taste	&		&		&		&		\\
\cellcolor[HTML]{cccccc} Online Order	&	\cellcolor[HTML]{cccccc} Pharmacy	&	\cellcolor[HTML]{dae8fc} Cost	&		&		&		&		&		&		&		\\
\cellcolor[HTML]{cccccc} Popular Times	&	\cellcolor[HTML]{cccccc} Popular Times	&	\cellcolor[HTML]{dae8fc} Tools/Utensils	&		&		&		&		&		&		&		\\
\cellcolor[HTML]{cccccc} Question and Answers 	&	\cellcolor[HTML]{cccccc} Product Categories	&		&		&		&		&		&		&		&		\\
\cellcolor[HTML]{cccccc} Serves Alcohol	&	\cellcolor[HTML]{cccccc} Question and Answers 	&		&		&		&		&		&		&		&		\\
\cellcolor[HTML]{cccccc} Smoking Allowed	&	\cellcolor[HTML]{cccccc} Serves Alcohol	&		&		&		&		&		&		&		&		\\
\cellcolor[HTML]{cccccc} Social Media	&	\cellcolor[HTML]{cccccc} Smoking Allowed	&		&		&		&		&		&		&	\textbf{Legend}	&		\\
\cellcolor[HTML]{cccccc} Take Out	&	\cellcolor[HTML]{cccccc} Social Media	&		&		&		&		&		&		&	\cellcolor[HTML]{cccccc}	&	Web Content	\\
\cellcolor[HTML]{cccccc} Videos	&	\cellcolor[HTML]{cccccc} Take Out	&		&		&		&		&		&		&		&	schema.org	\\
	&	\cellcolor[HTML]{cccccc} Videos	&		&		&		&		&		&		&	\cellcolor[HTML]{dae8fc}	&	Questionnaire	\\
    
    \end{tabular}
    }
    \caption{WFA Schema Components}
    \label{tab:wfa_schema_components}
\end{table*}

We draw up a list of competency questions that could potentially arise in the minds of the stakeholders.
We go one step further and partner with expert nutritionists, dietitians, and doctors from Stanford University in expanding the list by seeking their opinion about questions they ask their patients while examining them.
The detailed questionnaire is presented in Appendix ~\ref{appendix:wfa_questionnaire}.
The questions were mainly related to taste, cost, and health impact of food.
We combine our knowledge and their inputs to create a comprehensive requirements table for queries to the world food atlas.
We noticed that many requirements for different components such as recipes, menu items, and dietary supplements were satisfied by entity templates in \url{https://schema.org}. Though \url{https://schema.org} was originally used by internet giants such as Google for optimizing search on web data and for data indexing on web searches, their goal is to create a standard data representation language.
We, therefore, combine our world food atlas entity requirement table taking inputs from dietitians, nutritionists, and computer scientists with entity templates from \url{https://schema.org}.

We also take into consideration aspects not covered in the competency questionnaire and schema.org to make the entity template list as comprehensive as possible.
We account for events happening around the globe along with dynamic real-time information updated by different stakeholders while populating the schema entity tables.
For example, aspects such as the 'non-contact delivery' option for restaurants and grocery stores due to COVID-19 and 'popular times' an eatery or store is open based on real-time occupation level given by service providers.

The components within the schema are interconnected and have dependencies on each other as shown in the red rectangle.
The main components are eateries, stores, menu items, food items, dietary supplements, recipes, ingredients, compounds, and health effects.
Eateries such as restaurants, fast food joints, and food trucks have certain items on their menu.
Each item has a specific recipe and varies in preparation based on the eatery and the chef preparing the dish. 
Similarly, stores such as supermarkets, and retail stores have food items that are across multiple categories including but not limited to cereals, snacks, candies, etc.
Stores either physical or online also sell dietary supplements.
Food items, menu items, and dietary supplements are composed of ingredients, which are in turn constituted by compounds.
Dishes, dietary supplements, ingredients, and compounds have health effects depending on their consumption amount, time, body physiology, and genetics.

We detail the different schema entities and their components in Table ~\ref{tab:wfa_schema_components}.
The entities include 'Eatery', 'Store', 'Menu Item', 'Food Item', 'Dietary Supplement', 'Ingredient', 'Compound', 'Nutrition Information' and 'Health Effect'.
Most of the components between the 'Eatery' entity and the 'Store' entity are similar and are derived from the schema.org entity tables.
The components derived outside schema.org and competency questionnaires are - Curbside Pickup, Deals/Offers, Serves Alcohol, Smoking Allowed, Dine-in, Take Out, Non-Contact Delivery, Online Order, Popular Times, Question and Answers, Social Media, Events, and Videos.
Components that are included in the 'Eatery' entity and excluded in the 'Store' entity are - 'Menu Item', Reservation, and Cuisine.
Components that are included in the 'Store' entity and excluded in the 'Eatery' entity are - 'Dietary Supplement', Products/Food Items, Product Category, and Pharmacy.

Each eatery such as a restaurant has a 'Menu Item' and stores have a 'Food Item' and 'Dietary Supplement'.
We distinguish 'Menu Item' and 'Food Item' because 'Menu Items' because chefs in restaurants may have different recipes for each item they prepare and can vary across restaurants whereas food items are standardized with respect to their preparation process and have fixed ingredients, composition, and nutrient content.
Both 'Menu Items' and 'Food Items' comprise ingredients with associated nutrition information, a specific taste, allergens, and certain health effects. 
The 'food item' entity has a growing location, origin city, origin country, and shelf life in addition to all other components present in the menu item entity.
Most of the information for the 'Dietary Supplement' entity is derived from schema.org. 
Though more fields were present that captured various aspects of dietary supplements, we decide to provide a representative sample that captures most of the information and is simple enough to be uploaded by non-expert stakeholders and data contributors.

'Recipes', 'Menu Items', 'Food Items', and Dietary Supplements contain 'Ingredients'.
We expand the ingredient components of schema.org to include aspects such as insulin index, glycemic index, and origin information apart from standard information such as nutritional content and suitability for certain diets among others.
Each 'ingredient' has multiple 'compounds' and each compound has different properties mostly derived from schema.org such as molecular weight, molecular formula, IUPAC name, and so on as described in the table.
Each 'menu item', 'food item', 'dietary supplement', 'ingredient' and 'compound' consumed has a certain 'health effect'.

We understand different health effects by consulting with nutritionists, doctors, and medical experts. We also incorporate certain components of the health aspect entity table from schema.org such as symptoms, health benefits, and misconceptions about certain types of food.
We add a separate column related to nutrition information in Table ~\ref{tab:wfa_schema_components}.
Though the nutrition information is implicitly available for recipes, menu items, food items, dietary supplements, and ingredients, we wanted to prevent redundancy which would have occurred by listing individual nutrition components under each entity column, unnecessarily leading to increased table length.
We noticed that \% daily value, vitamin, mineral, and electrolyte content were not captured in schema.org and have included these components based on the competency questionnaires.

We are in the process of designing our own schema for allergens, taste, and health effects since we feel the existing information is inadequate.
We present a comprehensive list of entity components and more importantly analyze the interdependencies and intricacies of each component.
Our schema serves as a benchmark for capturing food-related data and its inter-dependencies and is instrumental in collecting data from multiple sources for building the World Food Atlas.


\section{Moving Forward}
We develop a world food atlas for food navigation to solve the person-entity-context food recommendation problem.
We review current food atlases and knowledge graphs and critically analyze their limitations.
Apart from critiquing existing work, we propose a novel world food atlas architecture platform that congregates data from multiple streams and multiple stakeholders.
Our novel world food atlas schema weaves a canvas around the interconnected web of food components for food-related user queries.
The atlas has the potential to generate personalized location-based food data that can be used by individual users, food services, and healthcare providers to improve personal health, increase food supply chain efficiency and allow targeted healthy food recommendations.

Over time, we seen potential for the schema to expand to include more interconnected components related to food and health.
While collecting data from multiple sources, we came across duplicate and redundant data. 
By incorporating and expanding our standardized data collection and specification pipeline, there is scope for decentralized data processing and management from different contributors.
In this way, it is possible to handle large-scale data and ensure scalability of the platform.
A vetting process for verifying the validity of data sources, processing user-generated data, and preserving privacy is essential to maintain and increase the robustness of the world food atlas platform.
Also, we have to deal with the problem of missing data through predictive analytic approaches using machine learning.
We strive towards automatic database expansions and decentralized blockchain approaches for empowering individual users, organizations, and different stakeholders.
Validating responses from the world food atlas is essential to ensure the accuracy of data provided to users. 
We also intend to expand the number of components of the world food atlas schema.
We are working towards creating taste schema and health effect schema through consultations with dietitians and nutritionists.
Our world food atlas will be a hub for all food-related atlases and be a one-stop solution for all food-related queries.



\chapter{Large Language Model based Food Recommendation}

\section{Background}

In an era of ubiquitous digital assistants and tailored experiences, personalized food recommendations have emerged as a potent force shaping our culinary journey. Beyond mere convenience, these systems hold immense potential to improve dietary choices \cite{tran2021recommender}, address nutritional deficiencies \cite{chen2021personalized}, and even combat chronic diseases \cite{agapito2018dietos}.
Personalization in food recommendations transcends mere taste preferences. By incorporating underlying food-specific data of ingredients and recipes, cultural factors, health data, and real-time context, these systems can foster healthier and more fulfilling culinary experiences \cite{Rostami21PPFM}. Studies show that personalized recommendations can encourage healthier food choices, increase dietary adherence, and improve overall user satisfaction \cite{Rostami20PFM}. This intersection of personal well-being and culinary delight underscores the profound societal and individual benefits of effective food recommendation systems.


Food recommendation systems encompass a diverse array of approaches, from content-based filtering relying on user-rated recipes to collaborative filtering leveraging shared preferences \cite{ornab2017empirical}. Recent advancements have demonstrated the integration of hybrid and recommendation fusion techniques to cater to the multi-dimensionality of food choices \cite{Rostami20PFM}. Moreover, the rise of LLMs has sparked immense interest in their potential to revolutionize food recommendations \cite{Aljbawi2020HealthawareFP}.
LLMs hold the promise of understanding the linguistic nuances of food descriptions, user preferences, and context, paving the way for highly personalized and contextually aware recommendations \cite{geng2022recommendation}. However, existing LLM-based recommendation systems often lack a holistic approach, struggling to seamlessly integrate the diverse components crucial for effective food recommendations \cite{Podszun2023}.


Despite significant advancements in food recommendation systems, there remains a pronounced gap between user expectations and the performance of existing technologies. These systems frequently fall short in offering personalized suggestions, failing to account for essential factors such as dietary restrictions, cultural preferences, and the real-time context of users. Current methods that leverage LLM-based algorithms face challenges in accurately interpreting the nuanced language of food, which is critical for generating meaningful and customized recommendations. This limitation undermines their ability to truly personalize the user experience and maximize the relevance of their suggestions \cite{Aljbawi2020HealthawareFP}.
This disconnect between potential and reality underscores the need for a paradigm shift in food recommendation systems, one that leverages the power of LLMs while addressing their limitations. Geng \textit{et al.} \cite{geng2022recommendation}  introduce a promising methodology to harness LLMs for recommendation purposes, termed Recommendation as Language Processing (RLP). While this represents a significant advancement in utilizing LLMs for recommendations, the demand for a food-centric approach to enhance food-specific recommendations remains pronounced. A system that successfully bridges this gap has the potential to transform our interactions with food, promoting healthier, more satisfying, and culturally informed culinary experiences.


This chapter proposes Food Recommendation as Language Processing, F-RLP, as a novel framework consisting of two primary elements: the first component is designed to aggregate and process food recommendation-specific data, employing a series of mathematical and logical operations to distill this information into a final vector. This vector is then inputted into a LLM. The second component, an LLM-based recommendation system, is adept at interpreting this contextual and personalized vector, utilizing it to generate accurate and tailored food recommendations.
Our contribution lies in the following key aspects:
\begin{itemize}
    \item Comprehensive food recommendation framework: Our hybrid framework integrates various personal and contextual data, with a specific focus on numerical data, which traditionally poses a challenge for LLMs. 
    \item Specialized LLM training: We propose a novel training regimen, centered around the use of enhanced counterfactual data allowing expert insights affect fine-tuning. 
    \item Novel context injection: F-RLP  introduces a context injection technique to the LLMs, streamlining the process of complex mathematical computations. This is achieved by supplying the LLM with a context-oriented list of options in conjunction with the user's query. 
\end{itemize}
\begin{figure}
    \centering
    \includegraphics[width=1\textwidth]{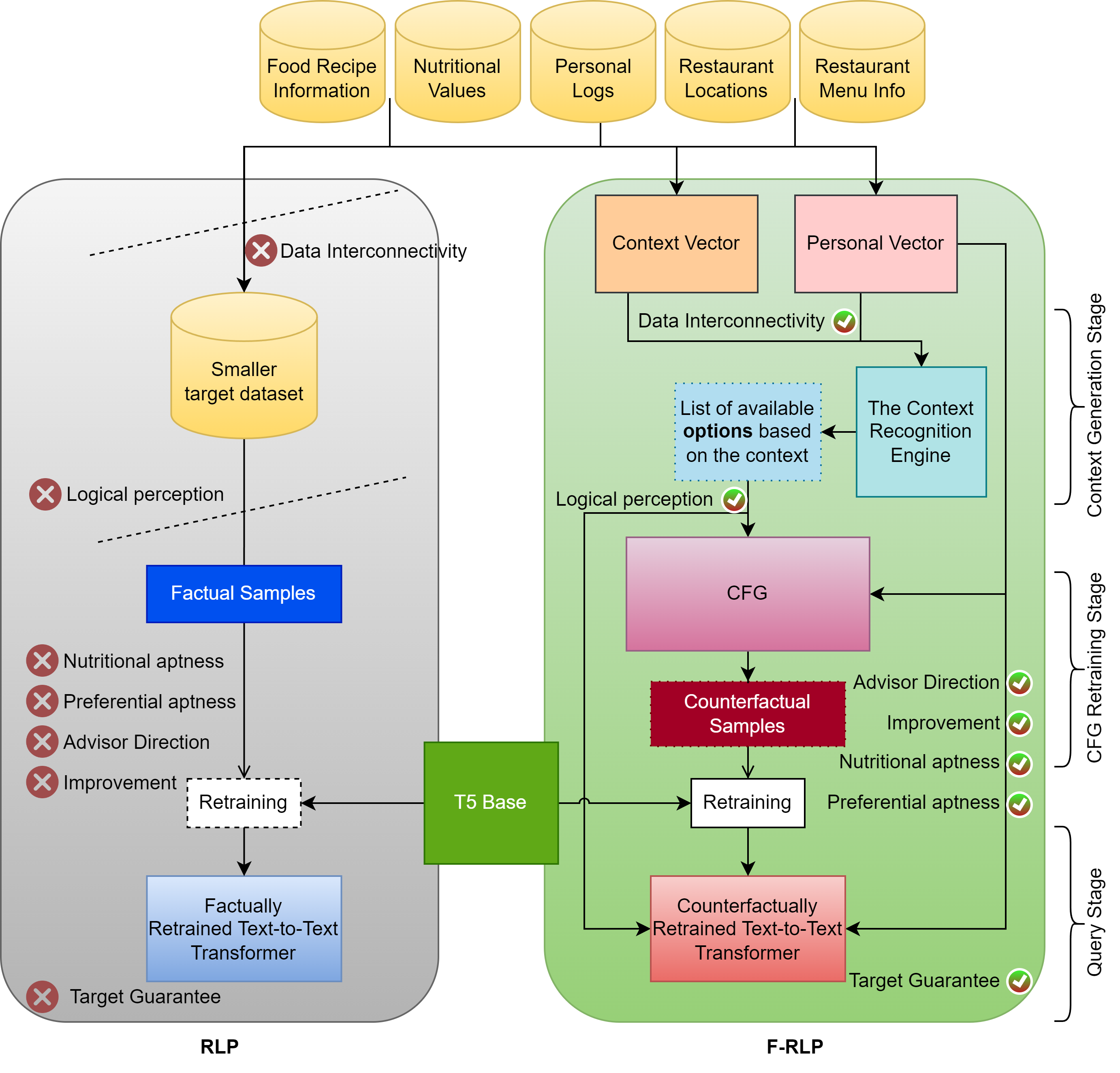}
    \caption{RLP vs F-RLP: On the left side we see a general propose LLM-based recommendation system, which is not ideal for food recommendation as it will lose inter connectivity of the food related data when it does not know how tie the food related data together. Furthermore as LLMs are by default trained using factual sample inputs the logical perception of the data and any improving reasoning does not exist on a general purpose LLM-based recommendation and even the target output being among an actual option is not guaranteed, however the F-RLP proposed in this work connects the data together, improves the sample data by effective counterfactual sample engineering and guarantees target by providing a list of options to the LLM }
    \label{fig:arch}
\end{figure}


\section{Personal Vector Generation}
In advancing personalized food recommendation systems, the delineation of user-specific vectors is instrumental in tailoring suggestions to individual dietary needs and preferences. This section introduces the concept of personal vector generation, comprising two distinct yet complementary components: the biological personal vector and the preferential personal vector. Each vector encapsulates unique facets of a user's food-related characteristics, serving as a cornerstone for personalized recommendation: Biological and Preferential Personal Vectors.
\subsubsection{Biological Personal Vector}
This vector encompasses restrictions and requirements tied directly to the user's health, as dictated by nutritional guidelines or medical advisories. For instance, it may include mandates such as a minimum meat consumption or avoidance of certain allergens, informed by a nutritionist's directives. The vector's construction is guided by the imperative to align with health objectives, encapsulating items that must be consumed or avoided.

\subsubsection{Preferential Personal Vector}
Contrary to the health-centric biological vector, the preferential vector captures the user's taste preferences and aversions. This vector is not governed by health considerations but by personal taste, including preferred ingredients and disliked food items.

\subsection{Temporal Considerations in Vector Construction}
A critical inquiry in vector generation pertains to the temporal window for analyzing user data to accurately construct these vectors. Given the dynamic nature of dietary habits and physiological responses, this analysis acknowledges the influence of external factors like location, stress levels, financial status, and environmental conditions on user behavior and physiology.
\subsubsection{Biological Vector Temporal Window}
Recognizing the direct correlation between a user's physiological state and recent dietary intake, this study proposes a dual timeframe for the biological vector. A "memory-enabled" approach is adopted, creating one vector based on the previous day's intake and another encompassing the last three days, thereby capturing the immediate impact of recent dietary habits as some research suggests that the most recent food intake has a high impact on the current physiological state of the user \cite{levitsky2022rise}.
\begin{figure}
    \centering
    \includegraphics[width=1\textwidth]{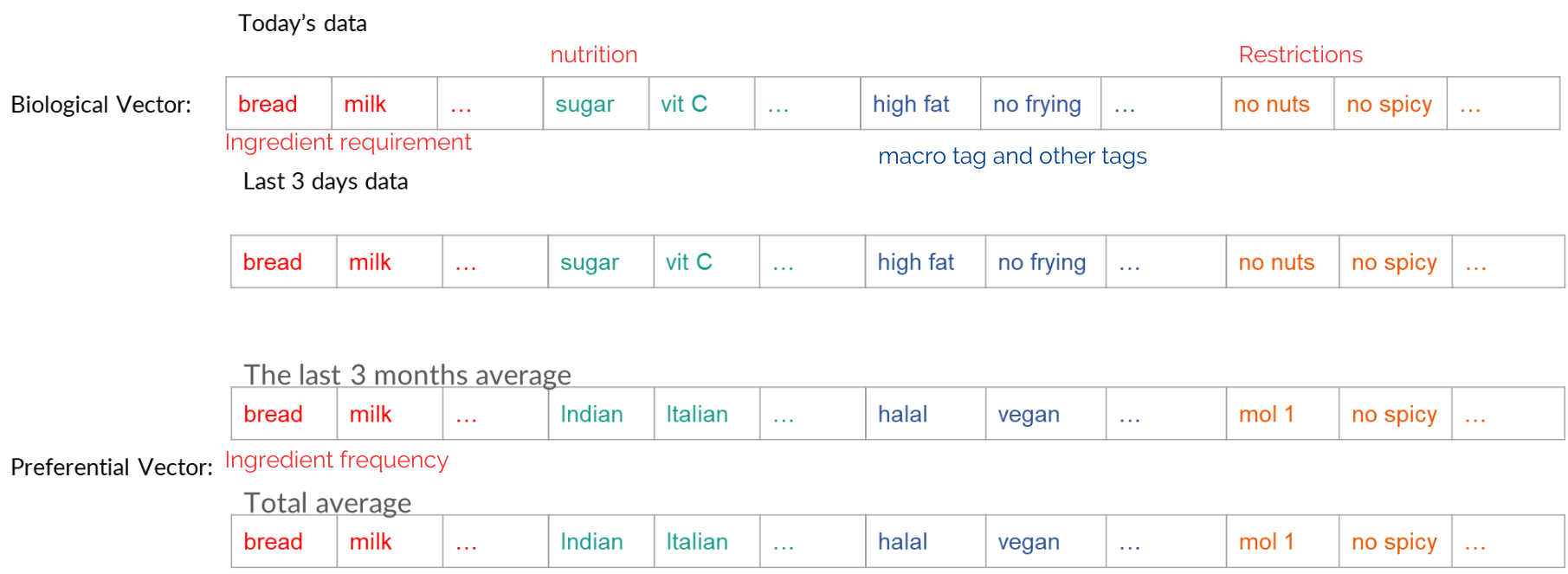}
    \caption{This figure illustrates the dual-component structure of personal vectors in food recommendation systems, differentiating between biological and preferential vectors. The biological vector, constructed based on the most recent dietary changes (the same day and the last three days), incorporates essential health-related instructions from health professionals. In contrast, the preferential vector, spanning broader periods (three and twelve months), captures the user's taste preferences and aversions based on historical dietary patterns. This visual representation underscores the tailored approach to capturing both immediate health requirements and longer-term dietary preferences to inform personalized food recommendations.}
    
    \label{fig:pfm_vec}
\end{figure}
\subsubsection{Preferential Vector Temporal Window}
Contrasting with the biological vector, the preferential vector acknowledges the longer duration required for shifts in dietary habits. Research indicates that forming a new eating habit can take, on average, 66 days, with variations extending up to eight months \cite{lally2010habits}. Accordingly, two timeframes are proposed: a three-month period reflecting the average duration for habit formation and a one-year period to accommodate the detection of infrequent preferences or aversions.

\section{The Feasibility Study Model}
The \ref{fig:pfm_vec} illustrates two primary vectors: the biological personal vector and the preferential personal vector. Each vector is constructed over two distinct periods, tailored to capture the dynamics of physiological needs and taste preferences separately.

\begin{enumerate}
    \item Biological Personal Vector: The biological vector is depicted focusing on a recent period of dietary change, emphasizing the immediate past: the same day and the last three days. This vector comprises items that are essential for the user's health, as mandated by health professionals or derived from automated dietary analysis. For example, it might include instructions like "Consume at least half a pound of meat daily" or "Avoid sugar to meet nutritional goals." These items are represented as points within the vector, each corresponding to a specific health-related directive or requirement.
    \item Preferential Personal Vector: In contrast, the preferential vector extends over longer, contextually relevant periods that reflect the slower pace of change in dietary preferences: three months and twelve months. This approach acknowledges the average time it takes for new eating habits to form and accommodates the detection of less frequent but significant preferences. Items within this vector illustrate the user's preferences, showing how they favor certain ingredients or dishes based on their consumption history. For instance, it might highlight a strong preference for spicy foods or a consistent avoidance of broccoli.
\end{enumerate}

While this feasibility study adopts time as the primary clustering factor for simplicity, it acknowledges the potential for more nuanced models that consider additional contextual factors. Future research could explore dynamic, personalized time segmentation models that account for a broader array of influences on dietary behavior and preferences. This preliminary case study lays the groundwork for more sophisticated approaches to personal vector generation in food recommendation systems, highlighting the necessity for continued innovation in capturing the evolving nature of individual dietary patterns.

\section{F-RLP Paradigm and Model}
The F-RLP paradigm harnesses the power of LLMs to infuse food-specific personalizing and context into the broader RLP framework. In this chapter, we adopt the T5 model introduced by \cite{t5} as our base RLP \cite{rostami2024food}. F-RLP preparation includes two main buildup steps and a utility step:

\begin{enumerate}
    \item Context Generation Stage: This stage involves supplying the LLM with a curated list of contextualized options within each query, ensuring that the responses are practical and appropriately aligned with the user's context and location.
    \item Counterfactual Generation (CFG) Retraining Stage:  In this stage, the foundational LLM is retrained using counterfactual data, tailored to be both personalized and improvement-oriented. This retraining process, facilitated by the CFG component's design, aims to imbue the model with the capability to offer added value to the user.
    \item Query Stage: Occurring after the LLM has been fully prepared, this stage is when the user seeks recommendations. Our framework enriches this process by incorporating specific auxiliary data into the query to refine the recommendation process, details of which will be explored subsequently.
\end{enumerate}

As shown in Figure 1, F-RLP consists of three stages, the first two are training and buildup steps and the last stage is a utility step. 
In comparing F-RLP with RLP, F-RLP distinguishes itself primarily during the Context Generation Stage by utilizing both personal and contextual vectors to curate a list of options. This list is subsequently utilized in the CFG Retraining Stage, and then tailored to individual queries during the Query Stage. A significant innovation in the CFG Retraining Stage under F-RLP involves the generation of counterfactual data, which enriches the training dataset with nutritional and preference-based enhancements—a feature RLP does not incorporate. Furthermore, each query in the Query Stage is crafted with a personal vector and a set of contextual options, ensuring targeted selection from the curated list. This mechanism represents another key advantage of F-RLP over RLP. The query process leverages a Counterfactually Retrained Text-to-Text Transformer, illustrating F-RLP's advanced approach to food recommendation. In the following, we will delve deeper into these stages. 

\subsection{Context Generation Stage}
In this stage, we convert the user's food log along with other personal data, such as location and sensory information gathered from wearables and smartphones, into a Context Vector and a Personal Vector. We employ a simplified model of the context and personal vector representation as introduced by \cite{Rostami21PPFM}. These vectors are then processed by the Context Recognition Engine (CRE) to generate a list of contextual food choice options available to the user, taking into account their current location and context.

It is crucial to highlight the significance of the CRE's existence and its role within the overall architecture. By considering location and context, the CRE plays a fundamental part in curating the contextual list of options. The design of the CRE draws upon our prior work. For further information on designing a CRE, readers are directed to \cite{Rostami21PPFM} and \cite{Rostami2022WFA}. 
Methods including query lookup on geospatial restaurant datasets, as introduced by \cite{Rostami21WFA} and \cite{Rostami2022WFA}, can be utilized to fulfill the implementation requirements of this component. In our experiment, we deploy a simplified version of the Context Recognition Engine (CRE), which selects twenty food choices at random from the \cite{marin2021recipe1m+} dataset. This dataset not only provides the nutritional content of each food choice but also details its ingredients. We employ these simplified CRE samples to demonstrate the effectiveness of the F-RLP methodology in practice.

\subsection{Counterfactual Generation (CFG) Retraining Stage}
Figure 1 illustrates the workflow of F-RLP, which benefits significantly from being trained with counterfactual data, in contrast to RLP. RLP, being trained solely on the user's factual data, is limited to predicting user behavior without offering any enhancements in terms of nutritional intake or preferences. Additionally, if a user receives dietary guidelines from a healthcare professional or advisor, there is no assurance that their historical data reflects adherence to these recommendations. Hence, RLP cannot guarantee compliance with such dietary instructions, as its operation is purely based on factual user logs.
Conversely, F-RLP incorporates a CFG component that utilizes expert advice to select the optimal choice, focusing on nutritional value and personal preferences, thus ensuring improvement. The CFG takes both the list of contextual options and the personal vector as inputs, generating counterfactual training data. This data is then used to retrain the text-to-text transformer, enhancing its ability to make recommendations that are not just based on past behavior but also aligned with expert dietary advice and improved nutritional goals.

\subsection{Query Stage}
As previously discussed, we utilize the T5 model as the foundational framework for our F-RLP system. After retraining the T5 model with counterfactual data, it's primed for querying. The training phase involved using a combination of the options list and the personal vector as inputs, with counterfactual data serving as the output. Consequently, to query the retrained model effectively, we require both a list of contextualized options and the personal vector at the query stage. This setup enables the retrained model to apply the insights gained from the counterfactual training phase, allowing it to select the most suitable option based on the comprehensive criteria it has learned.

\begin{figure}
    \centering
    \includegraphics[width=1.0\textwidth]{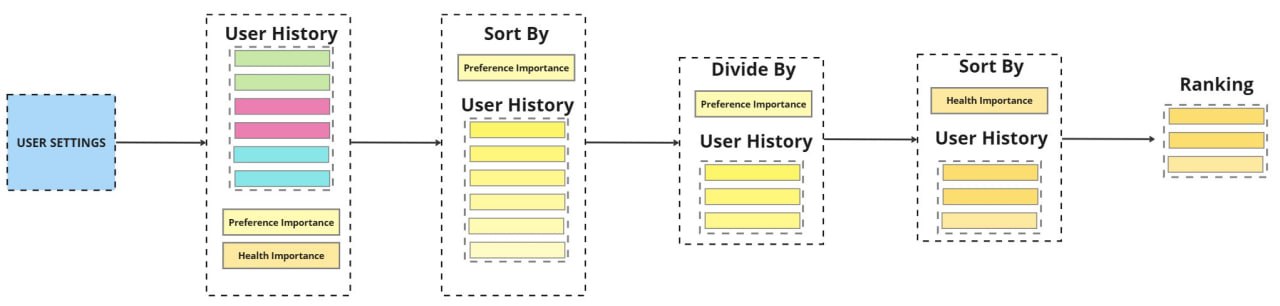}
    \caption{The algorithm for the counterfactual generation: This algorithm first sorts the option list based on the metric with highest priority in the given settings file. On the next step it divides the sorted list by the corresponding level of the highest priority metric and takes the batch with the highest score and sorts the new list again this time based on the second most important priority metric, and then picks the top option. This assures that both healthiness and preference are taken into consideration based on the priority they are given in the settings file.}
    \label{fig:algo}
    \vspace{-20pt}
\end{figure}

\section{CFG Setup and Configuration}
\subsection{CFG Input}

The CFG process necessitates two primary inputs: the Personal Vector, derived from the individual's personal dataset, and the option list, generated by the CRE.
We engage in an N-of-1 intensive longitudinal study of personal data collected by an adult male, encompassing his food log and sensory data (including sleep patterns, physical activity levels, and heart rate) over a two-year period. Such comprehensive data collection is made possible through the use of food logging applications \cite{Rostami2020FL}, which gather a wide range of food-related information. This dataset encompasses every food item and ingredient consumed by the individual daily, alongside biometric data obtained from an Oura ring.
The Personal Vector itself is divided into two segments: one segment includes straightforward numerical data representing the average biometric readings over the last three days, while the other segment compiles a list of the individual's most favored ingredients based on consumption patterns observed over the last thirty days.

The second input for the CFG process is a list of contextual options, produced by the CRE. While this chapter does not delve into the intricate implementation details of various CRE components, we simplify the approach by selecting twenty random food choices from the dataset referenced as \cite{marin2021recipe1m+} each time. These options are then prioritized based on the Personal Vector as well as any specified settings, ensuring that the recommendations are both relevant and personalized to the individual's preferences and context.

\begin{figure}
    \centering
    \includegraphics[width=0.72\textwidth]{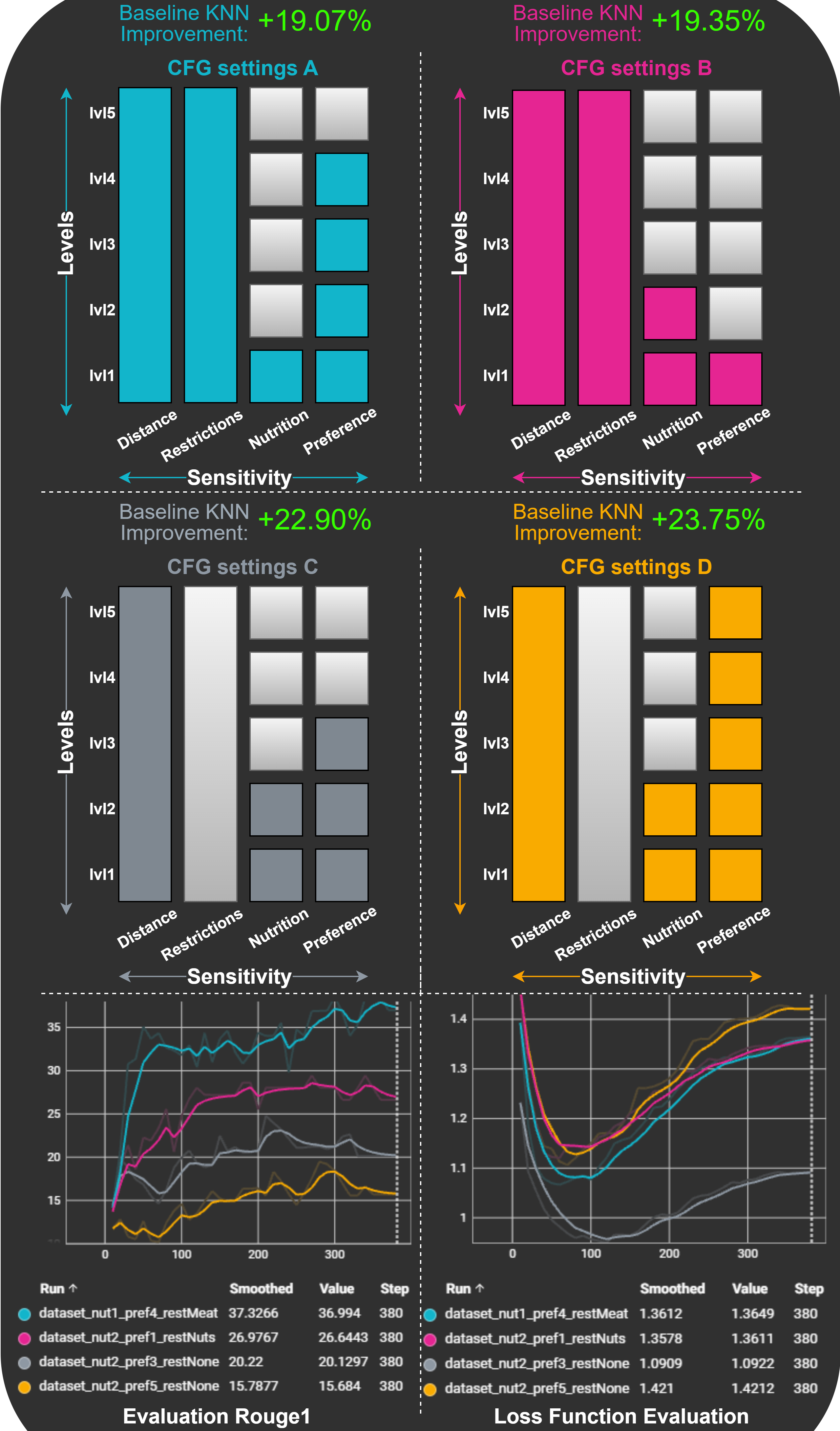}
    \caption{For each category, we utilize comparison metrics to assess the degree of improvement achieved by different setting classes relative to the baseline model without Counterfactual Generation (No CFG). The results indicate a positive enhancement across all categories. The horizontal axis of the charts represents the specific item selected to gauge sensitivity levels, while the vertical axis measures the extent of the sensitivity level itself.}
    \label{fig:eval}
    \vspace{-20pt}
\end{figure}

\subsection{CFG Settings}
The CFG process prioritizes four key factors when ranking the list of options, as follows:
\begin{itemize}
    \item Distance: This factor evaluates the significance of the proximity of the food choice (e.g., restaurant) to the user.
    \item Restrictions: This factor accounts for the presence of restricted ingredients, such as allergens or ingredients specifically advised against by a healthcare provider.
    \item Nutrition: This factor considers the nutritional content of the food in the decision-making process.
    \item Preference: This factor assesses the user's personal taste preferences in the final decision.
\end{itemize}
The CFG settings enable the assignment of different sensitivity levels to each of these factors, where a factor assigned a higher sensitivity level will exert a more substantial influence on the ranking of the options list as seen in \ref{fig:algo}. Except for the distance and restriction factors, which are binary, there are five sensitivity levels available for each factor.

In this experiment, the distance factor is always activated. This is because considering the proximity of food options is essential when generating the list in the CRE component. Given that we are working with a predetermined list of options for this study, we operate under the assumption that the distance factor has already been accounted for by providing a list of food choices that are within a reachable distance for the user.  

The restriction factor operates on a binary basis, indicating whether there is a specific list of ingredients that must be completely avoided or if no such restrictions exist. Figure 2 illustrates four sample setting profiles, along with the outcomes of each corresponding model trained under these conditions, presented on the left side.
In setting A, \textit{meat} was identified as a dietary restriction, whereas \textit{nuts} were the focus in setting B. The list of restricted items for setting A included "Beef", "Ham", "Cow", "Lamb", "Chicken", "Steak", "Burger", "Hotdog", "Goat", "Turkey", "Sausage", and "Rib". For setting B, the restrictions encompassed "Nuts", "Seeds", "Pecans", "Almonds", and "Pistachios". These settings illustrate how specific dietary constraints are applied and managed within the CFG process to tailor the recommendation system to individual dietary needs and preferences.

\subsection{CFG Output}
The distance factor is pre-considered within the CRE component, ensuring that only food options within an accessible range to the user are included. Meanwhile, the restriction factor is applied by excluding any food items that contain the specified restricted ingredients. The nutrition and preference factors, unlike the binary distance and restriction factors, are assigned integer values ranging from zero to five to determine their priority levels.
The algorithm employed to rank the remaining options list based on these non-binary factors initially prioritizes the factor with the higher sensitivity, organizing the list accordingly as seen in \ref{fig:algo}. Subsequent selection from the list depends on the actual level value assigned to each factor. For instance, if a factor's level value is two, the list is halved, and the top portion is selected; if the value is three, the list is divided by three, and so forth. Following this initial sorting, the list is then reorganized according to the second factor using the same method, resulting in a fully sorted list.
This final list, produced by the CFG, serves as the basis for retraining the F-RLP model \cite{rostami2024food}.

\begin{figure*}[t]
    \centering
    \includegraphics[width=\textwidth]{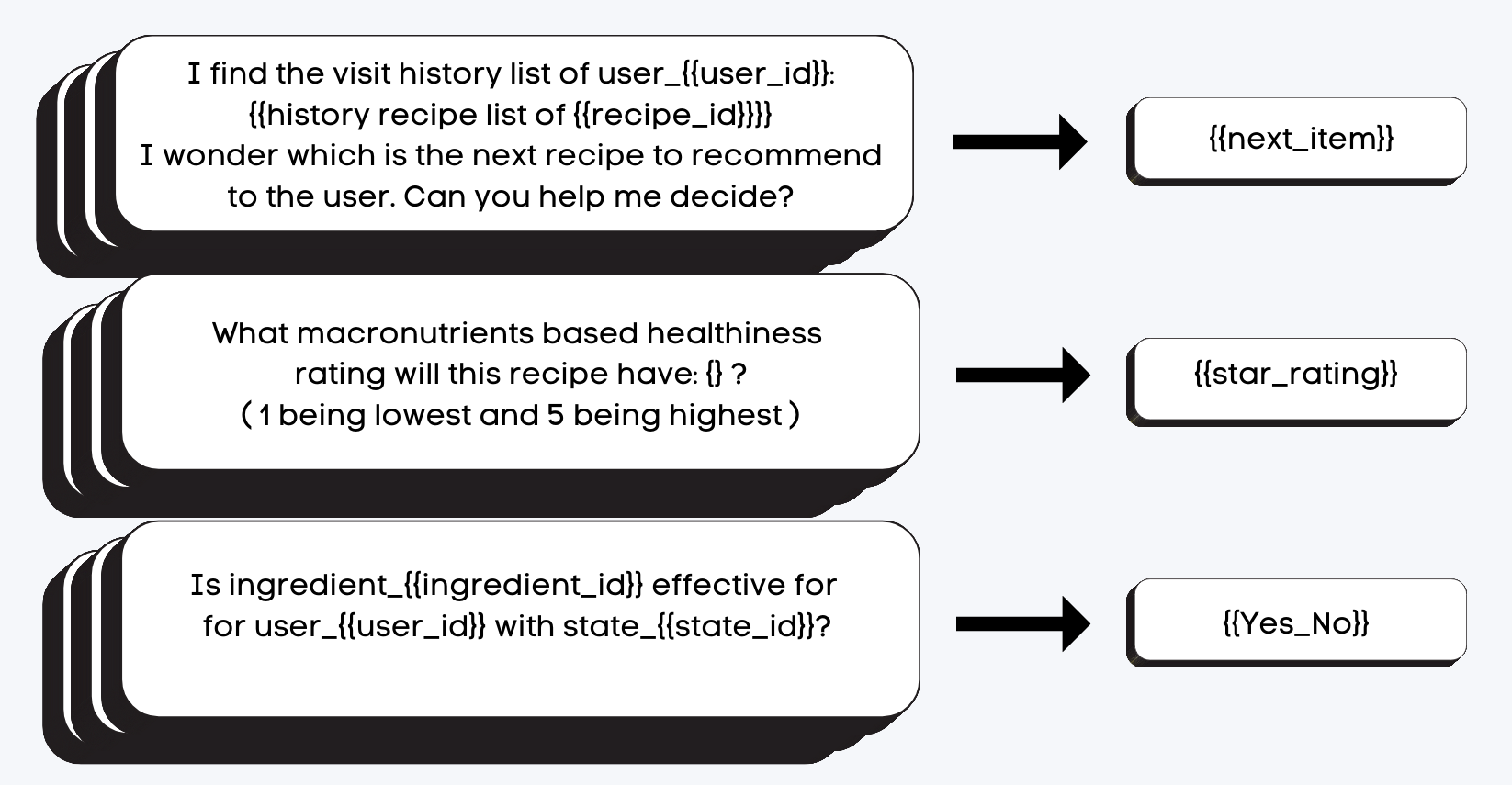}
    \caption{Showcasing a demo of template-based sample recommendation sample generation similar to the idea proposed by \cite{geng2022recommendation} but with completely food specific templates, enabling personalized and contextual data injection in the fine-tuning stage.
}
    \label{fig:demo_finetune}
\end{figure*}

\section{Additional Fine-Tuning}
In the domain of Large Language Model (LLM)-based food recommendation systems, refining pre-training and fine-tuning methodologies is critical for optimizing performance and accuracy. This section delineates advanced fine-tuning strategies that draw on state-of-the-art techniques in LLM recommendation systems to further enhance the current framework.

\subsection{ID-Based Item Representation}
Inspired by the technique proposed by \cite{Hua_2023}, this method involves abstracting complex food items and user interactions into unique identifiers (IDs). A corresponding mapping dataset retains the associations between these IDs and their real-world food item names and descriptions. This abstraction aims to streamline the model’s processing by reducing item complexity to numeric IDs, thus addressing and mitigating the challenge of hallucination in model outputs. \ref{fig:demo_finetune} is an exemplar on how the generated IDs will replace the users and items in the LLM pre-training samples. Implementing this ID-based representation within our framework could significantly improve the model's efficiency and accuracy by providing a more structured approach to understanding and generating food recommendations.
\begin{figure*}[t]
    \centering
    \includegraphics[width=\textwidth]{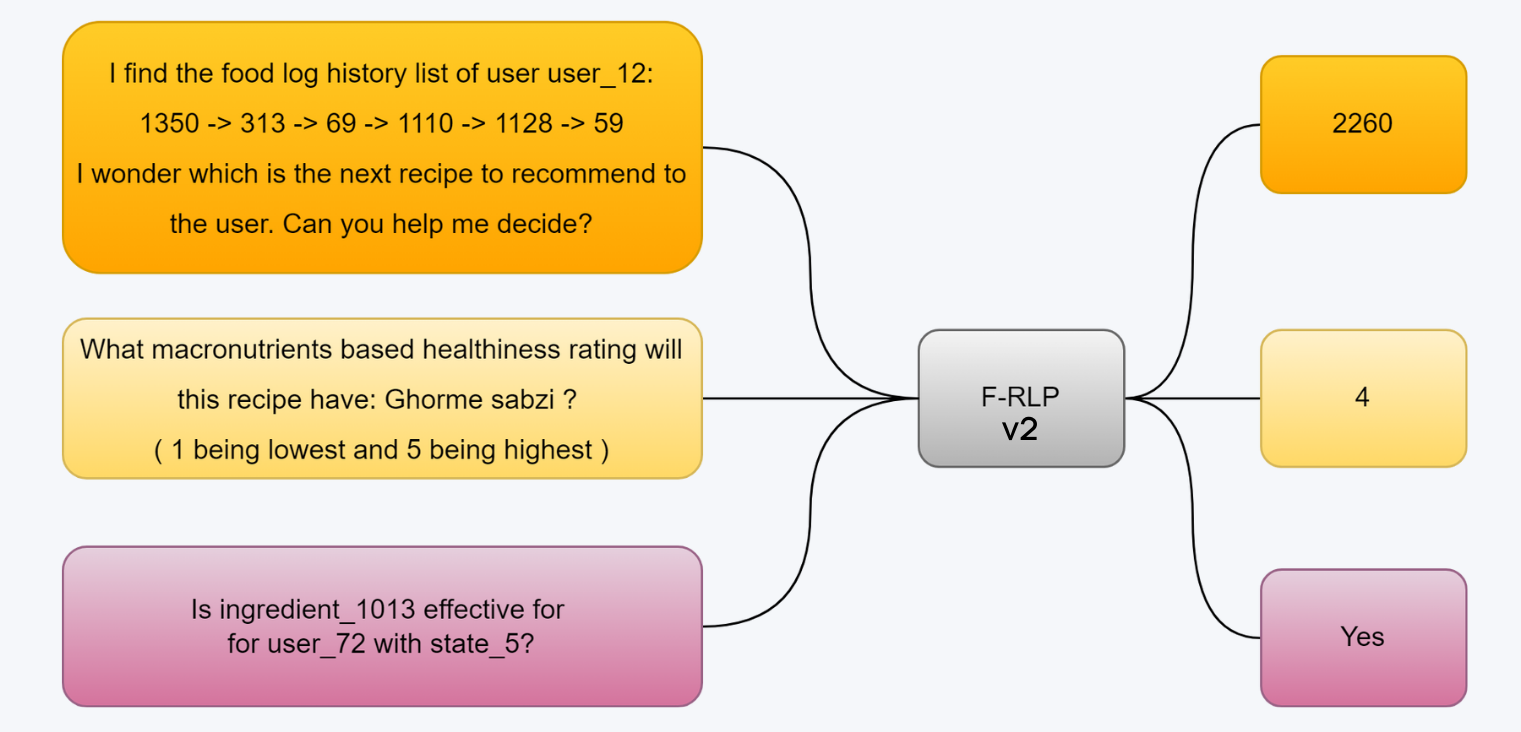}
    \caption{An exemplar of final result when training using the the ID representation of users and items, and fine-tuning based on the food-specific templates for each type of recommendation as you can see for in this figure, explored in the Additional Fine-Tuning section of this chapter and inspired by \cite{geng2022recommendation}.
}
    \label{fig:demo_result}
\end{figure*}
\subsection{Implementing Recommendation Training Templates}
Building upon insights from the \cite{geng2022recommendation} paper, we propose the adoption of custom-designed templates as a novel strategy for enriching the LLM’s training data. These templates are engineered to incorporate explicit recommendation contexts, such as nutritional information, into the training process, thereby enhancing the LLM’s ability to generate context-aware recommendations. \ref{fig:demo_finetune} Demonstrate three different categories of such templates for sequential recommendation, healthiness star rating recommendation and health effect Yes/No recommendation. For each category of recommendation, templates explicitly articulate key item relations, employing varied phrasings to ensure comprehensibility and retention by the LLM. Subsequent re-training of the LLM with these templates datasets aims to refine its output, focusing on the precision of dietary and nutritional advice.

The process of template development is iterative, focusing on the creation of diverse templates that cover a range of recommendation scenarios, including but not limited to nutritional content, dietary preferences, food intake history, and food health effect facts based on different sources of choice. The effectiveness of these templates is highlighted through the generation of training samples, each designed to mirror realistic user query scenarios and dietary considerations.

Through extensive template application and subsequent LLM re-training, we propose a significantly enhanced methodology which can be adopted in future studies in more detail to leverage the capacity for delivering personalized and contextually relevant food recommendations.

\begin{figure*}[t]
    \centering
    \includegraphics[width=\textwidth]{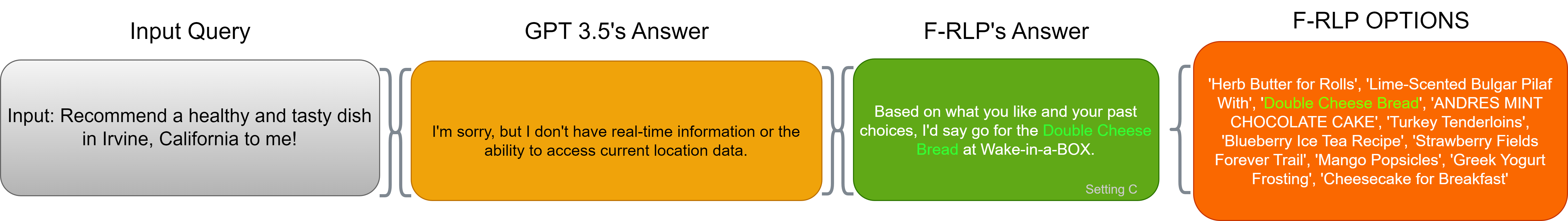}
    \caption{Showcasing F-RLP's meaningful and precise answer comparison with GPT 3.5 on our text food query.
}
    \label{fig:egg}
\end{figure*}

\section{Results}

The challenge of general food recommendation, hindered by an almost infinite number of classes, limited sample sizes, and unbalanced data distribution, renders most classification models ineffective. However, our proposed framework demonstrates significant promise over traditional approaches. This is substantiated by a comparative analysis against a KNN classifier using the same dataset. We evaluate the performance by measuring the deviation of the model's top recommendation from the CFG-sorted list and calculating the error rate in comparison to the baseline KNN model. \ref{fig:eval} offers crucial insights into the effectiveness of our methodology, highlighting its superior performance.
\begin{itemize}
    \item The sensitivity axis can be customized by each individual, leading to highly personalized results. This adjustment allows for a tailored experience that aligns closely with each user's unique preferences and needs.
    \item Contextual Relevance: By presenting a curated set of options to the LLM recommender with each query, we ensure that the context is integrated from the outset. This method acknowledges the significance of context, including location, in the selection process. 
    \item The model, retrained with counterfactual data, is designed to prioritize user satisfaction.
\end{itemize}
\ref{fig:egg} presents a sample query and response from F-RLP, illustrating its precision and relevance, a result of the effective integration between the option list and the LLM within our framework. To showcase this comparison we leverage from basic prompt-engineering like method to translate the LLM's output into a full sentence via a predefined template because our initial proof of study was solely re-training the dataset using Jason format but for future work the technique introduced in the Additional Fine-Tuning section could replace this step. \ref{fig:egg} also contrasts this query's outcome with those from GPT 3.5 \cite{NEURIPS2020_1457c0d6}. 

\section{Moving Forward}

This chapter presents a technically detailed exploration of Large Language Model (LLM) based food recommendation framework. The focus lies on the methodological implementation and the potential impact of these techniques on model performance. Future research will delve into refining these approaches, assessing their scalability, and exploring their applicability across different dietary contexts and user preferences. The ultimate goal is to advance the technical foundation of LLM-based food recommendation systems, ensuring they remain at the forefront of personalized dietary guidance.
In conclusion, while traditional food recommendation systems struggle with limitations of scale and data, the rise of LLMs introduces a promising solution. Yet, the general application of LLMs falls short in the specialized domain of food recommendations. Our F-RLP framework addressed this by tailoring LLM capabilities specifically for food, enhancing both accuracy and personalization. F-RLP represented a significant advancement in food recommendation technology, bridging the gap between generic algorithms and the specific needs of dietary guidance, and setting a new benchmark for precision in the field.
\chapter{Conclusion}
This thesis has sought to introduce a holistic framework for contextual and personalized food recommendations powered by Large Language Models (LLMs).  The emphasis has been placed on meticulously defining the essential components of effective food recommendation and developing an integrated approach that optimizes the interplay of these components within the LLM architecture.  This concluding chapter presents a synthesis of key findings,  highlights the study's unique contributions, acknowledges limitations, and provides recommendations for future avenues of research.

\section{Main Findings}

In this research, we introduced a comprehensive framework for food recommendation, articulating a detailed design for each component and the interactions among them. Our study underscored the necessity of incorporating essential components in any food recommendation system aimed at practical application in real-life scenarios, with a focus on recognizing context, personalization, and the vast domain of food items. Each component was thoroughly defined, detailing its role, position, and function within the overall system.

Significantly, we highlighted two innovative components often neglected in existing research: the multimedia food logger and the World Food Atlas (WFA). The multimedia food logger enhances the recording of food intake, offering a detailed schema to encapsulate various aspects of a food event, which proves challenging without a structured framework to represent the critical dimensions of food encounters.

Moreover, we delved into the importance of context representation and its impact on the causal analysis of food recommendations and decision-making processes related to food. Our empirical investigation into the influence of various causal factors on food choices yielded encouraging results, affirming the efficacy of our event representation model in identifying clear causal connections.

The introduction of the World Food Atlas represents a significant advancement in addressing the challenge of location-based food search queries. Unlike existing platforms that are limited in geographic coverage and lack strategies for expansion, our WFA framework is designed to aggregate data from diverse sources into a spatially-enabled database. It further empowers individual contributions, facilitating crowd-sourced expansion alongside automated data collection. This dual approach positions the WFA as a dynamic solution to the complex issue of global food item accessibility, setting a new standard for open-source, location-based food recommendation systems.

Finally, we integrated the various strands of our research by channeling all collected data into a Large Language Model (LLM) to serve as the backbone of our recommendation system. This step forward showcases a unique methodology not merely reliant on straightforward raw user data but also on generating enhanced counterfactual data. This data aims to foster healthier and more aligned choices with the user's preferences and settings. By amalgamating the counterfactual data with a personalized model vector—derived from the user's historical log—we crafted a comprehensive dataset for personal training. This research illustrates a sophisticated method to dynamically re-train the LLM, considering context as a pivotal query input. By integrating simple queries to location-based services like the World Food Atlas, we bestowed spatial awareness upon the LLM, thus facilitating context-aware recommendations. Moreover, we introduced several groundbreaking techniques that advance the frontier of food recommendation systems into new realms. This endeavor not only enhances the current state of the art but also charts a course towards a future where personal food recommendations reach unprecedented levels of precision and personalization, marking the onset of a new era in the journey towards personalized food recommendation systems.

\subsection{Strength}

This thesis offers significant advancements in the field of personalized food recommendation.  A key strength lies in its meticulous design of essential components tailored to the unique demands of food-related recommendations, addressing a shortcoming in general-purpose LLM-based approaches.  The framework provides detailed specifications for a comprehensive food logging platform designed specifically for personalization.  Furthermore, the novel World Food Atlas bridges the gap between food recommendations and location-based availability, ensuring practicality. The work introduces a holistic framework for LLM-based food recommendations, surpassing prior efforts that often focused on isolated components or relied on non-LLM training.  By integrating domain-specific models and mechanisms to ensure logical coherence, your model refines the reliability of LLM-generated recommendations. The framework's ability to access and process data from external food servers represents another significant contribution to real-world implementation.  Crucially, this thesis pioneers a new, LLM-driven research direction in food recommendation, leveraging the strengths of LLMs in managing a vast range of possibilities, aligning with the inherent complexity of food choices.

\textbf{\textit{Comprehensive Component Design:}} Detailed design of essential components specifically tailored for the food recommendation domain, addresses gaps in conventional LLM-based systems.

\textbf{\textit{Comprehensive Food Logging Platform:}} Introduction of a meticulously designed platform for capturing detailed food intake and event data, enhancing personalization capabilities.

\textbf{\textit{World Food Atlas (WFA):}} Development of a novel geospatial database to enable real-time, location-based food queries, facilitating access to a wide array of food options based on geographic location.

\textbf{\textit{Holistic Framework for LLM-Based Recommendation:}} Presentation of a first-of-its-kind holistic framework integrating specialized components and data streams for improved accuracy and reliability of food recommendations.

\textbf{\textit{Integration of Food-Specific Models into LLMs:}} Incorporation of food-specific models to ensure recommendations are grounded in gastronomic preferences and nutritional needs, overcoming the limitations of general-purpose LLM recommenders.

\textbf{\textit{Food Decoding and Option List Generation Interface:}} Introduction of an interface to enhance the relevance and context-appropriateness of LLM recommendations, reducing the risk of illogical suggestions.

\textbf{\textit{External Food Servers Integration:}} Overcoming technical challenges to integrate external food servers into the LLM-based recommendation process, enabling dynamic and responsive recommendations.

\textbf{\textit{Pioneering New Research Directions:}} Leveraging LLM's capabilities to handle a vast range of classes, setting new standards for personalized food recommendation systems and laying the groundwork for future research in the domain.

\subsection{Limitations}
The endeavor to revolutionize the landscape of food recommendations resembles the architectural marvel of Rome, a testament to the fact that greatness is not achieved overnight. This study, while pioneering in its scope and ambition, recognizes the vast expanse of the field and the incremental steps required to actualize the vision it proposes. Despite our efforts to delineate crucial components of personal food recommendations and illuminate the promising path of LLM-based food recommendation systems, the journey from conceptual framework to tangible reality remains long and fraught with challenges.

A primary area identified for future enhancement is the food logging platform. The current landscape of technology evolves at an unprecedented pace, heralding the advent of new sensors and methodologies that could significantly streamline the food logging process. The prospect of transitioning from active user involvement to a more passive surveillance-based approach, where food intake is logged automatically through advanced sensors, exemplifies the potential for profound advancements in this domain.

Moreover, the integration of food image recognition technology with the food knowledge graph holds immense promise for enriching the granularity and accuracy of food recommendations. However, the construction of a truly comprehensive food knowledge graph was beyond this project's scope. Instead, a simplified version of each component was employed to demonstrate the framework's viability within a research context, acknowledging the need for further development and refinement.

The LLM integration, heralded as a significant advancement in personalized and contextual food recommendation, similarly opens a vista of opportunities for future enhancements. The potential for augmenting the contextual and personal relevance of recommendations appears boundless, inviting ongoing exploration and innovation in harnessing LLM capabilities more effectively.

In sum, while this thesis lays foundational stones and charts a course toward revolutionizing personal food recommendations, it also underscores the vast potential for future advancements.

\section{Significance of the study}

\subsubsection{Advancements in Personalized Food Recommendation Technology}

\textbf{\textit{Comprehensive Food Logging Platform \& World Food Atlas (WFA):}}
This research introduces state-of-the-art technologies that significantly enhance artificial intelligence and personalization of food recommendations. The comprehensive food logging platform offers an innovative approach to capturing and analyzing individual dietary habits, preferences, and nutritional intake with unprecedented detail. Meanwhile, the World Food Atlas brings a groundbreaking solution to location-based food discovery, utilizing a novel geospatial database that aggregates diverse food data sources. Together, these technologies enable a more nuanced understanding of user preferences and the availability of food options, contributing to more accurate and geographically relevant food recommendations.

\subsubsection{Enhancements in Computational Food Recommendation Frameworks}

\textbf{\textit{Holistic Framework for LLM-Based Recommendation \& Integration of Food-Specific Models into LLMs:}}
This thesis pioneers a holistic framework that integrates LLMs with food-specific models, marking a significant leap forward in computational food recommendation systems. This integration allows for the leveraging of LLM's vast processing capabilities while ensuring the recommendations are deeply personalized and nutritionally considerate. By addressing the complex dimensions of personalization, including dietary restrictions, nutritional goals, and taste preferences, this framework sets a new benchmark for the precision and relevance of food recommendations.
\subsubsection{Technological Innovations for Context-Aware and Health-Conscious Recommendations}

\textbf{\textit{Food Decoding and Option List Generation Interface \& External Food Servers Integration:}}
The development of a food decoding interface and the successful integration of external food servers into the recommendation process represent major technological strides in making food recommendation systems more context-aware and health-conscious. These innovations ensure that recommendations are not only tailored to individual taste preferences but also to their current location, health goals, and available food options. This approach significantly improves the utility and effectiveness of food recommendations, fostering healthier eating habits and enhancing the overall dining experience.

Each of these contributions highlights the thesis's significant impact on advancing personalized food recommendation technologies, frameworks, and methodologies. Through its comprehensive approach and innovative solutions, this research paves the way for future developments that promise to transform how individuals make food choices, ultimately leading to improved dietary habits and enhanced quality of life.

\section{Future Research Direction}

As this thesis lays the groundwork for an innovative approach to personal food recommendations, it simultaneously opens several avenues for future scholarly inquiry. The multidisciplinary nature of this research, intersecting culinary science, technology, and health, invites a broad spectrum of exploratory paths. Below, we delineate key areas where future research could significantly advance the domain:

\begin{enumerate}
  \item Automated and Seamless Food Logging:
      \begin{itemize}
      \item Investigating the integration of emerging sensor technologies and surveillance approaches for effortless, passive food logging. This would minimize user burden and potentially capture a greater breadth of food-related data.
      \item Developing advanced image recognition algorithms specifically tailored to comprehensive food identification, including the analysis of complex and mixed dishes.
      \item Research to address ethical considerations related to data privacy and user autonomy in passive food logging.
    \end{itemize}
  \item Knowledge Graph Expansion and Integration:
        \begin{itemize}
      \item Construction of a comprehensive food knowledge graph that encompasses detailed nutritional information, intricate relationships between food items, and their impact on various physiological parameters.
      \item Development of methods for seamless integration of this knowledge graph with LLM architecture, enabling reasoning over food-related knowledge and further enhancing recommendation accuracy and personalization.
    \end{itemize}
  \item Advancements in LLM Contextualization and Personalization:
        \begin{itemize}
      \item Exploration of novel pre-training methods and fine-tuning techniques to tailor LLMs further to the specific nuances of personalized food recommendation.
      \item Designing innovative strategies for incorporating real-time user context, including dynamic physiological states and rapidly shifting preferences.
      \item Research into explainability techniques aimed at clarifying the reasoning behind LLM recommendations, fostering user trust and acceptance.
    \end{itemize}
\end{enumerate}
By pursuing these directions, future research has the potential to significantly advance the field of personalized food recommendations, offering more nuanced, health-conscious, and culturally aware dietary suggestions that cater to the global population's diverse needs and preferences.


\clearpage
\phantomsection

\bibliographystyle{abbrv}
\bibliography{thesis}

\begin{thebibliography}{100}

\bibitem{FoodLog:Applications}
{FoodLog: Multimedia Tool for Healthcare Applications}.
\newblock Technical report.

\bibitem{Motivate:Publication}
{Motivate: Towards context-aware recommendation mobile system for healthy
  living - IEEE Conference Publication}.

\bibitem{deepnetw}
Dser: Deep-sequential embedding for single domain recommendation, 2022.

\bibitem{classprob}
{\em Learning Multi-Subset of Classes for Fine-Grained Food Recognition}, 2022.

\bibitem{recsys2023}
page 177–192.
\newblock IGI Global, June 2023.

\bibitem{gnn}
The advance of recommendation system with graph neural network, 2023.

\bibitem{contentbased}
Content-based movie recommendation system: An enhanced approach to personalized
  movie recommendations, 2023.

\bibitem{probdeep}
Dietary behavior based food recommender system using deep learning and
  clustering techniques, 2023.

\bibitem{signals}
Improving implicit feedback-based recommendation through multi-behavior
  alignment, 2023.

\bibitem{colabfilter}
Trust enhanced collaborative filtering recommendation algorithm, 2023.

\bibitem{Rostami2020FL}
{A. Rostami et al.}
\newblock Multimedia food logger.
\newblock In {\em Proceedings of the 28th ACM International Conference on
  Multimedia}, MM ’20. ACM, Oct. 2020.

\bibitem{Rostami20PFM}
{A. Rostami et al.}
\newblock Personal food model.
\newblock In {\em Proceedings of the 28th ACM International Conference on
  Multimedia}, 2020.

\bibitem{Rostami2022WFA}
{A. Rostami et al.}
\newblock World food atlas for food navigation.
\newblock In {\em Proceedings of the 7th International Workshop on Multimedia
  Assisted Dietary Management on Multimedia Assisted Dietary Management}, MM
  ’22, 2022.

\bibitem{Abhari2019AAspects}
S.~Abhari, R.~Safdari, L.~Azadbakht, K.~B. Lankarani, S.~R. Niakan~Kalhori,
  B.~Honarvar, K.~Abhari, S.~M. Ayyoubza-Deh, Z.~Karbasi, S.~Zakerabasali, and
  Y.~Jalilpiran.
\newblock {A systematic review of nutrition recommendation systems: With focus
  on technical aspects}.
\newblock {\em Journal of Biomedical Physics and Engineering}, 9(6):591--602,
  12 2019.

\bibitem{Abhari2019_review_nutrition_recSys}
S.~Abhari, R.~Safdari, L.~Azadbakht, K.~B. Lankarani, S.~R. Niakan~Kalhori,
  B.~Honarvar, K.~Abhari, S.~M. Ayyoubza-Deh, Z.~Karbasi, S.~Zakerabasali, and
  Y.~Jalilpiran.
\newblock {A systematic review of nutrition recommendation systems: With focus
  on technical aspects}.
\newblock {\em Journal of Biomedical Physics and Engineering}, 9(6):591--602,
  12 2019.

\bibitem{abu2024supporting}
H.~Abu-Rasheed, M.~H. Abdulsalam, C.~Weber, and M.~Fathi.
\newblock Supporting student decisions on learning recommendations: An
  llm-based chatbot with knowledge graph contextualization for conversational
  explainability and mentoring.
\newblock {\em arXiv preprint arXiv:2401.08517}, 2024.

\bibitem{Adam2007StressSystemb}
T.~C. Adam and E.~S. Epel.
\newblock {Stress, eating and the reward system}.
\newblock {\em Physiology and Behavior}, 91(4):449--458, 7 2007.

\bibitem{AdomaviciusTowardExtensions}
G.~Adomavicius and A.~Tuzhilin.
\newblock {Toward the Next Generation of Recommender Systems: A Survey of the
  State-of-the-Art and Possible Extensions}.
\newblock Technical report.

\bibitem{Afaghi2008AcuteIndices}
A.~Afaghi, H.~O'Connor, and C.~M. Chow.
\newblock {Acute effects of the very low carbohydrate diet on sleep indices}.
\newblock {\em Nutritional Neuroscience}, 11(4):146--154, 2008.

\bibitem{ahmed2023better}
T.~Ahmed and P.~Devanbu.
\newblock Better patching using llm prompting, via self-consistency.
\newblock In {\em 2023 38th IEEE/ACM International Conference on Automated
  Software Engineering (ASE)}, pages 1742--1746. IEEE, 2023.

\bibitem{Aizawa2019FoodLog:Application}
K.~Aizawa.
\newblock {FoodLog: Multimedia Food Recording Platform and Its Application}.
\newblock In {\em Proceedings of the 5th International Workshop on Multimedia
  Assisted Dietary Management}, MADiMa '19, page~32, New York, NY, USA, 2019.
  Association for Computing Machinery.

\bibitem{al2022food}
H.~H. Al-Chalabi and M.~N. Jasim.
\newblock Food recommendation system based on data clustering techniques and
  user nutrition records.
\newblock In {\em International Conference on New Trends in Information and
  Communications Technology Applications}, pages 139--161. Springer, 2022.

\bibitem{ali2013classification}
A.~Ali, S.~M. Shamsuddin, and A.~L. Ralescu.
\newblock Classification with class imbalance problem.
\newblock {\em Int. J. Advance Soft Compu. Appl}, 5(3):176--204, 2013.

\bibitem{Aljbawi2020HealthawareFP}
B.~Aljbawi and W.~Laurier.
\newblock Health-aware food planner: A personalized recipe generation approach
  based on gpt-2.
\newblock In {\em Thesis}, 2020.

\bibitem{Apaolaza2018EatWellbeing}
V.~Apaolaza, P.~Hartmann, C.~D'Souza, and C.~M. L{\'{o}}pez.
\newblock {Eat organic – Feel good? The relationship between organic food
  consumption, health concern and subjective wellbeing}.
\newblock {\em Food Quality and Preference}, 63:51--62, 1 2018.

\bibitem{AsgariMehrabadi2020SleepPreprint}
M.~Asgari~Mehrabadi, I.~Azimi, F.~Sarhaddi, A.~Axelin, H.~Niela-Vil{\'{e}}n,
  S.~Myllyntausta, S.~Stenholm, N.~Dutt, P.~Liljeberg, and A.~M. Rahmani.
\newblock {Sleep Validation of Commercially Available Smart Ring and Watch
  Against Medical-Grade Actigraphy in Everyday Settings (Preprint)}.
\newblock {\em JMIR mHealth and uHealth}, 5 2020.

\bibitem{socialfoodatlas}
S.~F. Atlas.
\newblock Atlas - mammamiaaa, 2018-19.

\bibitem{Azimi2019PersonalizedStudy}
I.~Azimi, O.~Oti, S.~Labbaf, H.~Niela-Vilen, A.~Axelin, N.~Dutt, P.~Liljeberg,
  and A.~M. Rahmani.
\newblock {Personalized maternal sleep quality assessment: An objective
  iot-based longitudinal study}.
\newblock {\em IEEE Access}, 7:93433--93447, 2019.

\bibitem{tasteatlas}
M.~Babi\'{c}.
\newblock World food atlas: Discover 16493 local dishes \& ingredients, 2015.

\bibitem{BaHammam2013TheAssessment}
A.~S. BaHammam, A.~M. Alaseem, A.~A. Alzakri, and M.~M. Sharif.
\newblock {The effects of Ramadan fasting on sleep patterns and daytime
  sleepiness: An objective assessment}.
\newblock {\em Journal of Research in Medical Sciences}, 18(2):127--131, 2013.

\bibitem{bakker2022fine}
M.~Bakker, M.~Chadwick, H.~Sheahan, M.~Tessler, L.~Campbell-Gillingham,
  J.~Balaguer, N.~McAleese, A.~Glaese, J.~Aslanides, M.~Botvinick, et~al.
\newblock Fine-tuning language models to find agreement among humans with
  diverse preferences.
\newblock {\em Advances in Neural Information Processing Systems},
  35:38176--38189, 2022.

\bibitem{bao2020unilmv2}
H.~Bao, L.~Dong, F.~Wei, W.~Wang, N.~Yang, X.~Liu, Y.~Wang, J.~Gao, S.~Piao,
  M.~Zhou, et~al.
\newblock Unilmv2: Pseudo-masked language models for unified language model
  pre-training.
\newblock In {\em International conference on machine learning}, pages
  642--652. PMLR, 2020.

\bibitem{10.1145/3624918.3629550}
K.~Bao, J.~Zhang, Y.~Zhang, W.~Wenjie, F.~Feng, and X.~He.
\newblock Large language models for recommendation: Progresses and future
  directions.
\newblock In {\em Proceedings of the Annual International ACM SIGIR Conference
  on Research and Development in Information Retrieval in the Asia Pacific
  Region}, SIGIR-AP '23, page 306–309, New York, NY, USA, 2023. Association
  for Computing Machinery.

\bibitem{Barnard2002AData}
K.~Barnard, V.~Cardei, and B.~Funt.
\newblock {A comparison of computational color constancy algorithms - Part I:
  Methodology and experiments with synthesized data}.
\newblock {\em IEEE Transactions on Image Processing}, 11(9):972--984, 9 2002.

\bibitem{belzner2023large}
L.~Belzner, T.~Gabor, and M.~Wirsing.
\newblock Large language model assisted software engineering: prospects,
  challenges, and a case study.
\newblock In {\em International Conference on Bridging the Gap between AI and
  Reality}, pages 355--374. Springer, 2023.

\bibitem{Bonner2018CausalRecommendation}
S.~Bonner and F.~Vasile.
\newblock {Causal embeddings for recommendation}.
\newblock In {\em RecSys 2018 - 12th ACM Conference on Recommender Systems},
  pages 104--112, New York, NY, USA, 9 2018. Association for Computing
  Machinery, Inc.

\bibitem{Bossard2014Food-101Forests}
L.~Bossard, M.~Guillaumin, and L.~Van~Gool.
\newblock {Food-101 - Mining discriminative components with random forests}.
\newblock In {\em Lecture Notes in Computer Science (including subseries
  Lecture Notes in Artificial Intelligence and Lecture Notes in
  Bioinformatics)}, volume 8694 LNCS, pages 446--461. Springer Verlag, 2014.

\bibitem{boz2024improving}
A.~Boz, W.~Zorgdrager, Z.~Kotti, J.~Harte, P.~Louridas, D.~Jannach, and
  M.~Fragkoulis.
\newblock Improving sequential recommendations with llms.
\newblock {\em arXiv preprint arXiv:2402.01339}, 2024.

\bibitem{breneman2013food}
V.~Breneman.
\newblock Food environment atlas.
\newblock 2013.

\bibitem{DBLP:journals/corr/abs-2005-14165}
T.~B. Brown, B.~Mann, N.~Ryder, M.~Subbiah, J.~Kaplan, P.~Dhariwal,
  A.~Neelakantan, P.~Shyam, G.~Sastry, A.~Askell, S.~Agarwal,
  A.~Herbert{-}Voss, G.~Krueger, T.~Henighan, R.~Child, A.~Ramesh, D.~M.
  Ziegler, J.~Wu, C.~Winter, C.~Hesse, M.~Chen, E.~Sigler, M.~Litwin, S.~Gray,
  B.~Chess, J.~Clark, C.~Berner, S.~McCandlish, A.~Radford, I.~Sutskever, and
  D.~Amodei.
\newblock Language models are few-shot learners.
\newblock {\em CoRR}, abs/2005.14165, 2020.

\bibitem{t5}
{C. Raffel et al.}
\newblock Exploring the limits of transfer learning with a unified text-to-text
  transformer.
\newblock {\em J. Mach. Learn. Res.}, 21(1), 2020.

\bibitem{Chaix2019Time-RestrictedDiseases}
A.~Chaix, E.~N. Manoogian, G.~C. Melkani, and S.~Panda.
\newblock {Time-Restricted Eating to Prevent and Manage Chronic Metabolic
  Diseases}, 8 2019.

\bibitem{chen2017person}
C.-H. Chen, M.~Karvela, M.~Sohbati, T.~Shinawatra, and C.~Toumazou.
\newblock Person—personalized expert recommendation system for optimized
  nutrition.
\newblock {\em IEEE transactions on biomedical circuits and systems},
  12(1):151--160, 2017.

\bibitem{Chen2016Deep-basedRetrieval}
J.~Chen and C.~W. Ngo.
\newblock {Deep-based ingredient recognition for cooking recipe retrieval}.
\newblock In {\em MM 2016 - Proceedings of the 2016 ACM Multimedia Conference},
  pages 32--41. Association for Computing Machinery, Inc, 10 2016.

\bibitem{Chen2018SampleAhead:Data}
Q.~Chen, W.~Qiu, Y.~Zhang, L.~Xie, and A.~Yuille.
\newblock {SampleAhead: Online Classifier-Sampler Communication for Learning
  from Synthesized Data}.
\newblock {\em British Machine Vision Conference 2018, BMVC 2018}, 3 2018.

\bibitem{chen2021_pFoodReQ}
Y.~Chen, A.~Subburathinam, C.~H. Chen, and M.~J. Zaki.
\newblock {Personalized Food Recommendation as Constrained Question Answering
  over a Large-scale Food Knowledge Graph}.
\newblock In {\em WSDM 2021 - Proceedings of the 14th ACM International
  Conference on Web Search and Data Mining}, pages 544--552. Association for
  Computing Machinery, Inc, 8 2021.

\bibitem{chhipa2022recipe}
S.~Chhipa, V.~Berwal, T.~Hirapure, and S.~Banerjee.
\newblock Recipe recommendation system using tf-idf.
\newblock In {\em ITM Web of Conferences}, volume~44, page 02006. EDP Sciences,
  2022.

\bibitem{Chu2019RespirationSensors}
M.~Chu, T.~Nguyen, V.~Pandey, Y.~Zhou, H.~N. Pham, R.~Bar-Yoseph,
  S.~Radom-Aizik, R.~Jain, D.~M. Cooper, and M.~Khine.
\newblock {Respiration rate and volume measurements using wearable strain
  sensors}.
\newblock {\em npj Digital Medicine}, 2(1):1--9, 12 2019.

\bibitem{cui2023chatlaw}
J.~Cui, Z.~Li, Y.~Yan, B.~Chen, and L.~Yuan.
\newblock Chatlaw: Open-source legal large language model with integrated
  external knowledge bases.
\newblock {\em arXiv preprint arXiv:2306.16092}, 2023.

\bibitem{dettmers2024qlora}
T.~Dettmers, A.~Pagnoni, A.~Holtzman, and L.~Zettlemoyer.
\newblock Qlora: Efficient finetuning of quantized llms.
\newblock {\em Advances in Neural Information Processing Systems}, 36, 2024.

\bibitem{devlin2018bert}
J.~Devlin, M.-W. Chang, K.~Lee, and K.~Toutanova.
\newblock Bert: Pre-training of deep bidirectional transformers for language
  understanding.
\newblock {\em arXiv preprint arXiv:1810.04805}, 2018.

\bibitem{Dooley2018FoodIntegration}
D.~M. Dooley, E.~J. Griffiths, G.~S. Gosal, P.~L. Buttigieg, R.~Hoehndorf,
  M.~C. Lange, L.~M. Schriml, F.~S. Brinkman, and W.~W. Hsiao.
\newblock {Food on: A harmonized food ontology to increase global food
  traceability, quality control and data integration}.
\newblock {\em npj Science of Food}, 2(1):1--10, 1 2018.

\bibitem{dooley2018_FoodOn_Ontology}
D.~M. Dooley, E.~J. Griffiths, G.~S. Gosal, P.~L. Buttigieg, R.~Hoehndorf,
  M.~C. Lange, L.~M. Schriml, F.~S.~L. Brinkman, and W.~W.~L. Hsiao.
\newblock {FoodOn: a harmonized food ontology to increase global food
  traceability, quality control and data integration}.

\bibitem{Drescher2007AMeasures}
L.~S. Drescher, S.~Thiele, and G.~B. Mensink.
\newblock {A new index to measure healthy food diversity better reflects a
  healthy diet than traditional measures}.
\newblock {\em Journal of Nutrition}, 137(3):647--651, 3 2007.

\bibitem{duarte2022blending}
A.~Duarte and O.~Belo.
\newblock Blending case-based reasoning with ontologies for adapting diet menus
  and physical activities.
\newblock In {\em Proceedings of SAI Intelligent Systems Conference}, pages
  829--843. Springer, 2022.

\bibitem{Edwards1992SpicyThermoregulation}
S.~J. Edwards, I.~M. Montgomery, E.~Q. Colquhoun, J.~E. Jordan, and M.~G.
  Clark.
\newblock {Spicy meal disturbs sleep: an effect of thermoregulation?}
\newblock {\em International Journal of Psychophysiology}, 13(2):97--100, 1992.

\bibitem{eliyas2022recommendation}
S.~Eliyas and P.~Ranjana.
\newblock Recommendation systems: Content-based filtering vs collaborative
  filtering.
\newblock In {\em 2022 2nd International Conference on Advance Computing and
  Innovative Technologies in Engineering (ICACITE)}, pages 1360--1365. IEEE,
  2022.

\bibitem{Rostami21WFA}
A.~R. et~al.
\newblock World food atlas project.
\newblock In {\em Proceedings of the 13th International Workshop on Multimedia
  for Cooking and Eating Activities}, 2021.

\bibitem{fan2023recommender}
W.~Fan, Z.~Zhao, J.~Li, Y.~Liu, X.~Mei, Y.~Wang, J.~Tang, and Q.~Li.
\newblock Recommender systems in the era of large language models (llms).
\newblock {\em arXiv preprint arXiv:2307.02046}, 2023.

\bibitem{fang2024multi}
J.~Fang, S.~Gao, P.~Ren, X.~Chen, S.~Verberne, and Z.~Ren.
\newblock A multi-agent conversational recommender system.
\newblock {\em arXiv preprint arXiv:2402.01135}, 2024.

\bibitem{mainefoodatlas}
C.~for Community~GIS.
\newblock Maine food atlas, 2019.

\bibitem{friedman2015privacy}
A.~Friedman, B.~P. Knijnenburg, K.~Vanhecke, L.~Martens, and S.~Berkovsky.
\newblock Privacy aspects of recommender systems.
\newblock {\em Recommender systems handbook}, pages 649--688, 2015.

\bibitem{fu2023robust}
C.~Fu, W.~Wu, X.~Zhang, J.~Hu, J.~Wang, and J.~Zhou.
\newblock Robust user behavioral sequence representation via multi-scale
  stochastic distribution prediction.
\newblock In {\em Proceedings of the 32nd ACM International Conference on
  Information and Knowledge Management}, pages 4567--4573, 2023.

\bibitem{agapito2018dietos}
{G. Agapito et al.}
\newblock Dietos: A dietary recommender system for chronic diseases monitoring
  and management.
\newblock {\em Computer methods and programs in biomedicine}, 153:93--104,
  2018.

\bibitem{gao2020making}
T.~Gao, A.~Fisch, and D.~Chen.
\newblock Making pre-trained language models better few-shot learners.
\newblock {\em arXiv preprint arXiv:2012.15723}, 2020.

\bibitem{gao2023chat}
Y.~Gao, T.~Sheng, Y.~Xiang, Y.~Xiong, H.~Wang, and J.~Zhang.
\newblock Chat-rec: Towards interactive and explainable llms-augmented
  recommender system.
\newblock {\em arXiv preprint arXiv:2303.14524}, 2023.

\bibitem{Garg2018FlavorDB:Molecules}
N.~Garg, A.~Sethupathy, R.~Tuwani, R.~Nk, S.~Dokania, A.~Iyer, A.~Gupta,
  S.~Agrawal, N.~Singh, S.~Shukla, K.~Kathuria, R.~Badhwar, R.~Kanji, A.~Jain,
  A.~Kaur, R.~Nagpal, and G.~Bagler.
\newblock {FlavorDB: A database of flavor molecules}.
\newblock {\em Nucleic Acids Research}, 46(D1):D1210--D1216, 1 2018.

\bibitem{Garrido2013AAging.}
M.~Garrido, D.~Gonz{\'{a}}lez-G{\'{o}}mez, M.~Lozano, C.~Barriga, S.~D.
  Paredes, and A.~B. Rodr{\'{i}}guez.
\newblock {A Jerte valley cherry product provides beneficial effects on sleep
  quality. Influence on aging.}
\newblock {\em The journal of nutrition, health {\&} aging}, 17(6):553--60, 6
  2013.

\bibitem{geng2023hybrid}
H.~Geng, W.~Peng, X.~G. Shan, and C.~Song.
\newblock A hybrid recommendation algorithm for green food based on review text
  and review time.
\newblock {\em CyTA-Journal of Food}, 21(1):481--492, 2023.

\bibitem{geng2022recommendation}
{Geng et al.}
\newblock Recommendation as language processing (rlp): A unified pretrain,
  personalized prompt \& predict paradigm (p5).
\newblock In {\em Proceedings of the 16th ACM Conference on Recommender
  Systems}, pages 299--315, 2022.

\bibitem{ghazvininejad2019mask}
M.~Ghazvininejad, O.~Levy, Y.~Liu, and L.~Zettlemoyer.
\newblock Mask-predict: Parallel decoding of conditional masked language
  models.
\newblock {\em arXiv preprint arXiv:1904.09324}, 2019.

\bibitem{gomez2015netflix}
C.~A. Gomez-Uribe and N.~Hunt.
\newblock The netflix recommender system: Algorithms, business value, and
  innovation.
\newblock {\em ACM Transactions on Management Information Systems (TMIS)},
  6(4):1--19, 2015.

\bibitem{Gopalakrishnan2020AFR}
A.~K. Gopalakrishnan.
\newblock A food recommendation system based on bmi, bmr, k-nn algorithm, and a
  bpnn.
\newblock 2020.

\bibitem{greene2021electromagnetic}
J.~Greene.
\newblock {\em Electromagnetic Wearable Sensors: A Solution to Non-Invasive
  Real-Time Monitoring of Biological Markers during Exercise}.
\newblock Liverpool John Moores University (United Kingdom), 2021.

\bibitem{farmfreshatlaswisconsin}
R.~F. Group.
\newblock Farm fresh atlas, 2002.

\bibitem{gu2021ppt}
Y.~Gu, X.~Han, Z.~Liu, and M.~Huang.
\newblock Ppt: Pre-trained prompt tuning for few-shot learning.
\newblock {\em arXiv preprint arXiv:2109.04332}, 2021.

\bibitem{guntupalli2020high}
Y.~K. Guntupalli, V.~S. Saketh, S.~Amudheswaran, and D.~S. Vaishnav.
\newblock High-scale food recommendation built on apache spark using
  alternating least squares.
\newblock In {\em 2020 5th IEEE International Conference on Recent Advances and
  Innovations in Engineering (ICRAIE)}, pages 1--5. IEEE, 2020.

\bibitem{guo2024integrating}
N.~Guo, H.~Cheng, Q.~Liang, L.~Chen, and B.~Han.
\newblock Integrating large language models with graphical session-based
  recommendation, 2024.

\bibitem{harte2023leveraging}
J.~Harte, W.~Zorgdrager, P.~Louridas, A.~Katsifodimos, D.~Jannach, and
  M.~Fragkoulis.
\newblock Leveraging large language models for sequential recommendation.
\newblock In {\em Proceedings of the 17th ACM Conference on Recommender
  Systems}, pages 1096--1102, 2023.

\bibitem{Harvey2013YouPrediction}
M.~Harvey, B.~Ludwig, and D.~Elsweiler.
\newblock {You are what you eat: Learning user tastes for rating prediction}.
\newblock In {\em Lecture Notes in Computer Science (including subseries
  Lecture Notes in Artificial Intelligence and Lecture Notes in
  Bioinformatics)}, volume 8214 LNCS, pages 153--164. Springer Verlag, 2013.

\bibitem{Hashimoto1996VitaminHumans}
S.~Hashimoto, M.~Kohsaka, N.~Morita, N.~Fukuda, S.~Honma, and K.~I. Honma.
\newblock {Vitamin B12 enhances the phase-response of circadian melatonin
  rhythm to a single bright light exposure in humans}.
\newblock {\em Neuroscience Letters}, 220(2):129--132, 12 1996.

\bibitem{HaussmannFoodKG:Recommendation}
S.~Haussmann.
\newblock {FoodKG: A Semantics-Driven Knowledge Graph for Food Recommendation}.
\newblock Technical report.

\bibitem{Haussmann2019_food_kg}
S.~Haussmann, O.~Seneviratne, Y.~Chen, Y.~Ne’eman, J.~Codella, C.~H. Chen,
  D.~L. McGuinness, and M.~J. Zaki.
\newblock {FoodKG: A Semantics-Driven Knowledge Graph for Food Recommendation}.
\newblock In {\em Lecture Notes in Computer Science (including subseries
  Lecture Notes in Artificial Intelligence and Lecture Notes in
  Bioinformatics)}, volume 11779 LNCS, pages 146--162. 2019.

\bibitem{he2017neural}
X.~He, L.~Liao, H.~Zhang, L.~Nie, X.~Hu, and T.-S. Chua.
\newblock Neural collaborative filtering.
\newblock In {\em Proceedings of the 26th international conference on world
  wide web}, pages 173--182, 2017.

\bibitem{how2020predictive}
M.-L. How, Y.~J. Chan, and S.-M. Cheah.
\newblock Predictive insights for improving the resilience of global food
  security using artificial intelligence.
\newblock {\em Sustainability}, 12(15):6272, 2020.

\bibitem{hua2023tutorial}
W.~Hua, L.~Li, S.~Xu, L.~Chen, and Y.~Zhang.
\newblock Tutorial on large language models for recommendation.
\newblock In {\em Proceedings of the 17th ACM Conference on Recommender
  Systems}, pages 1281--1283, 2023.

\bibitem{Hua_2023}
W.~Hua, S.~Xu, Y.~Ge, and Y.~Zhang.
\newblock How to index item ids for recommendation foundation models.
\newblock In {\em Proceedings of the Annual International ACM SIGIR Conference
  on Research and Development in Information Retrieval in the Asia Pacific
  Region}, SIGIR-AP ’23. ACM, Nov. 2023.

\bibitem{Ito2018ModelingProcess}
T.~Ito, Y.~Fukazawa, D.~Zhu, and J.~Ota.
\newblock {Modeling Weather Context Dependent Food Choice Process}.
\newblock {\em Journal of Information Processing}, 26(0):386--395, 1 2018.

\bibitem{marin2021recipe1m+}
{J. Mar{\i}n et al.}
\newblock Recipe1m+: A dataset for learning cross-modal embeddings for cooking
  recipes and food images.
\newblock {\em IEEE Transactions on Pattern Analysis and Machine Intelligence},
  43(1):187--203, 2021.

\bibitem{jain2022personalized}
A.~Jain and A.~Singhal.
\newblock Personalized food recommendation—state of art and review.
\newblock {\em Ambient communications and computer systems: Proceedings of
  RACCCS 2021}, pages 153--164, 2022.

\bibitem{Jalali2016InteractiveStreams}
L.~Jalali.
\newblock {Interactive Event-driven Knowledge Discovery from Data Streams}.
\newblock 2016.

\bibitem{jasimfood}
M.~N. Jasim and A.~B. Hamid.
\newblock Food recommendation system based on nutritional needs of human beings
  and user preferences.

\bibitem{jeckmans2013privacy}
A.~J. Jeckmans, M.~Beye, Z.~Erkin, P.~Hartel, R.~L. Lagendijk, and Q.~Tang.
\newblock Privacy in recommender systems.
\newblock {\em Social media retrieval}, pages 263--281, 2013.

\bibitem{Ju2022MenuAIRF}
X.~Ju, F.~P.-W. Lo, J.~Qiu, P.~Shi, J.~Peng, and B.~P.~L. Lo.
\newblock Menuai: Restaurant food recommendation system via a transformer-based
  deep learning model.
\newblock {\em ArXiv}, abs/2210.08266, 2022.

\bibitem{KasaeyanNaeini2019AnMonitoring}
E.~Kasaeyan~Naeini, S.~Shahhosseini, A.~Subramanian, T.~Yin, A.~M. Rahmani, and
  N.~Dutt.
\newblock {An Edge-Assisted and Smart System for Real-Time Pain Monitoring}.
\newblock In {\em Proceedings - 4th IEEE/ACM Conference on Connected Health:
  Applications, Systems and Engineering Technologies, CHASE 2019}, pages
  47--52. Institute of Electrical and Electronics Engineers Inc., 9 2019.

\bibitem{kaur2023deepconn}
A.~Kaur, L.~Kaur, and A.~Singh.
\newblock Deepconn: patch-wise deep convolutional neural networks for the
  segmentation of multiple sclerosis brain lesions.
\newblock {\em Multimedia Tools and Applications}, pages 1--33, 2023.

\bibitem{khansili2018label}
N.~Khansili, G.~Rattu, and P.~M. Krishna.
\newblock Label-free optical biosensors for food and biological sensor
  applications.
\newblock {\em Sensors and Actuators B: Chemical}, 265:35--49, 2018.

\bibitem{kiddon2016globally}
C.~Kiddon, L.~Zettlemoyer, and Y.~Choi.
\newblock Globally coherent text generation with neural checklist models.
\newblock In {\em Proceedings of the 2016 conference on empirical methods in
  natural language processing}, pages 329--339, 2016.

\bibitem{grouplens}
J.~Konstan, J.~Riedl, A.~Borchers, and J.~Herlocker.
\newblock Recommender systems: A grouplens perspective.
\newblock 01 1998.

\bibitem{kumar2016survey}
A.~Kumar, P.~Tanwar, and S.~Nigam.
\newblock Survey and evaluation of food recommendation systems and techniques.
\newblock In {\em 2016 3rd International Conference on Computing for
  Sustainable Global Development (INDIACom)}, pages 3592--3596. IEEE, 2016.

\bibitem{lally2010habits}
P.~Lally, C.~H. Van~Jaarsveld, H.~W. Potts, and J.~Wardle.
\newblock How are habits formed: Modelling habit formation in the real world.
\newblock {\em European journal of social psychology}, 40(6):998--1009, 2010.

\bibitem{Lambiase2013TemporalWomen}
M.~J. Lambiase, K.~P. Gabriel, L.~H. Kuller, and K.~A. Matthews.
\newblock {Temporal relationships between physical activity and sleep in older
  women}.
\newblock {\em Medicine and Science in Sports and Exercise}, 45(12):2362--2368,
  12 2013.

\bibitem{law2010sub}
M.~K. Law, A.~Bermak, and H.~C. Luong.
\newblock A sub-backslash embedded cmos temperature sensor for rfid food
  monitoring application.
\newblock {\em IEEE journal of solid-state circuits}, 45(6):1246--1255, 2010.

\bibitem{lederer2022relation}
A.-K. Lederer and R.~Huber.
\newblock The relation of diet and health: You are what you eat, 2022.

\bibitem{lester2021power}
B.~Lester, R.~Al-Rfou, and N.~Constant.
\newblock The power of scale for parameter-efficient prompt tuning.
\newblock {\em arXiv preprint arXiv:2104.08691}, 2021.

\bibitem{levitsky2022rise}
D.~A. Levitsky, L.~Barre, J.~J. Michael, Y.~Zhong, Y.~He, A.~Mizia, and
  S.~Kaila.
\newblock The rise and fall of physiological theories of the control of human
  eating behavior.
\newblock {\em Frontiers in Nutrition}, 9:826334, 2022.

\bibitem{li2023personalized}
L.~Li, Y.~Zhang, and L.~Chen.
\newblock Personalized prompt learning for explainable recommendation.
\newblock {\em ACM Transactions on Information Systems}, 41(4):1--26, 2023.

\bibitem{li2023prompt}
L.~Li, Y.~Zhang, and L.~Chen.
\newblock Prompt distillation for efficient llm-based recommendation.
\newblock In {\em Proceedings of the 32nd ACM International Conference on
  Information and Knowledge Management}, pages 1348--1357, 2023.

\bibitem{li2018application}
X.~Li, W.~Jia, Z.~Yang, Y.~Li, D.~Yuan, H.~Zhang, and M.~Sun.
\newblock Application of intelligent recommendation techniques for consumers'
  food choices in restaurants.
\newblock {\em Frontiers in psychiatry}, 9:415, 2018.

\bibitem{Li2018ApplicationRestaurants}
X.~Li, W.~Jia, Z.~Yang, Y.~Li, D.~Yuan, H.~Zhang, and M.~Sun.
\newblock {Application of Intelligent Recommendation Techniques for Consumers'
  Food Choices in Restaurants}.
\newblock {\em Frontiers in Psychiatry}, 9(SEP):415, 9 2018.

\bibitem{lieber2021jurassic}
O.~Lieber, O.~Sharir, B.~Lenz, and Y.~Shoham.
\newblock Jurassic-1: Technical details and evaluation.
\newblock {\em White Paper. AI21 Labs}, 1:9, 2021.

\bibitem{Lin2011EffectProblems}
H.~H. Lin, P.~S. Tsai, S.~C. Fang, and J.~F. Liu.
\newblock {Effect of kiwifruit consumption on sleep quality in adults with
  sleep problems}.
\newblock {\em Asia Pacific Journal of Clinical Nutrition}, 20(2):169--174, 6
  2011.

\bibitem{DBLP:journals/corr/abs-2110-03888}
J.~Lin, A.~Yang, J.~Bai, C.~Zhou, L.~Jiang, X.~Jia, A.~Wang, J.~Zhang, Y.~Li,
  W.~Lin, J.~Zhou, and H.~Yang.
\newblock {M6-10T:} {A} sharing-delinking paradigm for efficient multi-trillion
  parameter pretraining.
\newblock {\em CoRR}, abs/2110.03888, 2021.

\bibitem{lin2023multi}
X.~Lin, W.~Wang, Y.~Li, F.~Feng, S.-K. Ng, and T.-S. Chua.
\newblock A multi-facet paradigm to bridge large language model and
  recommendation.
\newblock {\em arXiv preprint arXiv:2310.06491}, 2023.

\bibitem{linden2003amazon}
G.~Linden, B.~Smith, and J.~York.
\newblock Amazon. com recommendations: Item-to-item collaborative filtering.
\newblock {\em IEEE Internet computing}, 7(1):76--80, 2003.

\bibitem{liu2022few}
H.~Liu, D.~Tam, M.~Muqeeth, J.~Mohta, T.~Huang, M.~Bansal, and C.~A. Raffel.
\newblock Few-shot parameter-efficient fine-tuning is better and cheaper than
  in-context learning.
\newblock {\em Advances in Neural Information Processing Systems},
  35:1950--1965, 2022.

\bibitem{liu2023chatgpt}
J.~Liu, C.~Liu, R.~Lv, K.~Zhou, and Y.~Zhang.
\newblock Is chatgpt a good recommender? a preliminary study.
\newblock {\em arXiv preprint arXiv:2304.10149}, 2023.

\bibitem{Loprinzi2011Association2005-2006}
P.~D. Loprinzi and B.~J. Cardinal.
\newblock {Association between objectively-measured physical activity and
  sleep, NHANES 2005-2006}.
\newblock {\em Mental Health and Physical Activity}, 4(2):65--69, 12 2011.

\bibitem{Loprinzi2012TheWomen}
P.~D. Loprinzi, K.~L. Loprinzi, and B.~J. Cardinal.
\newblock {The relationship between physical activity and sleep among pregnant
  women}.
\newblock {\em Mental Health and Physical Activity}, 5(1):22--27, 6 2012.

\bibitem{Losso2018PilotMechanisms}
J.~N. Losso, J.~W. Finley, N.~Karki, A.~G. Liu, A.~Prudente, R.~Tipton, Y.~Yu,
  and F.~L. Greenway.
\newblock {Pilot Study of the Tart Cherry Juice for the Treatment of Insomnia
  and Investigation of Mechanisms}.
\newblock {\em American journal of therapeutics}, 25(2):e194--e201, 3 2018.

\bibitem{mahardikaa2020case}
I.~G.~T. Mahardikaa and I.~W. Suprianaa.
\newblock A case based reasoning system for recommendation of restaurant in
  jimbaran using k-nearest neighbor.
\newblock {\em Jurnal Elektronik Ilmu Komputer Udayana p-ISSN}, 2301:5373,
  2020.

\bibitem{Majjodi2022NudgingTH}
A.~E. Majjodi, A.~D. Starke, and C.~Trattner.
\newblock Nudging towards health? examining the merits of nutrition labels and
  personalization in a recipe recommender system.
\newblock {\em Proceedings of the 30th ACM Conference on User Modeling,
  Adaptation and Personalization}, 2022.

\bibitem{majumder-etal-2019-generating}
B.~P. Majumder, S.~Li, J.~Ni, and J.~McAuley.
\newblock Generating personalized recipes from historical user preferences.
\newblock In {\em Proceedings of the 2019 Conference on Empirical Methods in
  Natural Language Processing and the 9th International Joint Conference on
  Natural Language Processing (EMNLP-IJCNLP)}, pages 5976--5982, Hong Kong,
  China, Nov. 2019. Association for Computational Linguistics.

\bibitem{manoharan2020patient}
D.~S. Manoharan and A.~Sathesh.
\newblock Patient diet recommendation system using k clique and deep learning
  classifiers.
\newblock {\em Journal of Artificial Intelligence and Capsule Networks},
  2(2):121--130, 2020.

\bibitem{marin2019recipe1m+}
J.~Marin, A.~Biswas, F.~Ofli, N.~Hynes, A.~Salvador, Y.~Aytar, I.~Weber, and
  A.~Torralba.
\newblock Recipe1m+: A dataset for learning cross-modal embeddings for cooking
  recipes and food images.
\newblock {\em IEEE transactions on pattern analysis and machine intelligence},
  43(1):187--203, 2019.

\bibitem{Mehrabadi2020TheAnalysis}
M.~A. Mehrabadi, N.~Dutt, and A.~M. Rahmani.
\newblock {The Causality Inference of Public Interest in Restaurants and Bars
  on COVID-19 Daily Cases in the US: A Google Trends Analysis}.
\newblock 7 2020.

\bibitem{meimei1985personalized}
C.~Meimei and X.~Kangjie.
\newblock Personalized recommendation algorithm based on modified tensor
  decomposition model.
\newblock {\em Data Analysis and Knowledge Discovery}, 1(3):38--45, 1985.

\bibitem{melese2021food}
A.~Melese.
\newblock Food and restaurant recommendation system using hybrid filtering
  mechanism.
\newblock {\em Monthly Journal by TWASP}, 4(4):268--281, 2021.

\bibitem{min2019_food_recommendation}
W.~Min, S.~Jiang, and R.~Jain.
\newblock {Food Recommendation: Framework, Existing Solutions, and Challenges}.
\newblock {\em IEEE Transactions on Multimedia}, 22(10):2659--2671, 2020.

\bibitem{Min2019FoodChallenges}
W.~Min, S.~Jiang, and R.~C. Jain.
\newblock {Food Recommendation: Framework, Existing Solutions and Challenges}.
\newblock {\em IEEE Transactions on Multimedia}, pages 1--1, 12 2019.

\bibitem{Min2019AComputing}
W.~Min, S.~Jiang, L.~Liu, Y.~Rui, and R.~Jain.
\newblock {A survey on food computing}.
\newblock {\em ACM Computing Surveys}, 52(5), 9 2019.

\bibitem{min2019survey}
W.~Min, S.~Jiang, L.~Liu, Y.~Rui, and R.~Jain.
\newblock A survey on food computing.
\newblock {\em ACM Computing Surveys (CSUR)}, 52(5):1--36, 2019.

\bibitem{Min2021_foodKGReview}
W.~Min, C.~Liu, L.~Xu, and S.~Jiang.
\newblock {The Development and Applications of Food Knowledge Graphs in the
  Food Science and Industry}.
\newblock pages 1--45, 2021.

\bibitem{mooney2000content}
R.~J. Mooney and L.~Roy.
\newblock Content-based book recommending using learning for text
  categorization.
\newblock In {\em Proceedings of the fifth ACM conference on Digital
  libraries}, pages 195--204, 2000.

\bibitem{Musto2020TowardsAK}
C.~Musto, C.~Trattner, A.~D. Starke, and G.~Semeraro.
\newblock Towards a knowledge-aware food recommender system exploiting holistic
  user models.
\newblock {\em Proceedings of the 28th ACM Conference on User Modeling,
  Adaptation and Personalization}, 2020.

\bibitem{Naeini2019AInternet-of-Things}
E.~K. Naeini, I.~Azimi, A.~M. Rahmani, P.~Liljeberg, and N.~Dutt.
\newblock {A Real-time PPG Quality Assessment Approach for Healthcare
  Internet-of-Things}.
\newblock In {\em Procedia Computer Science}, volume 151, pages 551--558.
  Elsevier B.V., 1 2019.

\bibitem{Nag2019SynchronizingMonitoring}
N.~Nag, V.~Pandey, L.~Navali, P.~Mohan, and R.~Jain.
\newblock {Synchronizing Geospatial Information for Personalized Health
  Monitoring}.
\newblock 7 2019.

\bibitem{Namgung2019MenuChildren}
K.~Namgung, T.~H. Kim, Y.~S. Hong, and S.~Nazir.
\newblock {Menu Recommendation System Using Smart Plates for Well-balanced Diet
  Habits of Young Children}.
\newblock {\em Wireless Communications and Mobile Computing}, 2019, 2019.

\bibitem{NewtonAndersonEveryoneCulture}
E.~Newton~Anderson.
\newblock {Everyone Eats: Understanding Food and Culture}.
\newblock Technical report.

\bibitem{nikzamir2022highly}
A.~Nikzamir and F.~Capolino.
\newblock Highly sensitive coupled oscillator based on an exceptional point of
  degeneracy and nonlinearity.
\newblock {\em arXiv preprint arXiv:2206.04031}, 2022.

\bibitem{Nirmal2018OptimizationApproach}
I.~Nirmal, A.~Caldera, and R.~D. Bandara.
\newblock {Optimization Framework for Flavour and Nutrition Balanced Recipe: A
  Data Driven Approach}.
\newblock In {\em 2018 5th IEEE Uttar Pradesh Section International Conference
  on Electrical, Electronics and Computer Engineering, UPCON 2018}. Institute
  of Electrical and Electronics Engineers Inc., 12 2018.

\bibitem{Oh2017FromChronicles}
H.~Oh and R.~Jain.
\newblock {From multimedia logs to personal chronicles}.
\newblock In {\em MM 2017 - Proceedings of the 2017 ACM Multimedia Conference},
  pages 881--889, New York, New York, USA, 10 2017. Association for Computing
  Machinery, Inc.

\bibitem{Oh2018MultimodalJournaling}
H.~Oh, S.~Soundararajan, J.~Nguyen, and R.~Jain.
\newblock {Multimodal food journaling}.
\newblock In {\em HealthMedia 2018 - Proceedings of the 3rd International
  Workshop on Multimedia for Personal Health and Health Care, co-located with
  MM 2018}, pages 39--47. Association for Computing Machinery, Inc, 10 2018.

\bibitem{ornab2017empirical}
A.~M. Ornab, S.~Chowdhury, and S.~B. Toa.
\newblock {\em An empirical study of collaborative filtering algorithms for
  building a diet recommendation system}.
\newblock PhD thesis, BRAC University, 2017.

\bibitem{Pandey2020PersonalizedSleep}
V.~Pandey, D.~Deepak~Upadhyay, N.~Nag, and R.~Jain.
\newblock {Personalized User Modelling for Context-Aware Lifestyle
  Recommendations to Improve Sleep}.
\newblock Technical report, 2020.

\bibitem{PandeyUbiquitousHealth}
V.~Pandey, N.~Nag, and R.~Jain.
\newblock {Ubiquitous Event Mining to Enhance Personal Health}.

\bibitem{Pandey2018UbiquitousHealth}
V.~Pandey, N.~Nag, and R.~Jain.
\newblock {Ubiquitous event mining to enhance personal health}.
\newblock In {\em UbiComp/ISWC 2018 - Adjunct Proceedings of the 2018 ACM
  International Joint Conference on Pervasive and Ubiquitous Computing and
  Proceedings of the 2018 ACM International Symposium on Wearable Computers},
  pages 676--679, New York, New York, USA, 10 2018. Association for Computing
  Machinery, Inc.

\bibitem{Pandey2020ContinuousRetrieval}
V.~Pandey, N.~Nag, and R.~Jain.
\newblock {Continuous Health Interface Event Retrieval}.
\newblock 4 2020.

\bibitem{Pandey2021EventModelling}
V.~Pandey, A.~Rostami, N.~Nag, and R.~Jain.
\newblock {Event Mining Driven Context-Aware Personal Food Preference
  Modelling}.
\newblock pages 660--676. 2021.

\bibitem{Park2014AssociationsAdolescents}
S.~Park.
\newblock {Associations of physical activity with sleep satisfaction, perceived
  stress, and problematic Internet use in Korean adolescents}.
\newblock {\em BMC Public Health}, 14(1):1143, 11 2014.

\bibitem{Patel2015WearableChange}
M.~S. Patel, D.~A. Asch, and K.~G. Volpp.
\newblock {Wearable devices as facilitators, not drivers, of health behavior
  change}, 2 2015.

\bibitem{Patki2016TheVault}
N.~Patki, R.~Wedge, and K.~Veeramachaneni.
\newblock {The synthetic data vault}.
\newblock In {\em Proceedings - 3rd IEEE International Conference on Data
  Science and Advanced Analytics, DSAA 2016}, pages 399--410. Institute of
  Electrical and Electronics Engineers Inc., 12 2016.

\bibitem{Pearl2009CausalOverview}
J.~Pearl.
\newblock {Causal inference in statistics: An overview}.
\newblock {\em Statistics Surveys}, 3(0):96--146, 2009.

\bibitem{peng2023instruction}
B.~Peng, C.~Li, P.~He, M.~Galley, and J.~Gao.
\newblock Instruction tuning with gpt-4.
\newblock {\em arXiv preprint arXiv:2304.03277}, 2023.

\bibitem{Peuhkuri2012DietQuality}
K.~Peuhkuri, N.~Sihvola, and R.~Korpela.
\newblock {Diet promotes sleep duration and quality}, 5 2012.

\bibitem{Peuhkuri2012DietaryMelatonin}
K.~Peuhkuri, N.~Sihvola, and R.~Korpela.
\newblock {Dietary factors and fluctuating levels of melatonin}.
\newblock {\em Food {\&} Nutrition Research}, 56(1):17252, 1 2012.

\bibitem{Pigeon2010EffectsStudy}
W.~R. Pigeon, M.~Carr, C.~Gorman, and M.~L. Perlis.
\newblock {Effects of a tart cherry juice beverage on the sleep of older adults
  with Insomnia: A pilot study}.
\newblock {\em Journal of Medicinal Food}, 13(3):579--583, 6 2010.

\bibitem{Podszun2023}
M.~Podszun and B.~Hieronimus.
\newblock Can chatgpt generate energy, macro- and micro-nutrient sufficient
  meal plans for different dietary patterns?
\newblock {\em -}, May 2023.

\bibitem{POTTER2021579}
T.~Potter, R.~Vieira, and B.~{de Roos}.
\newblock Perspective: Application of n-of-1 methods in personalized nutrition
  research.
\newblock {\em Advances in Nutrition}, 12(3):579--589, 2021.

\bibitem{worldhungermap}
W.~F. Programme.
\newblock Hungermap live, 2022.

\bibitem{pu2023empirical}
G.~Pu, A.~Jain, J.~Yin, and R.~Kaplan.
\newblock Empirical analysis of the strengths and weaknesses of peft techniques
  for llms.
\newblock {\em arXiv preprint arXiv:2304.14999}, 2023.

\bibitem{qiao2022privacy}
Y.~Qiao, Q.~Sun, H.~Cao, J.~Wang, and T.~Hao.
\newblock Privacy-preserving dish-recommendation for food nutrition through
  edging computing.
\newblock {\em Transactions on Emerging Telecommunications Technologies},
  33(6):e3869, 2022.

\bibitem{raffel2020exploring}
C.~Raffel, N.~Shazeer, A.~Roberts, K.~Lee, S.~Narang, M.~Matena, Y.~Zhou,
  W.~Li, and P.~J. Liu.
\newblock Exploring the limits of transfer learning with a unified text-to-text
  transformer.
\newblock {\em The Journal of Machine Learning Research}, 21(1):5485--5551,
  2020.

\bibitem{Raghunathan2006TheProducts}
R.~Raghunathan, R.~W. Naylor, and W.~D. Hoyer.
\newblock {The Unhealthy = Tasty Intuition and Its Effects on Taste Inferences,
  Enjoyment, and Choice of Food Products}.
\newblock {\em Journal of Marketing}, 70(4):170--184, 10 2006.

\bibitem{rakhmawati2023halal}
N.~A. Rakhmawati, N.~I. Wibowo, T.~P. Rinjeni, S.~S. Indasari, A.~Indriawan,
  and R.~Indraswari.
\newblock Halal food products recommendation based on knowledge graphs and
  machine learning.
\newblock In {\em 2023 International Conference on Converging Technology in
  Electrical and Information Engineering (ICCTEIE)}, pages 65--70. IEEE, 2023.

\bibitem{rhone2017ers}
A.~Rhone and M.~Ver~Ploeg.
\newblock Ers’s updated food access research atlas shows an increase in
  low-income and low-supermarket access areas in 2015.
\newblock Technical report, 2017.

\bibitem{Risso2017ACulture}
D.~S. Risso, C.~Giuliani, M.~Antinucci, G.~Morini, P.~Garagnani, S.~Tofanelli,
  and D.~Luiselli.
\newblock {A bio-cultural approach to the study of food choice: The
  contribution of taste genetics, population and culture}.
\newblock {\em Appetite}, 114:240--247, 7 2017.

\bibitem{Romagnolo2017MediterraneanDiseases}
D.~F. Romagnolo and O.~I. Selmin.
\newblock {Mediterranean Diet and Prevention of Chronic Diseases}.
\newblock {\em Nutrition Today}, 52(5):208--222, 9 2017.

\bibitem{rostami2024food}
A.~Rostami, R.~Jain, and A.~M. Rahmani.
\newblock Food recommendation as language processing (f-rlp): A personalized
  and contextual paradigm, 2024.

\bibitem{rostami2020_personal_food_model}
A.~Rostami, V.~Pandey, N.~Nag, V.~Wang, and R.~Jain.
\newblock {Personal Food Model}.
\newblock In {\em MM 2020 - Proceedings of the 28th ACM International
  Conference on Multimedia}, pages 4416--4424, 2020.

\bibitem{rostami2020personal}
A.~Rostami, V.~Pandey, N.~Nag, V.~Wang, and R.~Jain.
\newblock Personal food model.
\newblock In {\em Proceedings of the 28th ACM International Conference on
  Multimedia}, pages 4416--4424, 2020.

\bibitem{Rostami2020PersonalModelb}
A.~Rostami, V.~Pandey, N.~Nag, V.~Wang, and R.~Jain.
\newblock {Personal Food Model}.
\newblock {\em Proceedings of the 28th ACM International Conference on
  Multimedia}, pages 4416--4424, 8 2020.

\bibitem{Rostami2021WorldProject}
A.~Rostami, Z.~Xie, A.~Ishino, Y.~Yamakata, K.~Aizawa, and R.~Jain.
\newblock {World Food Atlas Project}.
\newblock In {\em Proceedings of the 13th International Workshop on Multimedia
  for Cooking and Eating Activities}, CEA '21, pages 33--36, New York, NY, USA,
  2021. Association for Computing Machinery.

\bibitem{Rostami2021_}
A.~Rostami, Z.~Xie, A.~Ishino, Y.~Yamakata, K.~Aizawa, and R.~Jain.
\newblock {\em {World Food Atlas Project}}, volume~1.
\newblock Association for Computing Machinery, 2021.

\bibitem{Rostami2020}
A.~Rostami, B.~Xu, and R.~Jain.
\newblock {Multimedia Food Logger}.
\newblock In {\em MM 2020 - Proceedings of the 28th ACM International
  Conference on Multimedia}, pages 4548--4549, 2020.

\bibitem{Rostami2020MultimediaLogger}
A.~Rostami, B.~Xu, and R.~Jain.
\newblock {Multimedia Food Logger}.
\newblock In {\em Proceedings of the 28th ACM International Conference on
  Multimedia}, New York, NY, USA, 2020. ACM.

\bibitem{Roy2022}
D.~Roy and M.~Dutta.
\newblock A systematic review and research perspective on recommender systems.
\newblock {\em Journal of Big Data}, 9(1), May 2022.

\bibitem{Rubin2005CausalDecisions}
D.~B. Rubin.
\newblock {Causal inference using potential outcomes: Design, modeling,
  decisions}.
\newblock {\em Journal of the American Statistical Association},
  100(469):322--331, 3 2005.

\bibitem{Saha2017ModelingCampuses}
K.~Saha.
\newblock {Modeling Stress with Social Media Around Incidents of Gun Violence
  on College Campuses}.
\newblock {\em Proc. ACM Hum.-Comput. Interact}, 1, 2017.

\bibitem{Saha2019AUse}
K.~Saha, B.~Sugar, J.~Torous, B.~Abrahao, E.~Kıcıman, M.~De~Choudhury,
  G.~Tech, H.~M. School, N.~Shanghai, and M.~Research.
\newblock {A Social Media Study on the Effects of Psychiatric Medication Use}.
\newblock Technical report, 2019.

\bibitem{Sampasa-Kanyinga2018SleepAdolescents}
H.~Sampasa-Kanyinga, H.~A. Hamilton, and J.~P. Chaput.
\newblock {Sleep duration and consumption of sugar-sweetened beverages and
  energy drinks among adolescents}.
\newblock {\em Nutrition}, 48:77--81, 4 2018.

\bibitem{sanchez2020recommendation}
O.~R. Sanchez, I.~Torre, Y.~He, and B.~P. Knijnenburg.
\newblock A recommendation approach for user privacy preferences in the fitness
  domain.
\newblock {\em User Modeling and User-Adapted Interaction}, 30:513--565, 2020.

\bibitem{schaefer2023large}
M.~Schaefer, S.~Reichl, R.~ter Horst, A.~M. Nicolas, T.~Krausgruber, F.~Piras,
  P.~Stepper, C.~Bock, and M.~Samwald.
\newblock Large language models are universal biomedical simulators.
\newblock {\em bioRxiv}, pages 2023--06, 2023.

\bibitem{Schafer2017TowardsSystems}
H.~Sch{\"{a}}fer, S.~Hors-Fraile, R.~P. Karumur, A.~C. Valdez, A.~Said,
  H.~Torkamaan, T.~Ulmer, and C.~Trattner.
\newblock {Towards health (Aware) recommender systems}.
\newblock In {\em ACM International Conference Proceeding Series}, volume Part
  F128634, pages 157--161, New York, New York, USA, 7 2017. Association for
  Computing Machinery.

\bibitem{schafer2001commerce}
J.~B. Schafer, J.~A. Konstan, and J.~Riedl.
\newblock E-commerce recommendation applications.
\newblock {\em Data mining and knowledge discovery}, 5:115--153, 2001.

\bibitem{Selma2021DeepLF}
B.~A. Selma, B.~Narhimene, and R.~Nachida.
\newblock Deep learning for recommender systems: Literature review and
  perspectives.
\newblock {\em 2021 International Conference on Recent Advances in Mathematics
  and Informatics (ICRAMI)}, pages 1--7, 2021.

\bibitem{9742919}
M.~Shah, S.~Degadwala, and D.~Vyas.
\newblock Diet recommendation system based on different machine learners: A
  review.
\newblock In {\em 2022 Second International Conference on Artificial
  Intelligence and Smart Energy (ICAIS)}, pages 290--295, 2022.

\bibitem{Shi2019GutDiseases}
Z.~Shi.
\newblock {Gut Microbiota: An Important Link between Western Diet and Chronic
  Diseases}.
\newblock {\em Nutrients}, 11(10):2287, 9 2019.

\bibitem{shi2024don}
Z.~Shi and A.~Lipani.
\newblock Don’t stop pretraining? make prompt-based fine-tuning powerful
  learner.
\newblock {\em Advances in Neural Information Processing Systems}, 36, 2024.

\bibitem{Shivappa2019DietInflammation}
N.~Shivappa.
\newblock {Diet and Chronic Diseases: Is There a Mediating Effect of
  Inflammation?}
\newblock {\em Nutrients}, 11(7):1639, 7 2019.

\bibitem{siddik2023collaborative}
M.~B.~S. Siddik and A.~T. Wibowo.
\newblock Collaborative filtering based food recommendation system using matrix
  factorization.
\newblock {\em JURNAL MEDIA INFORMATIKA BUDIDARMA}, 7(3):1041--1049, 2023.

\bibitem{smith2022using}
S.~Smith, M.~Patwary, B.~Norick, P.~LeGresley, S.~Rajbhandari, J.~Casper,
  Z.~Liu, S.~Prabhumoye, G.~Zerveas, V.~Korthikanti, et~al.
\newblock Using deepspeed and megatron to train megatron-turing nlg 530b, a
  large-scale generative language model.
\newblock {\em arXiv preprint arXiv:2201.11990}, 2022.

\bibitem{Spence2015MultisensoryPerception}
C.~Spence.
\newblock {Multisensory Flavor Perception}, 3 2015.

\bibitem{St-Onge2016FiberSleep}
M.~P. St-Onge, A.~Roberts, A.~Shechter, and A.~R. Choudhury.
\newblock {Fiber and saturated fat are associated with sleep arousals and slow
  wave sleep}.
\newblock {\em Journal of Clinical Sleep Medicine}, 12(1):19--24, 2016.

\bibitem{Starke2021PromotingHF}
A.~D. Starke and C.~Trattner.
\newblock Promoting healthy food choices online: A case for multi-list
  recommender systems.
\newblock In {\em IUI Workshops}, 2021.

\bibitem{viennafoodatlas}
O.~Studio.
\newblock Food atlas vienna, 2021.

\bibitem{sun2019bert4rec}
F.~Sun, J.~Liu, J.~Wu, C.~Pei, X.~Lin, W.~Ou, and P.~Jiang.
\newblock Bert4rec: Sequential recommendation with bidirectional encoder
  representations from transformer.
\newblock In {\em Proceedings of the 28th ACM international conference on
  information and knowledge management}, pages 1441--1450, 2019.

\bibitem{NEURIPS2020_1457c0d6}
{T. Brown et al.}
\newblock Language models are few-shot learners.
\newblock In {\em Advances in Neural Information Processing Systems},
  volume~33, pages 1877--1901, 2020.

\bibitem{tran2021recommender}
{T. Tranet al.}
\newblock Recommender systems in the healthcare domain: state-of-the-art and
  research issues.
\newblock {\em Journal of Intelligent Information Systems}, 57:171--201, 2021.

\bibitem{Tanaka2013AssociationsWorkers}
E.~Tanaka, H.~Yatsuya, M.~Uemura, C.~Murata, R.~Otsuka, H.~Toyoshima,
  K.~Tamakoshi, S.~Sasaki, L.~Kawaguchi, and A.~Aoyama.
\newblock {Associations of protein, fat, and carbohydrate intakes with insomnia
  symptoms among middle-aged Japanese workers}.
\newblock {\em Journal of Epidemiology}, 23(2):132--138, 2013.

\bibitem{tao2020utilization}
D.~Tao, P.~Yang, and H.~Feng.
\newblock Utilization of text mining as a big data analysis tool for food
  science and nutrition.
\newblock {\em Comprehensive reviews in food science and food safety},
  19(2):875--894, 2020.

\bibitem{ThayerEnergyExercise}
R.~E. Thayer.
\newblock {Energy, Tiredness, and Tension Effects of a Sugar Snack Versus
  Moderate Exercise}.
\newblock Technical Report~1.

\bibitem{thongsri2022implementation}
N.~Thongsri, P.~Warintarawej, S.~Chotkaew, and W.~Saetang.
\newblock Implementation of a personalized food recommendation system based on
  collaborative filtering and knapsack method.
\newblock {\em Int. J. Electr. Comput. Eng}, 12(1):630--638, 2022.

\bibitem{vermontfoodatlas}
V.~F. to~Plate.
\newblock Vt food system atlas, 2022.

\bibitem{toledo2019food}
R.~Y. Toledo, A.~A. Alzahrani, and L.~Martinez.
\newblock A food recommender system considering nutritional information and
  user preferences.
\newblock {\em IEEE Access}, 7:96695--96711, 2019.

\bibitem{TrangTran2018AnDomain}
T.~N. Trang~Tran, M.~Atas, A.~Felfernig, and M.~Stettinger.
\newblock {An overview of recommender systems in the healthy food domain}.
\newblock {\em Journal of Intelligent Information Systems}, 50(3):501--526, 6
  2018.

\bibitem{trang2018overview}
T.~N. Trang~Tran, M.~Atas, A.~Felfernig, and M.~Stettinger.
\newblock An overview of recommender systems in the healthy food domain.
\newblock {\em Journal of Intelligent Information Systems}, 50:501--526, 2018.

\bibitem{Rostami21PPFM}
{V. Pandey et al.}
\newblock Event mining driven context-aware personal food preference modelling.
\newblock In {\em Pattern Recognition. ICPR International Workshops and
  Challenges}, 2021.

\bibitem{vairale2020recommendation}
V.~S. Vairale and S.~Shukla.
\newblock Recommendation of diet using hybrid collaborative filtering learning
  methods.
\newblock In {\em Advances in Computational Intelligence and Informatics:
  Proceedings of ICACII 2019}, pages 309--318. Springer, 2020.

\bibitem{Valtonen2005EffectsSubjects}
M.~Valtonen, L.~Niskanen, A.~P. Kangas, and T.~Koskinen.
\newblock {Effects of melatonin-rich night-time milk on sleep and activity in
  elderly institutionalized subjects}.
\newblock {\em Nordic Journal of Psychiatry}, 59(3):217--221, 2005.

\bibitem{van1985collaborative}
A.~van Duijnhoven, J.~Korst, and W.~Verhaegh.
\newblock Collaborative filtering with privacy.
\newblock {\em IEEE Transactions on Information Theory}, 31(4):469--472, 1985.

\bibitem{vanMeer2016FoodAge}
F.~van Meer, L.~Charbonnier, and P.~A. Smeets.
\newblock {Food Decision-Making: Effects of Weight Status and Age}, 9 2016.

\bibitem{von2023food}
J.~von Braun, K.~Afsana, L.~O. Fresco, and M.~H.~A. Hassan.
\newblock Food systems: seven priorities to end hunger and protect the planet.
\newblock In {\em Science and innovations for food systems transformation},
  pages 3--9. Springer International Publishing Cham, 2023.

\bibitem{wang2021market2dish}
W.~Wang, L.-Y. Duan, H.~Jiang, P.~Jing, X.~Song, and L.~Nie.
\newblock Market2dish: Health-aware food recommendation.
\newblock {\em ACM Transactions on Multimedia Computing, Communications, and
  Applications (TOMM)}, 17(1):1--19, 2021.

\bibitem{wang-etal-2023-rethinking-evaluation}
X.~Wang, X.~Tang, X.~Zhao, J.~Wang, and J.-R. Wen.
\newblock Rethinking the evaluation for conversational recommendation in the
  era of large language models.
\newblock In H.~Bouamor, J.~Pino, and K.~Bali, editors, {\em Proceedings of the
  2023 Conference on Empirical Methods in Natural Language Processing}, pages
  10052--10065, Singapore, Dec. 2023. Association for Computational
  Linguistics.

\bibitem{wei2021finetuned}
J.~Wei, M.~Bosma, V.~Y. Zhao, K.~Guu, A.~W. Yu, B.~Lester, N.~Du, A.~M. Dai,
  and Q.~V. Le.
\newblock Finetuned language models are zero-shot learners.
\newblock {\em arXiv preprint arXiv:2109.01652}, 2021.

\bibitem{wei2022chain}
J.~Wei, X.~Wang, D.~Schuurmans, M.~Bosma, F.~Xia, E.~Chi, Q.~V. Le, D.~Zhou,
  et~al.
\newblock Chain-of-thought prompting elicits reasoning in large language
  models.
\newblock {\em Advances in Neural Information Processing Systems},
  35:24824--24837, 2022.

\bibitem{wen2024hard}
Y.~Wen, N.~Jain, J.~Kirchenbauer, M.~Goldblum, J.~Geiping, and T.~Goldstein.
\newblock Hard prompts made easy: Gradient-based discrete optimization for
  prompt tuning and discovery.
\newblock {\em Advances in Neural Information Processing Systems}, 36, 2024.

\bibitem{Westermann2007TowardApplications}
U.~Westermann and R.~Jain.
\newblock {Toward a common event model for multimedia applications}.
\newblock {\em IEEE Multimedia}, 14(1):19--29, 1 2007.

\bibitem{WestermannTowardApplications}
U.~Westermann, R.~J.~I. multimedia, and u.~2007.
\newblock {Toward a common event model for multimedia applications}.
\newblock {\em ieeexplore.ieee.org}.

\bibitem{foodwasteatlas}
W.~F. WRAP, World Resources~Institute.
\newblock The food waste atlas, 2022.

\bibitem{wu2020ptum}
C.~Wu, F.~Wu, T.~Qi, J.~Lian, Y.~Huang, and X.~Xie.
\newblock Ptum: Pre-training user model from unlabeled user behaviors via
  self-supervision.
\newblock {\em arXiv preprint arXiv:2010.01494}, 2020.

\bibitem{https://doi.org/10.48550/arxiv.2305.19860}
L.~Wu, Z.~Zheng, Z.~Qiu, H.~Wang, H.~Gu, T.~Shen, C.~Qin, C.~Zhu, H.~Zhu,
  Q.~Liu, H.~Xiong, and E.~Chen.
\newblock A survey on large language models for recommendation, 2023.

\bibitem{wu2023computing}
X.~Wu, X.~Zhu, E.~Baralis, R.~Lu, V.~Kumar, L.~Rutkowski, and J.~Tang.
\newblock On computing paradigms-where will large language models be going.
\newblock In {\em 2023 IEEE International Conference on Data Mining (ICDM)},
  pages 1577--1582. IEEE, 2023.

\bibitem{wu2024personalized}
Y.~Wu, R.~Xie, Y.~Zhu, F.~Zhuang, X.~Zhang, L.~Lin, and Q.~He.
\newblock Personalized prompt for sequential recommendation.
\newblock {\em IEEE Transactions on Knowledge and Data Engineering}, 2024.

\bibitem{XieEventStreams}
L.~Xie, H.~Sundaram, M.~C. P. o.~t. IEEE, and u.~2008.
\newblock {Event mining in multimedia streams}.
\newblock {\em ieeexplore.ieee.org}.

\bibitem{xue2024repeat}
F.~Xue, Y.~Fu, W.~Zhou, Z.~Zheng, and Y.~You.
\newblock To repeat or not to repeat: Insights from scaling llm under
  token-crisis.
\newblock {\em Advances in Neural Information Processing Systems}, 36, 2024.

\bibitem{chen2021personalized}
{Y. Chenet al.}
\newblock Personalized food recommendation as constrained question answering
  over a large-scale food knowledge graph.
\newblock In {\em Proceedings of the 14th ACM International Conference on Web
  Search and Data Mining}, pages 544--552, 2021.

\bibitem{Yamaguchi2013RelationshipRegularity}
M.~Yamaguchi, H.~Uemura, S.~Katsuura-Kamano, M.~Nakamoto, M.~Hiyoshi,
  H.~Takami, F.~Sawachika, T.~Juta, and K.~Arisawa.
\newblock {Relationship of dietary factors and habits with sleep-wake
  regularity}.
\newblock {\em Asia Pacific Journal of Clinical Nutrition}, 22(3):457--465,
  2013.

\bibitem{Yang2019FederatedApplications}
Q.~Yang, Y.~Liu, T.~Chen, and Y.~Tong.
\newblock {Federated machine learning: Concept and applications}.
\newblock {\em ACM Transactions on Intelligent Systems and Technology},
  10(2):1--19, 1 2019.

\bibitem{yao2023beyond}
Y.~Yao, Z.~Li, and H.~Zhao.
\newblock Beyond chain-of-thought, effective graph-of-thought reasoning in
  large language models.
\newblock {\em arXiv preprint arXiv:2305.16582}, 2023.

\bibitem{ye2022unreliability}
X.~Ye and G.~Durrett.
\newblock The unreliability of explanations in few-shot prompting for textual
  reasoning.
\newblock {\em Advances in neural information processing systems},
  35:30378--30392, 2022.

\bibitem{yin2023heterogeneous}
B.~Yin, J.~Xie, Y.~Qin, Z.~Ding, Z.~Feng, X.~Li, and W.~Lin.
\newblock Heterogeneous knowledge fusion: A novel approach for personalized
  recommendation via llm.
\newblock In {\em Proceedings of the 17th ACM Conference on Recommender
  Systems}, pages 599--601, 2023.

\bibitem{zhang2023user}
G.~Zhang.
\newblock User-centric conversational recommendation: Adapting the need of user
  with large language models.
\newblock In {\em Proceedings of the 17th ACM Conference on Recommender
  Systems}, pages 1349--1354, 2023.

\bibitem{zhang2023instruction}
S.~Zhang, L.~Dong, X.~Li, S.~Zhang, X.~Sun, S.~Wang, J.~Li, R.~Hu, T.~Zhang,
  F.~Wu, et~al.
\newblock Instruction tuning for large language models: A survey.
\newblock {\em arXiv preprint arXiv:2308.10792}, 2023.

\bibitem{zhang2023gpt4roi}
S.~Zhang, P.~Sun, S.~Chen, M.~Xiao, W.~Shao, W.~Zhang, K.~Chen, and P.~Luo.
\newblock Gpt4roi: Instruction tuning large language model on
  region-of-interest.
\newblock {\em arXiv preprint arXiv:2307.03601}, 2023.

\bibitem{DBLP:journals/corr/ZhangYS17aa}
S.~Zhang, L.~Yao, and A.~Sun.
\newblock Deep learning based recommender system: {A} survey and new
  perspectives.
\newblock {\em CoRR}, abs/1707.07435, 2017.

\bibitem{10.1145/3285029}
S.~Zhang, L.~Yao, A.~Sun, and Y.~Tay.
\newblock Deep learning based recommender system: A survey and new
  perspectives.
\newblock {\em ACM Comput. Surv.}, 52(1), feb 2019.

\bibitem{zhang2022automatic}
Z.~Zhang, A.~Zhang, M.~Li, and A.~Smola.
\newblock Automatic chain of thought prompting in large language models.
\newblock {\em arXiv preprint arXiv:2210.03493}, 2022.

\bibitem{zheng2023large}
Y.~Zheng, H.~Y. Koh, J.~Ju, A.~T. Nguyen, L.~T. May, G.~I. Webb, and S.~Pan.
\newblock Large language models for scientific synthesis, inference and
  explanation.
\newblock {\em arXiv preprint arXiv:2310.07984}, 2023.

\bibitem{zhiyuli2023bookgpt}
A.~Zhiyuli, Y.~Chen, X.~Zhang, and X.~Liang.
\newblock Bookgpt: A general framework for book recommendation empowered by
  large language model.
\newblock {\em arXiv preprint arXiv:2305.15673}, 2023.

\bibitem{zhou2020laplacian}
J.~Zhou, Z.~Jiang, and S.~Wang.
\newblock Laplacian least learning machine with dynamic updating for imbalanced
  classification.
\newblock {\em Applied Soft Computing}, 88:106028, 2020.

\bibitem{Zulaika2018EnhancingGraphs}
U.~Zulaika, A.~Guti{\'{e}}rrez, and D.~L{\'{o}}pez-de Ipi{\~{n}}a.
\newblock {Enhancing Profile and Context Aware Relevant Food Search through
  Knowledge Graphs}.
\newblock {\em Proceedings}, 2(19):1228, 10 2018.

\end{thebibliography}

\appendix
\chapter{}
\section{Questionnaire to create World Food Atlas Schema} \label{appendix:wfa_questionnaire}
We present the list of potential questions nutritionists, physicians, and dietitians from Stanford University, USA have for the World Food Atlas platform. 

Category 1 - In this category, the main goal is to understand a user's physiology, psychology, their willingness to pay for food (healthy and unhealthy), and how much people enjoyed eating their food.
\begin{itemize}
    \item What did you eat?
    \item Did you cook it or did you buy it?
    \item From where did you buy it?
    \item Did you enjoy this dish? 
    \item Did you do any activity before or after eating this dish? How long and what kind of activity?
    \item Do you measure your blood glucose response to meals? If so, what was your blood sugar 2 hours after your meal?
    \item How many servings did you eat?
    \item How much did this dish cost?
    \item What else did you think about this food? What did you like or dislike about this food in particular?
    \item Did you take any medication with this meal? If yes, what kind and how much?
    \item Did you inject insulin before this meal? If yes, how many units?
    \item How are you feeling today? (Scale 1-5, perhaps)
    \item Are you experiencing any changes to your health? If so, are you experiencing any of the following?
        \begin{itemize}
            \item Numbness, tingling
            \item Frequent urination
            \item Blurry vision
            \item Sudden weight loss or weight gain
            \item Chest pain
            \item Back pain
            \item Dizziness
            \item Other (please explain)
        \end{itemize}
\end{itemize}

Category 2 - In this category, questions related to the nutrition content of food and associated health benefits are asked so as to empower researchers and food producers to make food that is more optimized for people's health.

\begin{itemize}
    \item What is the food’s shelf life?
    \item How is the food prepared? (How does steaming, baking, frying, etc. alter nutrition content?)
    \item Does this food have known health properties?
    \item What is this food’s pre-, pro-, and post-biotic content? (Important for those trying to improve gut health)
    \item What is this food’s fatty acid profile? (e.g., what is its ratio of omega 6 to omega 3 fatty acids?)
    \item Does this food have known disease prevention properties?
    \item Is it a known cause of food intolerance?
    \item Are there pickled/fermented versions of this food item, and what are its health benefits? 
    \item What impact does the food have on insulin levels? 
    \item What is the glycemic index of a food item?
    \item What is the food's insulin index?
    \item Effect on fasting (Does this item disrupt a fast? For instance, some foods like black coffee, unsweetened tea, and foods in the Fast Mimicking Diet do not break a fast)
    \item Effect on triglycerides (Does it lower or increase triglycerides in the body?)
    \item Effect on pre-existing diseases
    \item Distribution location (Are there pathogens/allergens on the site?)
    \item Ingredient interactions (Are there specific compounds in this item that interact with compounds in other ingredients or drugs? For example, grapefruits and statins.)
    \item What is the effect on your cardio-metabolic health? (Does this food item have a known link to heart disease, chronic heart failure, diabetes, kidney disease, etc.?)
    \item What is the vitamin, mineral, and electrolyte composition of the food? 
    \item What effect did the food have on your mental and physical health? 
    \item What effect did the food have on your sleep and stress?
    \item What effect did the food have on your skin? 
    \item What effect did the food have on your optical health?
    \item Where it was grown/raised/produced/obtained? (e.g., the effect of climate, soil content on nutritive content)
\end{itemize}



\end{document}